\documentclass[a4paper,11pt]{article}
\pdfoutput=1 % if your are submitting a pdflatex (i.e. if you have
\usepackage{graphicx}
\usepackage{jheppub} % for details on the use of the package, please
\usepackage{youngtab}
\usepackage{hyperref}
\usepackage{fancybox}
\usepackage{float}
\usepackage{amsfonts}
\usepackage{amsmath}
\usepackage{amsbsy}
\usepackage{graphicx}
\usepackage{amssymb}
\usepackage{amsthm}
\usepackage{bm}
\usepackage{comment}
\usepackage{epsfig}
\usepackage{latexsym}
\usepackage{pdflscape}
\usepackage{color}
\makeatletter
\def\@fpheader{\relax}
\makeatother

\renewcommand{\b}{\bar}

\DeclareSymbolFont{AMSa}{U}{msa}{m}{n}
\DeclareSymbolFont{AMSb}{U}{msb}{m}{n}
\DeclareMathSymbol{\fieldR}{\mathalpha}{AMSb}{"52}

\DeclareMathOperator{\tr}{tr}

\newcommand{\beq}{\begin{eqnarray}}
\newcommand{\eeq}{\end{eqnarray}}
\newcommand{\bea}{\begin{eqnarray}}
\newcommand{\eea}{\end{eqnarray}}
\newcommand{\be}{\begin{equation}}
\newcommand{\ee}{\end{equation}}
\newcommand{\bq}{\begin{equation}}
\newcommand{\eq}{\end{equation}}

\newcommand{\half}{\frac{1}{2}}
\newcommand{\nn}{\nonumber}
\newcommand{\no}{\nonumber}

\def\l{\lambda}

\def\z{\zeta}
\def\t{\tau}
\def\e{\epsilon}
\def\o{\omega}
\def\le{\left}
\def\ri{\right}

\def\half{\frac12}

\def\td{\tilde}
\def\d{\delta}
\def\6{\partial}

\def\a{\alpha}
\def\b{\beta}

\def\g{\gamma}
\def\tr{{\rm Tr}}

\def\le{\left}
\def\ri{\right}
\def\6{\partial}

\title{\boldmath FZZT branes and non-singlets of Matrix Quantum Mechanics}

\author[a]{Panagiotis Betzios,}
\affiliation[a]{Crete Center for Theoretical Physics, Institute for Theoretical and Computational Physics, Department of Physics, University of Crete 71003
Heraklion, Greece.}

\author[b]{Olga Papadoulaki}
\affiliation[b]{Mathematical Sciences and STAG Research Centre, University of Southampton, Highfield, Southampton SO17 1BJ, United Kingdom.}
\emailAdd{pbetzios@physics.uoc.gr}
\emailAdd{O.Papadoulaki@soton.ac.uk}
\abstract{We explore the non-singlet sector of matrix quantum mechanics dual to $c=1$ Liouville theory. The non-singlets are obtained by adding $N_{f}\times N$ bi-fundamental fields in the gauged matrix quantum mechanics model as well as a one dimensional Chern-Simons term. The present model is associated with a spin-Calogero model in the presence of an external magnetic field. In chiral variables, the low energy excitations-currents satisfy an ${SU(2N_{f})}_{\tilde{k}}$ K\v{a}c-Moody algebra at large $N$. We analyse the canonical partition function as well as two and four point correlation functions, discuss a Gross-Witten-Wadia phase transition at large $N, N_f$ and study different limits of the parameters that allow us to recover the matrix model of Kazakov-Kostov-Kutasov conjectured to describe a two dimensional black hole. The grand canonical partition function is a $\tau$- function obeying discrete soliton equations. We finally conjecture a possible dynamical picture for the formation of a black hole in terms of condensation of long-strings in the strongly coupled region of the Liouville direction.}

\keywords{{} \\Matrix Quantum Mechanics, Matrix Models, String Theory, Black holes}
\arxivnumber{1711.04369}

\begin{document} 
\preprint{[CCTP-2017-9, ICTP-IPP 2017/20]}
\maketitle
\flushbottom

\section{Introduction}\label{introduction}

The duality between the singlet sector of gauged Matrix Quantum Mechanics (MQM) and $c=1$ Liouville theory is long known to provide a very powerful complementary description of the physics of this low-dimensional version of string theory. A plethora of observables such as the partition function, scattering amplitudes of tachyons and loop operator correlators have been computed and shown to match in both sides of the correspondence, with the matrix model providing further predictions for the Liouville theory. Arguably the biggest difficulty was in establishing the dictionary between the two formalisms, a programme which has been brought close to completion at least for the singlet sector and the corresponding linear dilaton background\footnote{Some remaining questions related to the physical origin of the so-called leg-pole factors were recently answered~\cite{Balthazar:2017mxh}.}.

Much less is known though for the non-singlet sectors of MQM. The importance of understanding these sectors as well, comes from pertinent questions related to the presence of non-trivial states in the theory such as black holes. The existence of the 2-d black hole solution\cite{Witten:1991yr,Mandal:1991tz} and generalisations thereof~\cite{Mukherji:1991kz}, is long known from the string theory side and has provided a rare understanding of the role that stringy $\a'$ effects have in the physics of black holes. There exists also a proposal for a matrix model dual of the Euclidean black hole involving the addition of Polyakov-Wilson lines around the thermal circle~\cite{Kazakov:2000pm}. The main issue with this description is that it is inherently Euclidean and one does not have a control over the dynamical aspects in the physics of the Lorentzian black hole. In addition there is still some confusion about the thermodynamic interpretation of this model in comparison with results from the effective action of the 2-d string theory~\cite{Kazakov:2001pj}, and whether it can really describe 2-d black holes of arbitrary radius, since its derivation was based on the FZZ-duality~\cite{Fateev} which holds for the specific radius $R=3/2$ near the so-called black hole string correspondence point~\cite{Giveon:2005mi}.
 
In addition to possibly providing new input to these questions, a real time description would in principle serve as an arena for attacking problems related to the black hole formation and evaporation (information paradox)~\cite{Hawking:1976ra}, understanding the fast scrambling/chaotic behaviour of black holes, elucidating the nature of the black hole interior~\cite{Papadodimas:2015jra} and the role of state-dependence, and finally might even shed some light on spacelike singularities. We believe that even though we are quite far from achieving such a goal in the context of 2-d string theory, nevertheless the first steps have already been laid out in the works connecting the simplest non-trivial matrix model sectors such as the adjoint, with the presence of the so called FZZT branes~\cite{Fateev:2000ik,Teschner:2000md} that extend along the Liouville direction $\phi$. Some related quantum mechanical models that involve non trivial representations are of the spin-Calogero type~\cite{Minahan:1992ht,Minahan:1993mv} and have considerably enriched symmetries, spectrum and dynamics. 
An interesting fact about these models is that they can be derived by adding extra fundamental/antifundamental fermionic or bosonic degrees of freedom in addition to the $N \times N$ matrix $M_{i j}$ of MQM, and can thus also be thought of as models of open-closed string theory~\cite{Minahan:1992bz,Klebanov:2003km}. 

In this paper we analyze further the properties of this kind of models, containing a set of  $N_f \times N$ fermionic $\lbrace \psi_{\a  i}\, , \, \chi_{\beta j} \rbrace$, or bosonic $\lbrace V_{\a  i}\, , \, W_{\beta j} \rbrace$ matrices\footnote{The reason for two sets of fields is that they can transform either under the fundamental or anti-fundamental representation of $SU(N_f)$. They are also bi-fundamentals under the combined $SU(N_f) \times U(N)$ symmetry.}, interacting indirectly with the $N \times N$ matrix $M_{i j}$ of MQM through a non dynamical gauge field $A_{i j}$. These bifundamental matrices describe the degrees of freedom of open strings stretched between the FZZT and ZZ branes. In several steps, we will compare the physical properties of considering fermionic vs bosonic fields. We will also provide further evidence that one can indeed describe the 2-d black hole of~\cite{Kazakov:2000pm}, by resorting to a double scaling limit involving $N_f$ and the mass of the fundamentals/antifundamentals in addition to taking the usual double scaling limit of MQM.  

The main difficulty present, is in developing an appropriate dictionary between string theory and these matrix models that will allow for asking sharp questions for the dual emergent spacetime. For the usual MQM there are two main ways to connect with target space physics namely the collective field formalism~\cite{Das:1990kaa} which should be thought of as the string field theory of tachyons and the bi-local fermionic field theory formalism~\cite{Dhar:1992hr} that is in principle more powerful but harder for explicit computations. The collective field theory formalism for spin Calogero models and adjoint/fundamental Matrix models has been considered already in~\cite{Avan:1995sp,Avan:1996vi,Aniceto:2006rr} but is also quite complicated. In particular one should find new interesting states that solve the non-linear string field equations and then expand in fluctuations of the string fields. We believe that further work on this topic could give an impetus in answering the aforementioned questions from a target space point of view.

On the contrary, it is important to note that in order to answer questions related to thermalisation and a possible chaotic behaviour of such models, one does not need a fully developed dictionary and it is enough to compute appropriate two and four point correlators~\cite{Maldacena:2015waa} and analyse their properties along the lines of what has been achieved for the SYK model~\cite{Maldacena:2016hyu}. Even though the class of models we study are generically integrable, one can imagine a slight deformation of them that could result in a more complicated behaviour, such as a small anisotropy in the magnetic field of the spin-Calogero model we find in section~\ref{matrixmodel}. Nevertheless the integrable point of the parameter space is expected to be able to capture correctly the number of microstates of the dual geometry.

In relation to this point, we should also mention that similar models of adjoint interacting with fundamental degrees of freedom have been studied in the past and recent years~\cite{Iizuka:2008hg,Iizuka:2008eb,Michel:2016kwn} and the calculation of correlators in these models indicated that one needs a large number of flavors in order to have a chance to satisfy the refined criteria of chaotic behavior stemming from the four point function, a conclusion which seems in line with the scaling limit we take (where both $N\, , N_f \rightarrow \infty$), in order to have a strong backreaction effect that reproduces the matrix model partition function of the 2-d black hole and some generalisations thereof.

\paragraph{Structure of the paper and results -}  

In section~\ref{matrixmodel} we provide the Lagrangian and Hamiltonian description of our model and show how it can be related to a spin - Calogero model in the presence of an external constant magnetic field. This allows a complete characterisation of the specific MQM non-singlet representations that are activated due to the extra dynamical bi-fundamental fields. Introducing chiral variables and by defining currents associated to the symmetries present, one finds that at large $N$ they satisfy a K\v{a}c-Moody $SU(2N_{f})_{k}$ algebra, making a contact with WZW models, in accordance with the study of~\cite{Dorey:2016mxm}. This also allows to formulate a collective field theory description at large $N$. Since this discussion lies somewhat outside the main scope of the paper we relegate it to appendix~\ref{excitations}.

In section~\ref{partitionfunction}, we analyse the canonical partition function using different methods. The most compact expression is in terms of symmetric polynomials (see also the appendix~\ref{partitionssymmetric}). An equivalent description, is in terms of $U(N)$ characters. This later form makes it clear that the bi-fundamental fields effectively induce winding/vortex perturbations of arbitrary winding number. 

The expression of the partition function in terms of symmetric polynomials allows to directly take the large $N$ limit. One then finds that only a single rectangular representation contributes. More interesting physics is provided by the Veneziano scaling limit $N,  N_{f}\rightarrow\infty$ and $N_f/N = \text{const}$ that we analyse in subsection~\ref{LargeNlimit}. To study this limit, it is more convenient to introduce the density of eigenvalues on the unit circle in line with~\cite{Aharony:2003sx,Schnitzer:2004qt}. In this limit and keeping only the first winding mode, we find a Gross-Wadia-Witten~\cite{Gross:1980he,Wadia:1980cp} type phase transition (that is really a crossover for finite values of the parameters). This is expected to correspond to the string black hole crossover~\cite{AlvarezGaume:2005fv,AlvarezGaume:2006jg} in the Liouville dual of our model. 

In section~\ref{Correlationfunctions}, we continue with an analysis of two and four point correlation functions of bi-fundamental fields at finite temperature. This analysis is performed in the most interesting double scaling limit of large $N, N_f$. The retarded two point function exhibits the expected behaviour of early time exponential decay, with subsequent oscillatory behaviour in the high temperature phase. In the lower temperature phase the decay follows a slower power law. On the other hand for the four point function we cannot really probe the early time behaviour of the OTOC correlator. Nevertheless, the late time behaviour is captured by the fourier transform of two \emph{universal expressions} - the sine kernel - for really late times $t \gg N \beta$ and the - Airy kernel - for shorter timescales ($N \beta \gg t \gg \beta$). The sine kernel leads to a well studied ramp-plateau behaviour for the correlator (this is normally encountered in studies of the spectral form factor - here  it appears in the four point function of fundamentals that is related to a two-point function of the density of eigenvalues on the circle).

In section~\ref{Liouvilleconnections}, we discuss the relation of the present model with Liouville theory in the presence of boundaries and FZZT branes. A contact is made by an analysis of the winding modes on both sides; this then indicates that the mass $m$ of the bi-fundamentals is related to the Liouville parameter $\sigma$ parametrising the boundary cosmological constant via the formulae $\mu_B = \sqrt{\mu} \cosh \pi \sigma$ and $\sigma = 2 m$. In addition our analysis of the grand canonical partition function in subsection~\ref{grandcanonicalhierarchies} and appendix~\ref{hierarchies}, shows that one can define a complex string coupling parameter through the combination $\mu + i k$ ($k$ is a Chern-Simons level in the MQM side), that takes into account fluxes sourced by the FZZT branes. This concludes our preliminary connection between matrix model and Liouville theory/ FZZT brane parameters.

An important aspect of several sections of this work, is that one can perform a scaling limit that isolates the first winding modes in the expression for the partition function, see section~\ref{Liouvilleconnections}. This limit relates our model to the one of~\cite{Kazakov:2000pm} proposed to describe the physics of the two dimensional black hole~\cite{Witten:1991yr,Mandal:1991tz}. In subsection~\ref{grandcanonicalhierarchies} and appendix~\ref{hierarchies}, we show that the grand canonical partition function is a $\tau$ function of the Toda integrable hierarchy. While the most general partition function obeys discrete soliton equations, in the scaling limit that keeps the first winding modes these reduce to the simpler Toda differential equation. These equations need to be supplemented by Virasoro constraints. We discuss the relevant constraints in appendix~\ref{virasoroappendix}, but leave as a future work the task to solve them in conjuction with the discrete soliton equations.

\section{The Matrix Model}\label{matrixmodel}

In this section we briefly review the gauged version of matrix quantum mechanics (MQM) that describes a set of $N$ unstable ZZ branes~\cite{McGreevy:2003kb,McGreevy:2003ep}, before adding to the model $N_f$ FZZT branes described by the addition of extra fundamental/antifundamental fermionic or bosonic fields into the action.
\\
\\
The gauged MQM model is defined by
\begin{equation}\label{D0}
S_{MQM}=\int dt \, \tr \left(\frac{1}{2}\left(D_{t}M\right)^{2} - V(M) \right),
\end{equation}
where the covariant derivative with respect to the gauge group is $D_{t}M={\partial}_{t}M-i\left[A,M\right]\,$ with $V(M)= -\half M^2$ in the relevant case of $c=1$ Liouville theory described by the double scaling limit of MQM that focuses in the unstable inverted harmonic oscillator potential near the tip~\cite{Klebanov:1991qa,Boulatov:1991xz}. The theory has an $SU(N)$ gauge symmetry 
\begin{eqnarray}\label{gauge}
M(t)\rightarrow U(t)M(t)U^{\dagger}(t),\qquad A(t)\rightarrow  U(t)A(t)U^{\dagger}(t)+i  U(t){\partial}_{t}U^{\dagger}(t).
\end{eqnarray}
The conserved current associated to this symmetry is $J= - i [M, \, \dot M]$, note that it is traceless and its diagonal $U(1)$ part is trivial as one can also see from the transformation $M( t)\rightarrow U(t)M(t)U^{\dagger}(t)$.
One can set the gauge field to zero, but needs to impose the Gauss law constraint $\delta S/ \delta A=  -i[M, \, \dot M]=0$. This has the effect of projecting to the singlet sector where $J=0$.

In a more general context, the ungauged system decomposes into different irreducible representations of $SU(N)$ algebra classified by $J$~\cite{Polychronakos:2006nz}. More precicely, the allowed representations $\mathcal{R}$ should be such that they can be constructed taking products of the adjoint~\cite{Maldacena:2005hi}. A constraint on the possible representations comes from the fact that the diagonal part of $A$ does not appear in the action which leads to the selection rule that only zero weight states should be considered~\cite{Boulatov:1991xz}. This is again in accord with the fact that J is traceless. For the general representation, after diagonalising $M= U \Lambda U^\dagger$, $J= U [\Lambda, [\Lambda, \, A]] U^\dagger = U K U^\dagger$ with $A=i U \dot{U}^\dagger$, the Lagrangian becomes
\be
L= \half \sum_i \dot{\lambda}_i^2 +\half \sum_{i \neq j} \frac{K^\mathcal{R}_{i j} K^\mathcal{R}_{j i}}{(\lambda_i - \lambda_j)^2} - \sum_i V(\lambda_i)\, .
\ee
For the singlet sector ($K=0$) the second Calogero-type term is not there and the system reduces to N-free particles in the potential V. In the Hamiltonian picture, after redefining the wavefunction as $\tilde{\Psi}(\lambda)= \Delta(\lambda) \Psi(\lambda)$ one finds
\be
\hat{H} \tilde{\Psi}(\lambda)= \left[\sum_i^N -\half \frac{\partial^2}{\partial \lambda_i^2} + V(\lambda_i) + \half \sum_{i \neq j} \frac{K_{i j}^\mathcal{R} K_{j i}^\mathcal{R}}{(\lambda_i - \lambda_j)^2} \right]\hat{P} \tilde{\Psi}(\lambda)
\ee
with $\hat{P}$ a projector to zero weight\footnote{Equivalently to the zero $U(1)^{N-1}$ charge sector.} states. 

In particular the singlet sector describes N non interacting fermions in the potential $V(\lambda)$. The singlet sector has been shown to describe the physics of $c=1$ Liouville theory upon taking a special double scaling limit of WKB type~\cite{Kazakov:1988ch}. The bosonic string is described by an unstable cubic potential while the $0B$ superstring by a quartic potential. Taking the double scaling limit is equivalent to filling the fermi sea near the unstable tip of the potential and in this limit the model becomes Gaussian and exactly solvable. For more details we refer the reader to the excellent existing reviews~\cite{Klebanov:1991qa,Ginsparg:1993is}.
\\
\\
We will now consider an extension of this model. We want to ``feed in'' non-singlet representations while still keeping the model gauged. To achieve this we can add extra fields into the action. In particular we will be interested into adding two extra fermionic or bosonic matrices, $\lbrace \psi_{\alpha i}\, , \, \chi_{\beta i} \rbrace$ or $\lbrace V_{\a  i}\, , \, W_{\beta j} \rbrace$ with  $\alpha \in [1, N_f],\, \beta \in [1, \bar{N}_f]$, $i \in [1, N]$, transforming under the fundamental/anti-fundamental representation of $U(N)$ ($\psi \rightarrow U \psi, \, \, \chi  \rightarrow  \chi U^\dagger$) , ($V \rightarrow U V, \, \, W \rightarrow  W U^\dagger$), as well as under a global $SU(N_f)\times SU(\bar{N}_f)$-flavor like symmetry, which we symbolise with the Greek indices. The relevant fermionic action is
\be
S_{f} = \int dt \sum_a^{N_f} \tr \left( i \psi_\a^\dagger D_t \psi_\a -m_\a \psi_\a^\dagger \psi_\a  \right) + \sum_\beta^{\bar{N}_f}  \tr \left( i \chi_\beta^\dagger D^\dagger_t \chi_\beta - m_\b \chi_\b^\dagger \chi_\b \right) \, ,
\ee
with $D_t  \psi_\a = \partial_t  \psi_\a - i A  \psi_\a$ the covariant derivative for the fundamental. Similarly the bosonic action ($V, W$ are complex) is
\be
S_{b} = \int dt \sum_a^{N_f} \tr \left( i V_\a^\dagger D_t V_\a -m_\a V_\a^\dagger V_\a \right) +  \sum_\b^{\bar{N}_f} \tr \left( i W_\b^\dagger D^\dagger_t W_\b - m_\b W_\b^\dagger W_\b \right) \, .
\ee
Note that the fundamentals/antifundamentals are coupled only indirectly through the gauge field $A$ to the matrix $M$. We keep in mind the general possibility of having different masses for each flavor and different number of flavors for the (anti)-fundamentals but we will mostly focus on the equal number of flavors/equal masses case. This affects the global symmetry of the system which varies from $SU(N_f) \times SU(\bar{N}_f)$ when all masses are equal to $U(1)^{ N_f+ \bar{N}_f}$ in case that all the masses are different. Although it is possible to add terms of the form $\psi^\dagger F(M) \psi$, in the double scaling limit where we focus at the tip of the potential, they would lead to a renormalization of the mass term of the $N_f \times N$ matrix fields. Nevertheless it might be interesting to study such extensions in more detail.  We are now also in the position to augment the $SU(N)$ symmetry to $U(N)$, by adding the 1d Chern-Simons term $S_{CS} = k \int dt \tr A$~\cite{Polychronakos:1991bx}\footnote{This term will add topological charge into the system. See~\ref{partitionfunction} for more details.}. It is inconsistent to add this term in the usual MQM action, since one cannot fullfill the Gauss-law constraint $J+ k \mathbb{I}=0$ with finite dimensional matrices. In contrast, adding extra fermionic/bosonic degrees of freedom into the action makes it possible~\cite{Polychronakos:1991bx,Polychronakos:2001mi} as we will soon describe.
 
Similar models have been proposed in the past for the description of open-closed string theory~\cite{Minahan:1992bz} (albeit the single flavored non-gauged version) as well as in the context of the matrix model description of FZZT-branes~\cite{Gaiotto:2005gd} (where again the single flavored model is studied in some detail). The extra Chern-Simons term is not encountered in these studies\footnote{Exept in the case of type $0A$ string theory, which we will briefly discuss in chapter~\ref{Canonical}.}, but it can certainly be present and we would like to understand its physical implications. This term recently appeared in~\cite{Dorey:2016mxm,Dorey:2016hoj} and we will comment on the connection with these works in section~\ref{chiralvariables}. Finally, let us also note that $S_f$ arises in more modern studies of Grassmann matrix models~\cite{Anninos:2015eji,Anninos:2016klf,Tierz:2017nvl} which consider quartic interactions between the Grassmann matrices. In these later models after performing a Hubbard-Stratonovich transformation, the theory becomes quadratic in the presence of an ``effective" gauge field $A^{eff}_{ij}$ and thus similar to the fermionic sector of our model\footnote{One should note though that the ``effective'' gauge field is massive in that case.}. We thus conclude that the present model is a generalization of such models studied previously in the literature and blends several of their features. In the end of this section we will find out that it belongs to the general category of spin-Calogero models which are reviewed in~\cite{Polychronakos:2006nz} with extra kinetic terms for the fundamentals and in the presence of an external magnetic field.
\\
\\
We will now consider the total action with the fermionic and Chern-Simons term  $S_{MQM}+S_{CS}+S_{f}$ and keep an equal number of flavors for simplicity. The bosonic case can be treated in the same way just by replacing anti-commutators with commutators in the appropriate formulae. One can immediately find the Gauss-law constraint to be
\be
i[M, \, \dot{M}] = \sum_\a^{N_f} \psi_\a \psi^\dagger_\a + \chi_\a^\dagger \chi_\a  + k \mathbb{I} \, .
\ee
where in the quantum version one can interpret this constraint as projecting to the singlet representation of the global Hilbert space. Upon quantizing the theory the fermions acquire the usual anticommutation relations $[\psi_{\a i}, \, \psi^\dagger_{\b j} ]_+ = \delta_{\a \b} \delta_{i j}$ and $[\chi_{\a i}, \, \chi^\dagger_{\b j} ]_+ = \delta_{\a \b} \delta_{i j}$. One then also needs to consider the normal-ordered form of the constraint. By taking its trace, it is easy to see that the normal ordering ambiguity can shift/renormalise the value of the Chern-Simons level. Our normal ordering is defined according to the discussion in~\cite{Minahan:1992bz}. Hence, in the quantum version of the model we explicitly have
\be 
\label{constraint}
:J_{i j}: =  -k \delta_{i j} + \sum_{\a}^{N_f} \left[ \psi^\dagger_{\a j} \psi_{\a i} - \chi_{\a i}^\dagger \chi_{\a j}  \right]\, .
\ee
Taking its trace one finds
\be 
\label{contrace}
\sum_i^N  :J_{ii}:=0 \quad \Rightarrow \sum_\a^{N_f} \sum_i^N  \psi^\dagger_{\a i}  \psi_{\a i} - \chi_{\a i}^\dagger \chi_{\a i}  =  k N 
\ee
which means that the parameter $k$ should be quantised. This equation implies that the number of boxes in the Young tableaux should be an integer multiple of N. In addition we also have the stronger form
\be
\label{levelmatching} 
\sum_\a^{N_f}   \psi^\dagger_{\a i}  \psi_{\a i} - \chi_{\a i}^\dagger \chi_{\a i}  =  k \, , \quad \quad  \forall i\, .
\ee
We will now show how one can connect this model with the well studied spin Calogero models~\cite{Polychronakos:2006nz}. 
Using fermionic bilinears, one can define the following $SU(N_f)$ flavor-current operators ($i$ is like a ``lattice site''- not to be summed over)
\bea
(T^\psi_i)^A &=& \psi^\dagger_{\a i} T^A_{\a \b} \psi_{\b i}\, \, \Rightarrow \quad (T^{\psi}_i)_{\ a \b} =\half \left( \psi^\dagger_{\a i} \psi_{\b i} - \frac{\delta_{\a \b}}{N_f} \sum_\g  \psi^\dagger_{\g i} \psi_{\g i} \right)\, , \nn \\
(T^\chi_i)^A &=& \chi_{\a i} T^A_{\a \b} \chi^\dagger_{\b i}\, \, \Rightarrow \quad(T^{\chi}_i)_{\ a \b} = \half \left(  \chi_{\a i} \chi^\dagger_{\b i} - \frac{\delta_{\a \b}}{N_f} \sum_\g  \chi_{\g i} \chi^\dagger_{\g i} \right)\, , 
\eea
which satisfy the $SU(N_f)$ algebra $[T_{\a \b}\, , T_{\g \d}] = T_{\a \d} \delta_{\b \g} - T_{\g \b} \delta_{\a \d}\,$ independently and transform in the same way under $U(N)$. One can also combine them in a single operator $(T_i)^A = \Psi^\dagger_{m \a i} T^A_{\a \b} \Psi_{m \b i}$ with $\Psi^\dagger_m = (\psi^\dagger, \chi)$. %By acting with $\psi^\dagger_{\a i}\, , \chi^\dagger_{\a i}$ on the vacuum $|0 \rangle$, one can generate totally antisymmetric representations of $SU(N_f)\times SU(N_f)$ subject to the level-matching constraint~\ref{levelmatching}. For example if $k=0$ then some possible allowed states have the form $  (\psi^\dagger_{\a_1 i_1}\, \chi^\dagger_{\a_1 i_1})^{l_1} ...(\psi^\dagger_{\a_k i_k}\, \chi^\dagger_{\a_k i_k})^{l_k} |0 \rangle $.
Using a Schwinger-Wigner representation, one then defines also the following spin operators which satisfy an $SU(2)$ algebra for each lattice site and flavor  
\be
S^a =  \Psi^\dagger_m \frac{\sigma^a_{m n}}{2} \Psi_n \quad \Rightarrow \quad \begin{aligned} S^1_{\a i} & = \half \left(\psi^\dagger_{\a i} \chi^\dagger_{\a i} + \chi_{\a i} \psi_{\a i} \right) \\ S^2_{\a i} & = \frac{i}{2} \left(\chi_{\a i} \psi_{\a i} -\psi^\dagger_{\a i} \chi^\dagger_{\a i} \right) \\ S^3_{\a i} & = \half \left(\psi^\dagger_{\a i} \psi_{\a i} - \chi_{\a i} \chi^\dagger_{\a i} \right)
\end{aligned}
\ee
with $\sigma^a_{m n}, \, a= (1,2,3)$ the three Pauli matrices and $\Psi^\dagger_m = (\psi^\dagger, \chi)$\footnote{One could form different realisations of the algebra, but this choice of basis makes more natural the transformation of operators under $U(N)$.}. The nomenclature stems from the fact that this $SU(2)$ can also be thought of as a spin $SU(2)$ for which $\Psi_{\uparrow} = \psi$ and $\Psi_{\downarrow} = \chi^\dagger$. Finally, one can define the spin-flavor operators
\be
G^{a A} = \Psi^\dagger_{m \a} \frac{\sigma^a_{m n}}{2} T^A_{\a \b} \Psi_{n \b}\, .
\ee
The set of all these operators satisfies an $SU(2N_f)$ algebra, much similarly as in the non-relativistic quark model~\cite{Manohar:1998xv}, in the form
\bea
\left[G^{a A}\, , G^{b B}\right] & = & \frac{i}{2 N_f} \e_{a b c} \delta_{A B} S^{c} + \frac{i}{4} f_{A B C} \delta_{a b} T^C + \frac{i}{2} \e_{a b c} d_{A B C} G^{c C}, \nn \\
\left[ S^{a} \, , G^{b B}\right] & = & i \e_{a b c} G^{c B}, \nn \\
\left[T^{A} \, , G^{b B}\right] & = & i f_{A B C} G^{b C}\, ,
\eea
with $f_{A B C}$ the totally antisymmetric structure constants and $d_{A B C}$ the totally symmetric ones.
\\
\\
Integrating out the gauge field in the path integral after completing squares~\cite{Minahan:1992bz}, or passing again to the Hamiltonian picture and imposing the constraint, one can then write the Calogero-type interaction term in terms of the above operators.
In particular using the constraint~\ref{constraint} the form of the Calogero interaction term is found to be 
\be
\hat{H}^f_C \, = \, \half \sum_{i \neq j} \frac{\le(  \psi^\dagger_{\a j} \psi_{\a i} - \chi^\dagger_{\a i} \chi_{\a j} \ri) \le(  \psi^\dagger_{\b i} \psi_{\b j} - \chi^\dagger_{\b j} \chi_{\b i} \ri)}{(\lambda_i - \lambda_j)^2} \, .
\ee
Then, using the operators above, one can also write this expression as follows ($\frac{1}{N_f}\sum_\a^{N_f} S^a_{\a i}  = \bar{S}^a_i$)
\be
\hat{H}^f_C \, = \, \half \sum_{i \neq j} \frac{\frac{1}{2N_f}(N_f+k)^2 -\bar{S}^a_{i} \bar{S}^a_{j} -  T^A_i T^A_j - 4 G^{a A}_i G^{a A}_j }{(\lambda_i - \lambda_j)^2} \, ,
\ee
in terms of spin, flavor and spin-flavor operators with ferromagnetic long-range couplings. As a comparison the bosonic model carries through with very similar steps, the difference being in that the Calogero interaction term the long-range couplings are now anti-ferromagnetic. This part of the Hamiltonian has an $SU(2N_f)$  symmetry, which is broken through the mass terms $m_\a \tr\left( \psi^\dagger_\a\psi_\a + \chi_\a^\dagger \chi_\a\right)$. Since this form of the Calogero Hamiltonian is a bit non standard, it is quite illuminating to note that one can transform it in a more conventional spin Calogero form\footnote{We wish to thank A. Polychronakos for mentioning this possibility to us.} using the observation that we can equivalently define it in terms of $2N_f$-flavors of the fundamental fields $\Psi^\dagger_{\tilde{\a} i}= (\psi^\dagger_{\a i}, \chi_{\a i})$ by renormalizing both the vacuum energy and the Chern-simons level of the coupling as $\tilde{k} = k + N_f$ so that the constraint reads 
\be 
\label{constraint2}
:J_{i j}: =  -\tilde{k} \delta_{i j} + \sum_{\tilde{\a}}^{2N_f}  \Psi^\dagger_{\tilde{\a} j} \Psi_{\tilde{\a} i} \, .
\ee
This needs also to be supplemented with the condition $m_{\tilde{\a}} = (m_\a , - m_\a)$ for the mass term $m_{\tilde{\a}}  \Psi_{\tilde{\a} i}^\dagger \Psi_{\tilde{\a i}}$. The transformation in the bosonic case differs in the shift $\tilde{k}= k - N_f$. After this transformation the relevant terms in the Hamiltonian become
\bea
H_b = \half \sum_{i \neq j} \frac{\tilde{k}(\tilde{k}+2N_f)/2N_f + 2 S^{\tilde{A}}_{i} S^{\tilde{A}}_{j} }{(\lambda_i - \lambda_j)^2} \, + \sum_{i \tilde{A}} B^{\tilde{A}} S^{\tilde{A}}_{i} \nn \\
H_f = \half \sum_{i \neq j} \frac{\tilde{k}(2N_f- \tilde{k})/2N_f - 2 S^{\tilde{A}}_{i} S^{\tilde{A}}_{j} }{(\lambda_i - \lambda_j)^2} \, + \sum_{i \tilde{A}} B^{\tilde{A}} S^{\tilde{A}}_{i}\, ,
\label{Spinchain}
\eea
which is an $SU(2N_f)$ spin-Calogero model in the presence of a non-abelian ``magnetic'' field $B= \sum_{\tilde{A}} B^{\tilde{A}} T^{\tilde{A}}$ with $T^{\tilde{A}},\, \tilde{A}=1,...(2N_f)^2-1$ the $SU(2N_f)$ generators. In our specific case, $B= \text{diag} \lbrace m_{\tilde{a}}\rbrace$ is a diagonal matrix.  The mass term-magnetic field partially lifts the degeneracy of the energy states of the spin Calogero model, for more details see~\cite{Polychronakos:1993wc}. These models although integrable, have a highly intricate spectrum. In terms of allowed representations, eqn.~\ref{constraint2} indicates that one should consider representations having $\tilde{k} \times N$ boxes. Then one should decompose the $\tilde{k}$-fold symmetric/anti-symmetric tensor product rep of the N spins into $SU(2N_f)$ irreps,  having at most $2N_f$ rows/columns for the bosonic/fermionic case.  The full model then decomposes into a direct sum of standard spin Calogero models, where the spins carry different admissible representations according to the previous rules. We also conclude that in terms of the Young tableaux that characterize the allowed representations of $U(N)$, the bosonic/fermionic cases are simply related by flipping the tableaux along the diagonal. %The previous viewpoint in terms of boxes/anti-boxes for fundamentals/antifundamentals is fully equivalent to this.
Let us also finally mention that one can also take and combine symmetric reps for boxes with antisymmetric ones for anti-boxes or equivalently taking bosonic fundamentals with fermionic antifundamentals. One then has a supersymmetric Calogero model. In appendix~\ref{excitations} we elaborate more on the simplest gauge invariant excitations-currents that are found to obey an ${SU(2N_{f})}_{\tilde{k}}$ K\v{a}c-Moody algebra at large $N$. Based on this symmetry of the currents, we also provide a short description of the appropriate collective field theory describing their large $N$ dynamics.

\section{Partition function}\label{partitionfunction}

\subsection{Canonical ensemble}\label{Canonical}

In this section we will analyze the partition function of the model with $S=S_{MQM}+S_{CS}+S_f$.
We chose to integrate out the Grassmann matrices $\psi, \chi$ as well as the matrix $M$ to derive an effective action for the gauge field $A$ to be reduced to eigenvalues. Equivalently, one might try to compute the partition function with Hamiltonian methods as in~\cite{Dorey:2016hoj}. We will need to work in Euclidean time $\tau$ with period $\beta$. The gauge field $A$ and the Matrix $M$ are periodic functions of Euclidean time, while the fermions antiperiodic. We will choose to consider a refinement of the partition function where each fermion flavor has its own mass $m_\a$. In the end we can always restrict to the same mass case where $m_\a = m, \, \forall \a$. Integrating out the fermions we get
\bea
Z_f [A] &=& \int \mathcal{D} \psi^\dagger \mathcal{D} \chi^\dagger \mathcal{D} \psi \mathcal{D} \chi e^{- \oint d\tau \tr \left( \psi_\a^\dagger ( D_{\tau}+ m_\a ) \psi_\a + \chi_\a^\dagger ( D_{\tau}+m_\a ) \chi_\a  \right)} \, , \nn \\
&=& \prod_{\a=1}^{N_f} {\det} \left[\partial_{\tau} + i A  + m_\a \ri] {\det} \left[\partial_{\tau} - i A + m_\a \ri] \, .
\eea
The case of complex bosons similarly gives
\be
Z_b [A] =\prod_{\a=1}^{N_f}  {\det}^{-1} \left[\partial_{\tau} + i A  + m_\a \ri] {\det}^{-1} \left[\partial_{\tau} - i A + m_\a \ri] \, ,
\ee
These are functional determinants in the space of anti-periodic/periodic functions on $S^1$ respectively.
One can also integrate out the matrix $M$ to find
\be
Z_{MQM} [A] = \int \mathcal{D} M e^{- \oint \half \tr \le((D_{\tau}M)^2 + \omega^2 M^2 \ri) } = {\det}^{-\half} \le(-D_{\tau}^2 + \omega^2 \ri)\, .
\ee
Explicit computations with similar functional determinants can be found in~\cite{Aharony:2003sx,Betzios:2016lne}. For completeness we also carry out their computation in appendix~\ref{functdet}. Let us also note that all the resulting functional determinants are invariant under the original gauge symmetry $A(\tau)\rightarrow  U(\tau)A(\tau)U^{\dagger}(\tau)+i  U(\tau){\partial}_{\tau}U^{\dagger}(\tau)$. The total partition function is computed via
\be
Z_N^{(N_f)}=  \int \mathcal{D} A Z_{f/b}[A] Z_{MQM}[A] e^{-i k \oint \tr A}\, .
\ee
This has the schematic form $\int \mathcal{D} A f(A)$. It can be reduced to an integral over the zero modes of $A$. A thorough derivation can be found in~\cite{Osbornnotes}. Upon diagonalising $A=U \a  U^\dagger + i U \dot{U}^\dagger$, with $\a= diag(\a_1, ..\a_N)$ one finds
\be
\int \mathcal{D}A f(A) =\frac{1}{\beta^{N^2}} \int  d\mu_{U(N)}(\beta \a) f(\a)
\ee
It is then convenient to use the angle variables $\theta _{i}=\alpha_{i}\beta$, with $\theta_i \in [0, 2\pi]$. The appropriate $U(N)$ measure $d\mu_{U(N)}(\theta)$ is
\be
 d\mu_{U(N)}(\theta)= \frac{1}{(2\pi)^N N!}\prod_i^N d \theta_i \prod_{i<j}^N |e^{i \theta_i} - e^{i \theta_j}|^2 
\ee
with the $1/N!$ corresponding to the discrete Weyl-group $\mathcal{S}_N$ of permuting the N eigenvalues and the $(2\pi)^N$ coming from the stability group $U(1)^{\otimes N}$.

Upon assembling the various terms from appendix~\ref{functdet}, one finds the following form of the partition function as an angular integral in the case of fermions (one simply obtains the bosonic expression by replacing the cosines with sines in the denominator)
\be
Z_N^{(N_f)}=\frac{q^{\frac{N^2}{2}}}{N!}\int_0^{2\pi} \prod_i^N \frac{d \theta_i}{2 \pi} e^{-i k \theta_i} \prod_{i<j} |e^{i \theta_i} - e^{i \theta_j}|^2 \frac{ \prod_{i, a}^{N, N_f} 4 {\cos}  \le(\half(\theta_i -i \beta m_\a) \ri) {\cos} \le(\half(\theta_i +i \beta m_\a) \ri) }{\prod_{i , j}(1- q e^{i \theta_i - \theta_j})}
\ee
where $q=e^{- \beta \omega}$\footnote{This allows to describe both the normal and the inverted harmonic oscillator by setting $\omega=i$.} and the normalization now reproduces the correct normalization of the matrix harmonic oscillator~\cite{Boulatov:1991xz} if we neglect the fermionic fields $\psi, \, \chi$. One notices that the integral is symmetric in $m_\a \leftrightarrow - m_\a$. We can further massage this expression into ($M= \sum_\a m_\a$)
\be\label{fullpartitionfunction}
Z_N^{(N_f)}=\frac{e^{\beta M N}}{N!}\int_0^{2\pi} \prod_i^N \frac{d \theta_i}{2 \pi} e^{-i k \theta_i} \prod_{i<j} |e^{i \theta_i} - e^{i \theta_j}|^2 \frac{q^{\frac{N^2}{2}} \prod_{i, \a}   \le(1 + e^{- \beta m_\a}e^{i \theta_i} \ri) \le(1+ e^{- \beta m_\a}e^{- i \theta_i}  \ri)}{\prod_{i , j}(1- q e^{i \theta_i - \theta_j})}\, .
\ee
\\
One can also arrive to the same expression via a different route, that provides a good check for this expression and gives a more clear basis-independent interpretation of the various terms. In particular one trades the integration over the gauge field to a $U(N)$ twist of the boundary conditions of the fields and an integral over the $U(N)$ twist~\cite{Boulatov:1991xz,Betzios:2016lne}. For example the MQM partition function can be written in terms of the matrix H.O. propagator $G_E(M, M')= \langle M' ; \beta | M ; 0 \rangle_E$ as $Z_{MQM}(\beta)= \int dM d U G_E (M, U M U^\dagger)$. We now treat all the terms in a similar fashion and find for the full partition function ($\det$ is now a matrix determinant)
\bea\label{fullpartitionfunctionabstract}
Z_N^{(N_f)} &\sim& \int \mathcal{D} M \int_{U(N)} \mathcal{D} U G_E (M, U M U^\dagger) \det U^{-k} \prod_\a^{N_f} {\det} (1+e^{-\beta m_\a} U) {\det} (1+e^{- \beta m_\a}U^\dagger)\,  \nn  \\
&=& \int_{U(N)} \mathcal{D} U  \det U^{-k} \frac{\prod_\a^{N_f} {\det} (1+e^{-\beta m_\a} U) {\det} (1+e^{- \beta m_\a}U^\dagger)}{\det(q^{-\half} I \otimes I - q^\half U \otimes U^\dagger)}\, ,
\eea
which matches~\ref{fullpartitionfunction} upon diagonalising $U$ (in the denominator we have the determinant in the tensor product space).
The first interesting fact to notice is that the fermions induce winding perturbations, whose strength is governed by $m_\a$ and $\beta$, in the form of determinant operators. This can be made even more explicit by exponentiating
\be\label{fermionwinding}
\prod_{\a=1}^{N_f}{\det} (1+e^{-\beta m_\a} U) {\det} (1+e^{-\beta m_\a} U^\dagger) = \exp \left[ \sum_{l=1} \frac{(-1)^{l+1}}{l} z_w(l \beta) \left[\tr U^l + \tr U^{-l} \right]  \right]\, ,
\ee
with $z_w(\beta) = \sum_\a^{N_f} e^{- \beta m_\a}$, the single winding mode fugacity\footnote{
A similar winding perturbation involving all winding modes in the context of Liouville theory was also recently found in~\cite{Betzios:2016lne}, having only a $\beta$ dependence.}. Fundamental/antifundamental complex bosons can be treated similarly and give some inverse determinant factors of the form
\be\label{bosonwinding}
\prod_{\a=1}^{N_f}{\det}^{-1} (1-e^{-\beta m_\a} U) {\det}^{-1} (1-e^{-\beta m_\a} U^\dagger) = \exp \left[ \sum_{l=1} \frac{1}{l} z_w(l \beta) \left[\tr U^l + \tr U^{-l} \right]  \right]\, .
\ee
These terms can be interpreted as a grand canonical partition function of positive/negative winding Wilson lines with fugacities $e^{- \beta m_\a}$, in a similar spirit to that of section~\ref{longliouville}. In the first case the statistics is fermionic whereas in the latter bosonic as expected from the nature of the fundamentals.
Moreover if one tunes the total number of bosons and fermions to be equal with equal masses, one finds that the vacuum energy term $e^{\beta M N}$ from the numerator cancels with a similar term in the denominator which is a hint for a supersymmetric point in the moduli space of these kind of models and thus of the supersymmetric version open-closed string theory~\footnote{One could also enhance the adjoint degrees of freedom using the matrix model of~\cite{McGreevy:2003dn} to describe $\mathcal{N}=2$ supersymmetry.}. From a more formal standpoint, these expressions are also the generating functions of characters of the fundamental/antifundamental representation of $U(N)$ , $\tr U^n = \chi_f (U^n), \,  \tr (U^\dagger)^n = \chi_{af}( U^n)$, see appendix~\ref{characterssymmetric}. In the same spirit one can also exponentiate the term in the denominator of~\ref{fullpartitionfunctionabstract} using~\cite{Aharony:2003sx}
\be\label{adjointmatter}
\det(I \otimes I - q U \otimes U^{-1})^{-1} = \exp \left(\sum_{l=1}^\infty \frac{q^l}{l} \chi_{adj}(U^l) \right), \, \quad  \chi_{adj}(U^l) = \tr(U^l) \tr((U^{-1})^l) \, . \nn \\
\ee
This is to be expected since the matrix $M_{i j}$ is just another matter field that transforms in the adjoint representation of $U(N)$. This formalism allows to treat matter fields with Gaussian action in any representation, by just replacing with the appropriate character. Similar models with fundamental and adjoint characters can be found in QCD studies~\cite{Bhanot:1981eb,Chen:1981ut} and more recently in~\cite{Schnitzer:2004qt} and display a quite interesting phase structure at large $N$. We provide more details in the next subsection.
\\
\\
Finally we now turn to the interesting factor $\det U^k$ arising due to the topological Chern-Simons term. This term has already appeared in the Leutwyler Smilga matrix model~\cite{Leutwyler:1992yt} describing the IR physics of finite volume QCD~\cite{Verbaarschot:2000dy}. Similarly to that case, one should interpret the parameter $k$ as labelling different superselection/topological sectors. In addition it is also possible to sum over distinct topological sectors to derive a partition function with $\theta$ angle dependence by $Z(\theta) = \sum_k e^{i \theta k} Z_k$. To understand the string theory meaning of this $\theta$ angle, let us note that in the case of type $0A$ supersymmetric string theory a similar Chern-Simons term has been discussed in~\cite{Douglas:2003up}. In that work it was understood that it arises from the presence of a different number of $D0$-$\bar{D0}$ branes/anti-branes. From the point of view of supersymmetric 2-d string theory it has the interpretation of adding $k$-units of RR electric flux to the system through a bulk term of the form $k \int d \tau d \phi F_{\tau \phi}$. It is natural for us to expect that our term has a similar string theory interpretation, albeit our theory is bosonic or $0B$ and the flux should be understood to be sourced by the spacetime filling FZZT branes. This is in accord with the constraint whose trace measures the difference between the number of fundamental/anti-fundamental fields that describe the FZZT branes/anti-branes. The $\theta$ angle corresponds to the flux, which can be changed continuously in the $Z(\theta)$ ensemble.

\subsubsection{Partition function and symmetric polynomials}

One can rewrite the partition function in terms of symmetric polynomials. The relevant computation is carried out in appendix~\ref{canonicalsymmetricpolynomials}. The computation follows~\cite{Dorey:2016mxm} with a slight extension.
The result for bosons reads
\be\label{fullcanonicalb}
\mathcal{Z}_N^{b}  = q^{N^2/2} \prod_{j=1}^N \frac{1}{1-q^j} \sum_\l K_{\l, (\tilde{k}^N)}(q) s_\l (X) \, .
\ee
with a similar result for fermions given in appendix~\ref{canonicalsymmetricpolynomials}. The sum is over all partitions/representations $\lambda$, $K_{\l, (\tilde{k}^N)}(q)$ are Kostka polynomials and  $s_\l (X)$ are Schur-polynomials of the variables $x_{\tilde{a}} = e^{-\beta m_{\tilde{a}} }$. One should also keep in mind the relations  for bosons/fermions $\tilde{k}=k \mp N_f$ and that $\tilde{m}_{a} = (m_a, - m_a)$. For more details see also the appendix~\ref{partitionssymmetric}.

\subsubsection{Large $N, N_f$ limits}\label{LargeNlimit}

\begin{figure}[t!]
\begin{center}
\includegraphics[scale=0.65]{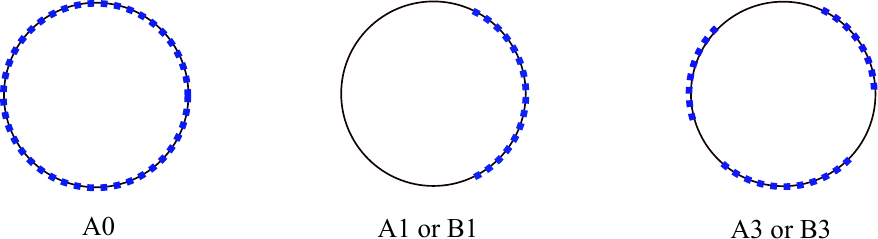}
 \end{center}
 \caption[]{For the $A_{0}$ case the density of eigenvalues has support on the full circle. For $A_{1}$ the support is on an arc. More complicated cases exist with support on many arcs.}
\label{19}
\end{figure}

One can take a large $N$ limit of eqn.~\ref{fullcanonicalb}. Then one finds a single rectangular partition/representation to contribute. The result is
\be
\mathcal{Z}^b_{N \rightarrow \infty} = q^{N^2/2} \prod_{j=1}^N \frac{1}{1-q^j} \chi_{R_{\tilde{k}, C}} (q, X)
\ee
where $\chi_{R_{\tilde{k}, C}}$ is the character of the $\hat{A}_{2 N_f}$ affine Lie algebra at level $\tilde{k}$ associated to the rep $R_{\tilde{k}, C}$ of $SU(2N_f)$. This is the $\tilde{k}$-fold symmetrization of the $C^{th}$ antisymmetric rep. The partition function for the fermions is quite similar and contains the character of the $\tilde{k}$-fold antisymmetrization of the $C^{th}$ symmetric rep. 
\\
\\
This result is nevertheless too simple to exhibit interesting phenomena such as phase transitions. One would expect that there should be a competition between different representations as was found in the Douglas-Kazakov phase transition~\cite{Douglas:1993iia}. In that context one finds the Young tableaux to acquire a continuous shape for large reps and a phase transition when the rectangular shape with a corner ``melts'' to a smooth shape with no sharp edges.
\\
\\
To this end, another very interesting limit that one can examine in our construction is the following (of Veneziano type)
\begin{eqnarray}\label{venezianolimit}
N \quad\text{and}\quad N_{f}\rightarrow\infty \qquad \text{with}\qquad N_f/N = 2 L \quad \text{fixed}
\end{eqnarray}
In such a limit one expects a large number of FZZT branes to condense causing a large backreaction on the closed string background of the ZZ branes. It would be very interesting to study this limit, since it corresponds to the limit of large partitions for the expression~\ref{fullcanonicalb}, when both $\tilde{k}, N$ are large. The results of~\cite{Barns-Graham:2017zpv} could be of use, since they allow a representation for the partition function in which $N$ and $N_f$ are treated in a similar fashion. Instead we will follow a more standard procedure that involves the density of eigenvalues on the circle $\rho(\theta) = \frac{1}{N} \sum_{i=1}^N \delta(\theta - \theta_i)$. The normalization is $\int_{-\pi}^\pi \rho(\theta) = 1$. We also define the moments
\be
\rho_n = \int_{-\pi}^\pi d \theta \rho(\theta) \cos n \theta \, , \qquad |\rho_n| \leq 1 \, .
\ee
Following~\cite{Aharony:2003sx,Schnitzer:2004qt} one can show that the adjoint characters in this limit behave in a similar fashion to the fundamental ones. In particular one needs to find the equilibrium configuration based on the action
\bea 
S(\theta) &=& N \sum_{n=1}^\infty \frac{1}{n} \left[a_n \rho_n + \frac{N_f}{N} b_n \right] \left(\tr U^n + \tr U^{-n} \right)\, , \nn \\
a_n &=& q^n \, , \qquad b_n = (\pm 1)^{n+1} x^n  \, , \qquad x = e^{-\beta m} \, ,
\eea
where we consider the case of equal masses and the $\pm$ refer to bosons/fermions.
The saddle point is given by self-consistently solving
\bea
\frac{\partial S(\theta)}{\partial \theta} &=& 2 \sum_{n=1}^\infty c_n \sin (n \theta) = - \int_{\mathcal{C}} d \theta' \rho(\theta') \cot \half (\theta - \theta') \, , \nn \\
c_n &=& q^n \rho_n + (\pm 1)^{n+1} \frac{N_f}{N} x^n  
\eea
where now all expressions have support on a contour $\mathcal{C}$, since the eigenvalues can generically have support on several segments of the unit circle. In this equation, the Vandermonde makes the eigenvalues to spread, while the potential terms clump them. The result depends on this competition. The two simplest cases are to have support on the full circle ($A_0$ case) or in an arc of the circle ($A_1$ case), see fig.~\ref{19}. It could also saturate in some arcs. The most general case is described in~\cite{Jurkiewicz:1982iz}, with a support on multiple arcs. If one keeps only the first winding mode, one finds a GWW phase transition~\cite{Gross:1980he,Wadia:1980cp} between the $A_0$ and $A_1$ phase. This third order phase transition is encountered before one reaches the Hagedorn temperature. From the dual string theory point of view according to the analysis of~\cite{AlvarezGaume:2005fv,AlvarezGaume:2006jg}, it should correspond to a string-black hole transition that is really a crossover for finite values of the parameters $(N, N_f)$.
In order to suppress the presence of any multi-critical phases and keep only the first winding mode we approach the phase transition from low enough temperatures in the scaling regime (the analogous scaling limit in the grand-canonical ensemble is discussed in section~\ref{2dBH})
\be
 x \ll 1 , \quad  \text{with} \quad \frac{N_f}{N} \, x \, \sim  \, \text{O(1)}\, .
\ee
Using this simplification, the resulting GWW transition happens for~\cite{Schnitzer:2004qt}
\be
q + \frac{2 N_f}{N} x =   1  \, .
\label{phaseboundary}
\ee

An interesting extension would be to analyse all the winding modes together to understand better the global phase structure of the model in the original Veneziano limit of eqn.~\ref{venezianolimit}, when $x = e^{- \beta \mu}$ can also be of $O(1)$. Since multiple arcs can then merge and split as the temperature is varied, one can expect the presence of novel multicritical phases as described in~\cite{Jurkiewicz:1982iz}. 

\section{Correlation functions at finite temperature}\label{Correlationfunctions}

In this section we analyse two and four point correlation functions in our model. This study can be performed independently of the precise dictionary to Liouville theory, since it is of generic interest whether the correlation functions of such models can exhibit characteristics resembling those expected from systems dual to black holes. One might try to perform this computation either at zero or finite temperature. At zero temperature the real time correlators will involve the Calogero model dynamics. Here we will briefly discuss the finite temperature case. We first describe how to compute Euclidean time correlators, the real time ones can be obtained through Wick rotating $t = - i \tau$, keeping in mind that one can obtain both a retarded and advanced correlator depending on the precise rotation. Next we focus on the two and four point functions and obtain explicit results in the large $N, N_f$-limit.

\subsection{Generating functional}

We will now describe the generating functional of correlators for the complex fundamental fields $\psi_{\alpha i}(\tau)$.
We introduce the complex sources $J_{\alpha i}(\tau)$ and define the generating functional of connected correlation functions $W(J)$ through (the trace is over the $U(N)$ indices)
\be
e^{-W(J)} = \mathcal{N} \int \mathcal{D} \Psi^\dagger \mathcal{D} \Psi \mathcal{D} M \mathcal{D} A \, e^{- \oint \half \tr \le((D_{\tau}M)^2 + \omega^2 M^2 \ri) } e^{- \oint d\tau \tr \left( \psi_\a^\dagger ( D_{\tau}+ m_\a ) \psi_\a  +  J^\dagger_{\alpha} \psi_{\alpha} +  \psi_\a^\dagger J_\alpha  \right)} 
\ee
with $\mathcal{N} = e^{- W(0)}$. One can obtain any $\psi_{\alpha i}(\tau)$ correlator by taking appropriate derivatives with respect to $J_{\alpha i}(\tau)$. It is now convenient to integrate out all the Gaussian fields $\psi_{\alpha i}(\tau), M_{i j}(\tau)$ to obtain the following expression as an integral over the non-dynamical gauge field $A_{i j}(\tau)$ and the external sources
\be
e^{-W(J)} = \mathcal{N} \int\mathcal{D} A \, \left[ {\det}\le(-D_{\tau}^2 + \omega^2 \ri) \right]^{- \half} \prod_{\a=1}^{N_f} {\det} \left[\partial_{\tau} - i A  + m_\a \ri]  e^{ \oint d \tau d \tau' \tr \left( J^\dagger_{\a}(\tau) \left[\mathcal{G}^f\right]_{\a  \b} (\tau , \tau ')  J_{\b}(\tau ') \right) } \, .
\ee
This is a gauge invariant functional expression for fermionic fundamentals, it is easy to generalise it for anti-fundamentals or complex bosons by using the appropriate determinants ( i.e. ${\det} \left[\partial_{\tau} + i A  + m_\a \ri]^{-1} $ for fundamental bosons). 

Taking the appropriate derivatives we can compute a generic correlator, and since the theory is gaussian one just needs to consider a product of propagators inserted in the Euclidean path integral and appropriately contracted. We see that the computation boils down to inverting the differential operator~\cite{Anninos:2016klf}
%The result for (anti)-fundamental fermions reads
%\be
%e^{-W(J)} = \mathcal{N} \int_{U(N)} \mathcal{D} U  \frac{\prod_\a^{N_f} {\det} (1+e^{-\beta m_\a} U) {\det} (1+e^{- \beta m_\a}U^\dagger)}{\det(q^{-\half} I \otimes I - q^\half U \otimes U^\dagger)} 
%\ee
%A Chern-Simons term $k \oint A$ can be trivially added to this result by introducing an extra factor $\det U^{-k}$ in the path integral.
\be
\left[\mathcal{G}^f\right]^{-1}_{\a i , \b j} (\tau , \tau') = \delta_{\alpha \beta} \delta(\tau - \tau') \left( \partial_\tau  \delta_{i j} - i A_{i j}(\tau) + m_\alpha \delta_{i j} \right)\, .
\ee 
In the end the result can be written in terms of closed Wilson lines $U =\mathcal{P} e^{i \oint d \tau A}$, as well as open ones with their endpoints contracted with sources since they are the appropriate gauge invariant combinations leaving the endpoints in time fixed. From now on, we focus in the case of equal masses $m_\alpha = m$, to simplify the formulae.
The result for the propagator depending on the time ordering is~\cite{Anninos:2016klf} (we symbolise with $\delta\tau =\tau -\tau'$)
\be
\mathcal{G}^f_{\a i , \b j} (\tau , \tau') =\left\{\begin{array}{lc}  \delta_{\a \b} e^{-m \vert\delta \tau\vert } \mathcal{P} e^{i \int_{\tau'}^{\tau} d \tau A} \left(I + e^{- \beta m} \mathcal{P} e^{ i\oint d \tau A} \right)^{-1} \, , \quad \text{for} \, \, \tau > \tau'\, ,  \\
  - \delta_{\a \b} e^{-m \vert\delta \tau\vert } \mathcal{P} e^{i \int_{\tau'}^{\tau} d \tau A} e^{- \beta m} \mathcal{P} e^{ i\oint d \tau A} \left(I + e^{- \beta m} \mathcal{P} e^{ i\oint d \tau A} \right)^{-1} \, , \quad \text{for} \, \, \tau < \tau'\, . 
 \end{array}\right.
\ee
The result for anti-fundamentals is quite similar, one has to replace with the oppositely oriented Wilson lines $U^\dagger = \mathcal{P} e^{- i \oint A}$.

\subsection{Two-point function}

For two point functions, one just needs to use one propagator of either fundamental or anti-fundamental fields.  As an example we will study the correlator (for $\tau > \tau'$)
\be
\langle \psi_{\alpha i}(\tau) \psi^\dagger_{\beta j}(\tau') \rangle = \langle \mathcal{G}^f_{\a i , \b j} (\tau , \tau') \rangle_A = \delta_{\a \b} e^{-m \delta \tau } \langle \mathcal{P} e^{i \int_{\tau}^{\tau'} d \tau A} \left(I + e^{- \beta m} \mathcal{P} e^{ i\oint d \tau A} \right)^{-1} \rangle_A 
\ee

Passing in the diagonal basis via a gauge transformation
$A(\tau) \rightarrow U(\tau) \alpha U^\dagger(\tau) - i U(\tau) \partial_\tau U^\dagger(\tau) $ one finally finds an integral over the gauge field zero modes $\theta_k = \alpha_k \beta$
\be\label{GP}
\langle \mathcal{G}^f_{\a i , \b j} (\tau , \tau') \rangle = \delta_{\a \b}  \langle U_{i k}(\tau) \frac{ e^{-m \vert\delta \tau\vert } e^{i \vert\delta \tau\vert \theta_k/\beta}}{1+ e^{- \beta m} e^{i \theta_k}} U^\dagger_{k j}(\tau') \rangle_{\theta, U}\,, \quad \tau'<\tau
\ee

\be\label{GN}
\langle \mathcal{G}^f_{\a i , \b j} (\tau , \tau') \rangle =- \delta_{\a \b}  \langle U_{i k}(\tau) \frac{ e^{-m \vert\delta \tau\vert } e^{i \vert\delta \tau\vert \theta_k/\beta}e^{- \beta m} e^{i \theta_k}}{1+ e^{- \beta m} e^{i \theta_k}} U^\dagger_{k j}(\tau') \rangle_{\theta, U}\,, \quad \tau<\tau'
\ee
We now notice that the angular average over $\theta$ is reminiscent of the Chern-Simons deformation, with the role of the C-S level played by $\delta \tau/\beta$. %This indicates that if we want to study the large Euclidean time behavior of the correlator, it might be helpful to analyse the large $k$ physics. 
Upon analytically continuing time $\delta  \tau = i t$, this term would result in an exponential suppression of the correlator, but there is also an interplay with the average over $\theta$. Each term in the denominator will cancel one fundamental contribution (for $\theta_k$) from the partition function integrand. Therefore this correlator can be also thought of as a deformed version of the partition function integral.

We cannot compute this correlator analytically for finite $N, N_f$, but we are mainly interested in the large $N, N_f$ limit where the model exhibits a Gross-Witten-Wadia phase transition. In particular once we introduce the density of states $\rho(\theta) = \frac{1}{N} \sum_{i=1}^N \delta(\theta - \theta_i)$ we can express the $\theta$ part of the integral as 
\be
 \langle \frac{e^{-m |\delta \tau |} e^{i |\delta \tau | \theta_k/\beta}}{1+ e^{- \beta m} e^{i \theta_k}} \rangle_{\theta} \quad \rightarrow \quad \int_{\mathcal{C}} d \theta \rho(\theta) \frac{e^{-m |\delta \tau | } e^{i |\delta \tau | \theta/\beta}}{1+ e^{- \beta m} e^{i \theta}}
\ee
One then expects a possible transition in the properties of this integral due to the different support of the density of states (denoted by  the contour $\mathcal{C}$) among the two phases. 

\begin{figure}[t!]
\begin{center}
\includegraphics[scale=0.50]{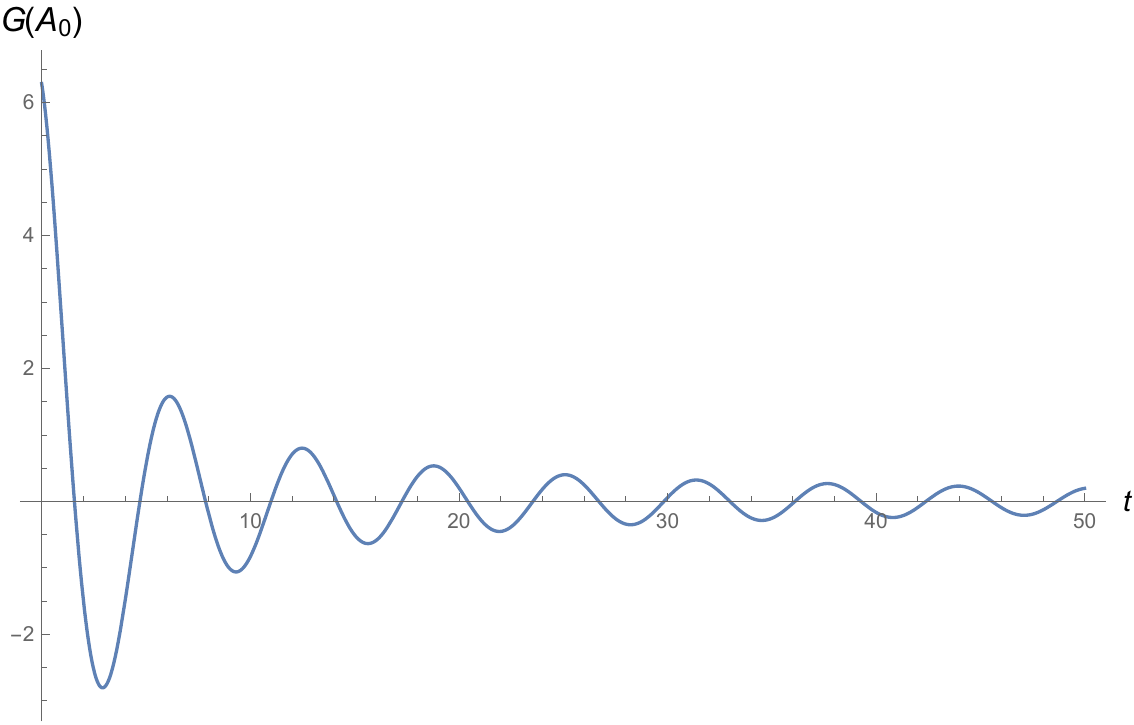}
 \end{center}
 \caption[]{A plot of the two point function in the $A_{0}$ phase ($\mu = 1, \beta = 10, L=1$). It oscillates and decays with a power law.}
\label{G2A0}
\end{figure}

Nevertheless before performing this integral, there is still a problem to be solved. We immediately notice that this correlator is not gauge invariant as it describes an open Wilson line connecting two generic $\psi$ fields and therefore one has to perform extra unitary averages, as can be seen from~\ref{GP} and~\ref{GN}. Notwithstanding this, one can define gauge invariant correlation functions that do not have such an ambiguity by using composite operators, for example $\mathcal{O}_{\alpha \beta}(\tau) =\frac{1}{N} \tr  \psi_\alpha \psi_\beta^\dagger (\tau)$. For such operators the correlators are gauge invariant~\footnote{They still rotate over the global $SU(N_f)$.} and can be reduced by Wick's theorem to products of the elementary field correlator, the final result involving only a single $\theta$ average i.e.
\bea
\langle \mathcal{O}_{\alpha \beta}(\tau) \mathcal{O}_{\gamma \delta}(\tau') \rangle &=&  \frac{\delta_{\alpha \delta } \delta_{\beta \gamma}}{N} \sum_k  \langle \frac{ e^{- m \vert\delta \tau\vert }e^{-i \vert\delta \tau\vert \theta_k/\beta}} {1+ e^{- \beta m} e^{i \theta_k}} \frac{ e^{- m \vert\delta \tau\vert } e^{-\beta m}e^{i\theta_k} e^{-i \vert\delta \tau\vert \theta_k/\beta}}{1+ e^{- \beta m} e^{i \theta_k}} \rangle_{\theta} \nn  \\ 
&=& \delta_{\alpha \delta } \delta_{\beta \gamma}  \int_{\mathcal{C}} \rho(\theta) \frac{ e^{-2 m \vert\delta \tau\vert}e^{-2i \vert\delta \tau\vert \theta/\beta}e^{-\beta m}e^{i\theta}} {\left(1+ e^{- \beta m} e^{i \theta}\right)^{2}}
\eea
We will now need to use the large-$N$ limits of the eigenvalue density in the two phases of the GWW transition.
In the $A_0$ phase one finds the eigenvalue distribution
\be
\rho_{A_0}(\theta) = \frac{1}{2\pi} \left(1+ \sum_{n=1}^\infty 2 \frac{N_f}{N} \frac{(\pm 1)^{n+1} x^n}{1- q^n} \cos n \theta \right)
\ee
If one keeps the first winding mode contribution one finds
\be
\rho_{A_0}^{1\text{st}}(\theta) = \frac{1}{2\pi} \left(1+  2 \frac{N_f}{N} \frac{x}{1-q} \cos \theta \right)
\ee
The density of states for the $A_1$ phase can only be described analytically for the simplified case of a single winding mode. One finds~\cite{Schnitzer:2004qt}
\bea\label{A1phase}
\rho_{A_1}^{1\text{st}}(\theta) &=& \frac{1}{\pi s^2}\left(s^2 - \sin^2 \frac{\theta}{2} \right)^\half  \cos \frac{\theta}{2}\, , \nn \\
s^2 &=&  \left( 1 + \frac{N_f}{N} \frac{x}{q} \right) - \left[\left( 1 + \frac{N_f}{N} \frac{x}{q} \right)^2 - \frac{1}{q} \right]^\half
\eea
with $s = \sin \theta_0/2$,  giving the opening of the gap $\theta \in (- \theta_0, \theta_0)$. One also finds the density of state moments in the two phases
\bea
(\rho_1)_{A_0} = \frac{N_f}{N} \frac{x}{1-q}  \, , \quad (\rho_1)_{A_1} = \left( 1 - \half s^2 \right) 
\eea
In the phase boundary defined by~\ref{phaseboundary} one finds $\theta = \pi$ and therefore $(\rho_1)_{A_0} = (\rho_1)_{A_1} = \half$, $s^2 =1$.
\\
\\
For the $A_0$ phase the integral can be performed exactly in terms of Hypergeometric functions $_2F_1$. The behaviour of the retarded correlator ($\tau = i t$) is oscillating and decays to zero for large real times $t$. The envelope of the decay is power law. A plot can be seen in fig.~\ref{G2A0}.

\begin{figure}[t!]
\begin{center}
\includegraphics[width=0.49\textwidth]{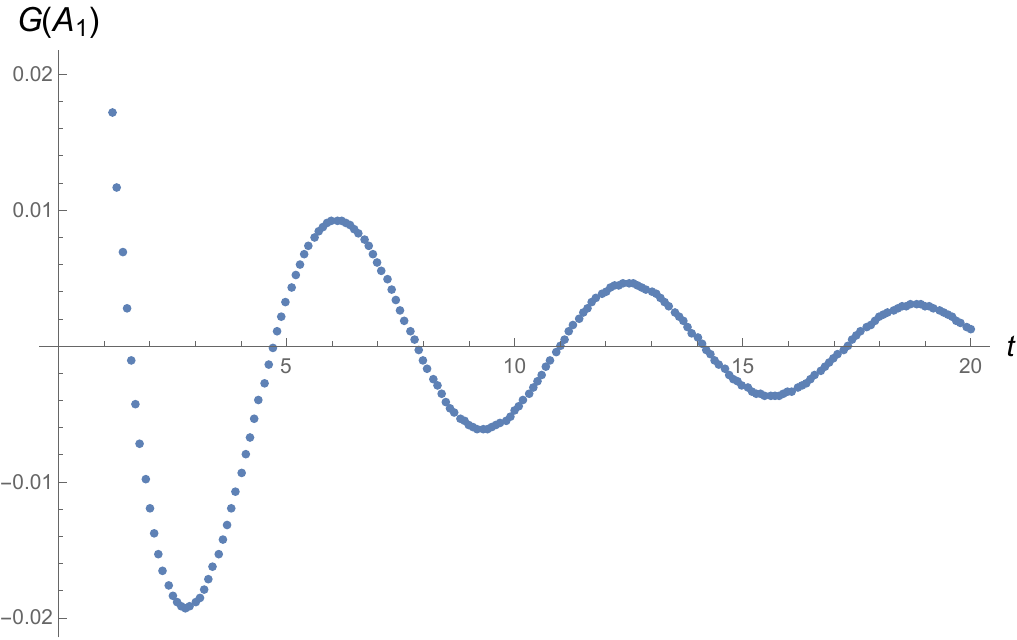}%
\hspace{2mm}\includegraphics[width=0.49\textwidth]{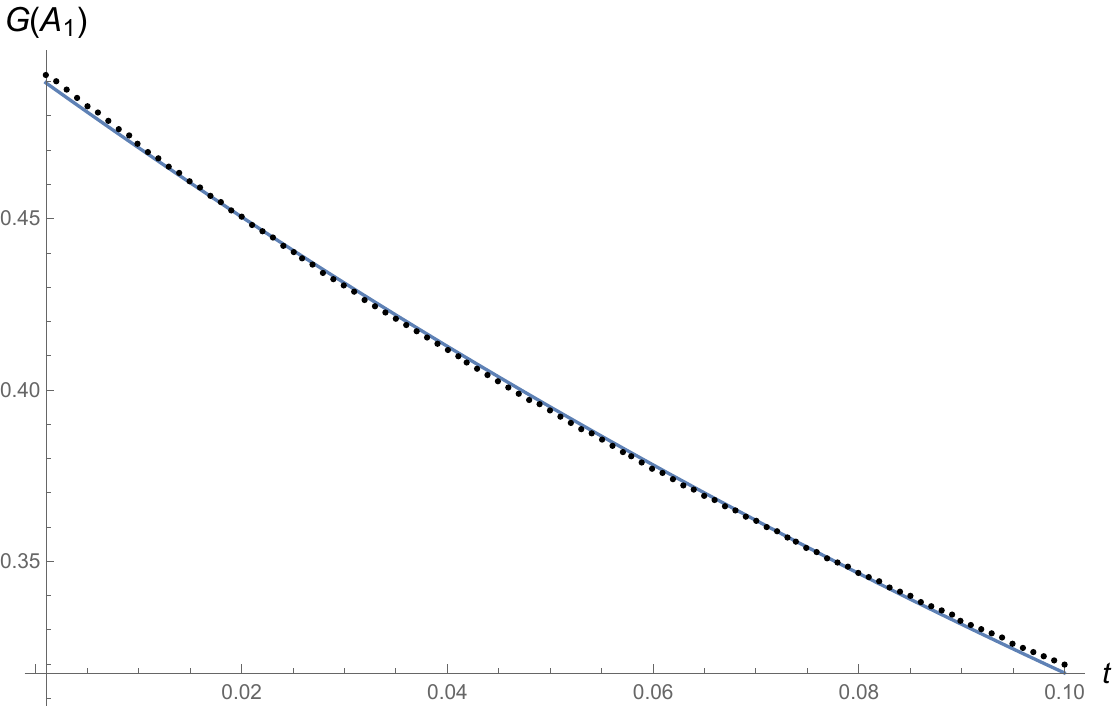}
 \end{center}
 \caption[]{A plot of the two point function in the $A_{1}$ phase ($\mu = 1, \beta = 0.1, L=1$). Left: The late time behaviour is oscillating with a power law decaying envelope. Right: The early time behaviour is very well approximated by a decaying exponential function $G(t) \sim \exp (- 4.38 t)$ drawn with a continuous curve (logarithmic plot).}
\label{G2A1}
\end{figure}

In the $A_1$ phase one can numerically compute the retarded correlator and again find an oscillating-decaying behaviour. In contrast with the $A_0$ phase the decay is now exponential for small times. A plot can be seen in fig.~\ref{G2A1}. The early time curve best fit is an exponentially decaying function. Such a behaviour is reminiscent of two point functions in thermalising systems in a state whose dual gravity description is a black hole.

We conclude with a summary of the results of this section. For low temperatures the retarded large-$N$ scaled correlator oscillates and decays to zero in a power law fashion, while above the GWW phase transition, the decay is much faster and has an exponential behaviour for early times. Finite-$N$ effects are expected to restore the unitarity of the correlator with the presence of persistent $O(1)$ fluctuations. All the integrals involved for finite $N, N_f$, are both compact and finite.
%\end{comment}
%In the next subsection, we proceed to compute the out of time four-point correlation function and see whether the system exhibits any signs of approximate chaotic behaviour, in the large $N, N_f$ limit. %For that one will need to use the density correlation function between the eigenvalues $\rho(\theta, \theta')$ in the two phases. As we mentioned at several times one could expect possible non-trivial behaviour only in the large $N,N_f$ limit where the phase transition is possible. 

\subsection{Four-point function OTOC}
In this subsection, we compute the out of time four point correlation function (OTOC) in order to see if our model exhibits any approximate signs of chaotic behaviour at large $N, N_f$. This computation will thus be performed both in the $A_0$ and $A_{1}$ phases.

For this computation it turns out that we need to use the density correlation function between the eigenvalues $\rho(\theta, \theta')$ in the two phases. As we have mentioned several times one could only expect a possible non-trivial behaviour in the large $N, N_f$ limit where there is a phase transition from $A_{0}$ to $A_{1}$, since for finite values of parameters the results are those of a unitary integrable model. 

One possibility is to compute first the Euclidean four point function for finite $N,N_f$ then analytically continue the exact result using the prescription from~\cite{larkin:1964zz,Roberts:2014ifa,Anninos:2018svg} and finally take the large $N,N_f$ limit. The issue that arises in our case is that we can only perform the integration numerically so we do not have an analytic control of the result of the integral, only of the integrand. A second possibility is to first take the large $N,N_f$ limit on the Euclidean four point function, obtain an analytic result and then continue that in real time. Of course generically we do not expect these two routes to give the same result always. Nevertheless since we are only interested in the large-N dynamics we will follow the second route and find that it produces well defined results.

First we define the appropriate density correlation function between the eigenvalues $\rho(\theta, \theta')$ for $A_{0}$ and $A_{1}$. To achieve this we remember the expression for the density of eigenvalues $\rho(\theta) = \frac{1}{N} \sum_{i=1}^N \delta(\theta - \theta_i)$. For the present Gross-Witten-Wadia model, it can also be expressed in terms of orthonormal Mathieu functions $M_i(\theta)$ as~\cite{Wadia:1980cp} (see appendix~\ref{OrthoMathieu} for the Mathieu equation)
\be
\rho(\theta) = \frac{1}{N} \sum_{i=1}^N M_i(\theta)^2 \, .
\ee
The two point correlation function can be expressed in terms of a connected and a disconnected part
\be\label{thetacorrelation}
\rho(\theta, \theta') = \frac{N}{N-1} \left(\rho(\theta) \rho(\theta') - \rho_c(\theta, \theta')^2 \right)\, , \qquad  \rho_c(\theta, \theta') = \frac{1}{N} \sum_{i=1}^N M_i(\theta) M_i(\theta') \, .
\ee
The connected part of interest is a Christoffel Darboux kernel constructed out of the orthonormal Mathieu functions. For more details see~\cite{Wadia:1980cp}. In the present case we will be interested in the large $N$ limit of these formulae. There exist different ways to take the large $N$ limit depending on the support of eigenvalues. A general discussion for the Christoffel Darboux kernel and its various scaling limits can be found in~\cite{Simon} and references within. In particular if we focus in the bulk of the spectrum we can take the large-$N$ limit as
\bea\label{BulkScaling}
\rho_c(x_1, x_2) &=& \lim_{N \rightarrow \infty}  \rho_c \left( \theta + 2 \pi \frac{x_1}{N}, \theta + 2 \pi \frac{x_2}{N} \right) =   \frac{\sin \left[ \pi (x_1 - x_2)  \right]}{\pi (x_1 - x_2)} \, , \nn \\
&=& \lim_{N \rightarrow \infty}  \frac{N \rho_c(\theta + 2 \pi \frac{x_1}{\rho(\theta)}, \theta + 2 \pi \frac{x_2}{\rho(\theta)})}{\rho(\theta)}
\eea
This is called the \emph{sine-kernel} and governs the eigenvalue level-spacings/correlations in the bulk of the spectral density. Since, in the case of the $A_{0}$ phase we have a distribution of eigenvalues around the whole unit circle and we are always in the bulk of the spectrum, it is the sine-kernel that governs the physics of the appropriate large-$N$ connected correlator.

On the other hand for the $A_{1}$ case the spectral support has two edges. One then finds that the appropriate density correlation function between eigenvalues depends on the way that one takes the large-$N$ limit. Essentially one can again ``zoom" in the bulk of the spectrum to find that the eigenvalue correlations are governed by the sine-kernel, but there is now the option to focus at the edge of the support of the spectrum. Which of the two limits is appropriate will depend on the time scale that we probe the system as we will soon describe. In the second type of scaling though, the resulting kernel depends on the behaviour of the density near the edge of the support for the spectral curve. Typically it takes the form of an Airy or Bessel kernel, in the present case we find from~\ref{A1phase} that the spectral density behaves as $\rho(\theta) \sim \sqrt{\theta_0 - \theta}$. This is a smooth edge behaviour characteristic of the Airy kernel. The appropriate scaling variables in this case are defined by $\theta_0 - \theta \sim N^{- 2/3} x$ and the edge limit of the correlation function is
\be\label{EdgeScaling}  
\rho_{A_{1}}^{edge}(x_1, x_2) = \lim_{N \rightarrow \infty} N^{1/3} \rho_c(\theta_0 +  \frac{x_1}{N^{2/3}}, \theta_0 +  \frac{x_2}{N^{2/3}}) =   \frac{Ai'(x_1) Ai( x_2) - Ai(x_1 ) Ai'(x_2)}{x_1 - x_2}
\ee
where $Ai (x)$ is the Airy function and with primes we denote the derivatives with respect to the argument. The correction terms to this formula are known to be of the order of $N^{-2/3} e^{c(x_1+x_2)}$ with $c$ an $O(1)$ constant.

Most importantly, we would like to stress here that since the occurence of the Sine and Airy kernels is universal and depends only on the generic behaviour of the spectrum in the bulk and edge of its support, our results have a much wider range of validity and our construction can be readily applied to a plethora of time dependent matrix models, for which the partition function and correlators can be reduced to unitary integrals such as~\cite{Aharony:2003sx,Anninos:2016klf}. 

We now turn to the appropriate observables. The simplest possibility is to compute the gauge variant four point function
\bea\label{non_inv}
&&\left\langle \Psi_{\alpha i}(\tau_{1})\bar{\Psi}_{\beta j}(\tau_{2})\Psi_{\gamma k}(\tau_{3})\bar{\Psi}_{\delta l}(\tau_{4}) \right\rangle_{c}=\\
&& \left\langle G^{P}_{\alpha i,\beta j}(\tau_{1}-\tau_{2})G^{P}_{\gamma k,\delta l}(\tau_{3}-\tau_{4})\right\rangle-\left\langle G^{P}_{\alpha i,\delta l}(\tau_{1}-\tau_{4})G^{P}_{\gamma k,\beta j}(\tau_{3}-\tau_{2})\right\rangle\nn
\eea
where the superscript $P$ denotes that the argument in the propagator is positive.

Nevertheless, in order again to surpass such obstacles of gauge-invariance, the connected four-point function that we shall compute is
\be\label{4pointE}
\left\langle \mathcal{O}_{\alpha\beta}(\tau_{1})\mathcal{O}_{\gamma\delta}(\tau_{2})\mathcal{O}_{\epsilon\zeta}(\tau_{3})\mathcal{O}_{\eta\theta}(\tau_{4}) \right\rangle_{c}
\ee
For the OTOC, the Euclidean times are ordered as $\tau_{4}<\tau_{2}<\tau_{3}<\tau_{1}$ and the subscript $c$ symbolises that we are interested in the connected part of the four point function. A way to rotate to the real time configuration respecting the above ordering is to choose $\tau_{i}$ to be equally spaced points on the Euclidean thermal circle and then to evolve only the ones that are at anti-diametrical points i.e 
\bea\label{4point}
&\left\langle \mathcal{O}_{\alpha\beta}(\beta/4)\mathcal{O}_{\epsilon\zeta}(it)\mathcal{O}_{\gamma\delta}(-\beta/4)\mathcal{O}_{\eta\theta}(-\beta/2 + it) \right\rangle_{c}=\nn\\
&\tr\, e^{- \beta H} \left[e^{-\beta H/4}\mathcal{O}_{\alpha\beta}(0) e^{-\beta H/4}\mathcal{O}_{\epsilon\zeta}(t)e^{-\beta H/4}\mathcal{O}_{\beta\gamma}(0) e^{-\beta H/4}\mathcal{O}_{\delta\epsilon}(t)\right]
\eea

Starting from \eqref{4pointE} and taking all possible contractions remembering to anti-commute any time that a fermion is exchanged, we have 

\bea\label{4p}
&&\left\langle \mathcal{O}_{\alpha\beta}(\tau_{1})\mathcal{O}_{\gamma\delta}(\tau_{2})\mathcal{O}_{\epsilon\zeta}(\tau_{3})\mathcal{O}_{\eta\theta}(\tau_{4}) \right\rangle_{c} = \\
&& \left\langle G_{\gamma j, \beta i}^{N}\left(\tau_{2}-\tau_{1}\right)G_{\alpha i, \delta j}^{P}\left(\tau_{1}-\tau_{2}\right)G_{\eta l, \zeta k}^{N}\left(\tau_{4}-\tau_{3}\right)G_{\epsilon k, \theta l}^{P}\left(\tau_{3}-\tau_{4}\right)\right\rangle\nn+\\
&& \left\langle G_{\gamma j, \theta l}^{P}\left(\tau_{2}-\tau_{4}\right)G_{\eta l, \delta j}^{N}\left(\tau_{4}-\tau_{2}\right)G_{\epsilon k, \beta i}^{N}\left(\tau_{3}-\tau_{1}\right)G_{\alpha i, \zeta k}^{P}\left(\tau_{1}-\tau_{3}\right)\right\rangle\nn+\\
&& \left\langle G_{\gamma j, \zeta k}^{N}\left(\tau_{2}-\tau_{3}\right)G_{\epsilon k, \delta j}^{P}\left(\tau_{3}-\tau_{2}\right)G_{\alpha i, \theta l}^{N}\left(\tau_{1}-\tau_{4}\right)G_{\eta l, \beta i}^{P}\left(\tau_{4}-\tau_{1}\right)\right\rangle\nn-\\
&& \left\langle G_{\alpha i, \delta j}^{P}\left(\tau_{1}-\tau_{2}\right)G_{\gamma j, \zeta k}^{N}\left(\tau_{2}-\tau_{3}\right)G_{\epsilon k, \theta l}^{P}\left(\tau_{3}-\tau_{4}\right)G_{\eta l, \beta i}^{N}\left(\tau_{4}-\tau_{1}\right)\right\rangle\nn-\\
&& \left\langle G_{\alpha i, \zeta k}^{P}\left(\tau_{1}-\tau_{3}\right)G_{\epsilon k, \theta l}^{P}\left(\tau_{3}-\tau_{4}\right)G_{\eta l, \delta j}^{N}\left(\tau_{4}-\tau_{2}\right)G_{\gamma j, \beta i}^{N}\left(\tau_{2}-\tau_{1}\right)\right\rangle\nn-\\
&& \left\langle G_{\alpha i, \theta l}^{P}\left(\tau_{1}-\tau_{4}\right)G_{\eta l, \delta j}^{N}\left(\tau_{4}-\tau_{2}\right)G_{\gamma j, \zeta k}^{N}\left(\tau_{2}-\tau_{3}\right)G_{\epsilon k, \beta i}^{N}\left(\tau_{3}-\tau_{1}\right)\right\rangle\nn-\\
&& \left\langle G_{\alpha i, \theta l}^{P}\left(\tau_{1}-\tau_{4}\right)G_{\eta l, \zeta k}^{N}\left(\tau_{4}-\tau_{3}\right)G_{\epsilon k, \delta j}^{P}\left(\tau_{3}-\tau_{2}\right)G_{\gamma j, \beta i}^{N}\left(\tau_{2}-\tau_{1}\right)\right\rangle\nn-\\
&& \left\langle G_{\alpha i, \zeta k}^{P}\left(\tau_{1}-\tau_{3}\right)G_{\epsilon k, \delta j}^{P}\left(\tau_{3}-\tau_{2}\right)G_{\gamma j, \theta l}^{P}\left(\tau_{2}-\tau_{4}\right)G_{\eta l, \beta i}^{N}\left(\tau_{4}-\tau_{1}\right)\right\rangle\nn-\\
&& \left\langle G_{\alpha i, \delta j}^{P}\left(\tau_{1}-\tau_{2}\right)G_{\gamma j, \theta l}^{P}\left(\tau_{2}-\tau_{4}\right)G_{\eta l, \zeta k}^{N}\left(\tau_{4}-\tau_{3}\right)G_{\epsilon k, \beta i}^{N}\left(\tau_{3}-\tau_{1}\right)\right\rangle\nn \, ,
\eea
where we use the appropriate propagator depending on the sign of the arguments denoted by $P,N$ for positive or negative respectively.

As we have stated above for the appropriate analytic continuation, we make the following identifications for the times: $\tau_{1}=\beta/4, \tau_{2}=-\beta/4, \tau_{3}= it,  \tau_{4}= -\beta/2 + it$. We will also label as $\tau = i t$ the Euclidean counterpart of the real time evolution $t$.

\begin{figure}[t!]
\begin{center}
\includegraphics[scale=0.50]{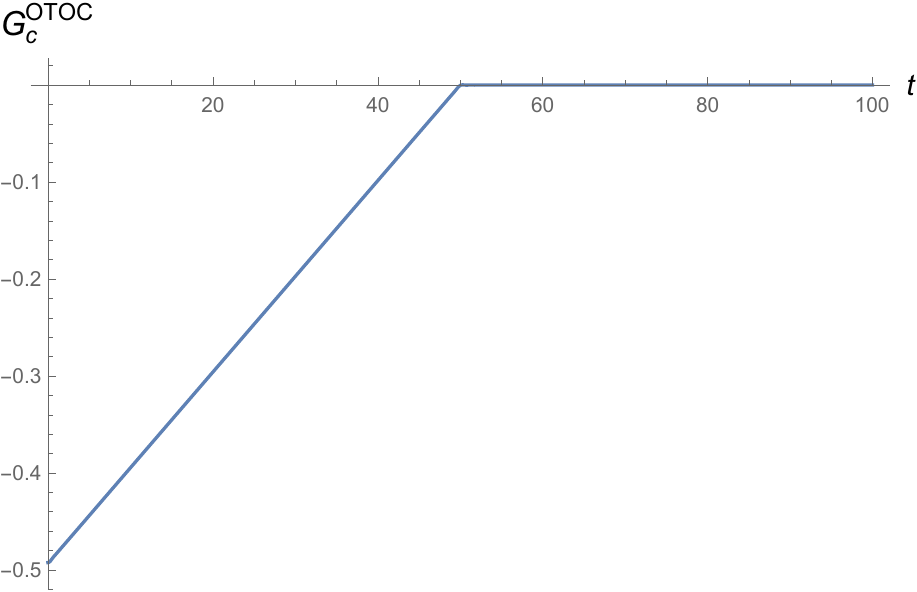}
 \end{center}
 \caption[]{A plot of the late time ($t \gg \beta $) connected OTOC at large $N,N_f$, $\beta = 0.1$. It exhibits a ramp-plateau behaviour.}
\label{G4ramp}
\end{figure}

From \eqref{4p} the part that gives the connected OTOC is the third term which using the aforementioned identifications reads

\bea\label{OTOC}
F(t)&=& \left\langle G_{\gamma j, \zeta k}^{N}\left(\tau_{2}-\tau_{3}\right)G_{\epsilon k, \delta j}^{P}\left(\tau_{3}-\tau_{2}\right)G_{\alpha i, \theta l}^{N}\left(\tau_{1}-\tau_{4}\right)G_{\eta l, \beta i}^{P}\left(\tau_{4}-\tau_{1}\right)\right\rangle \nn \\
&=& \frac{\delta_{\gamma\zeta}\delta_{\epsilon\delta}\delta_{\alpha\theta}\delta_{\eta\beta}}{N^{2}}\left\langle\frac{e^{-\beta m}e^{i\theta_k} e^{-2m(\tau_{3}-\tau_{2})}e^{2i\theta_k \frac{(\tau_{3}-\tau_{2})}{\beta}}}{\left(1+e^{-\beta m}e^{i\theta_k}\right)^2}\frac{e^{-\beta m}e^{i\theta_l'} e^{-2m(\tau_{1}-\tau_{4})}e^{2i\theta_l' \frac{(\tau_{1}-\tau_{4})}{\beta}}}{\left(1+e^{-\beta m}e^{i\theta_l'}\right)^2}\right\rangle_{\theta,\theta'} \nn \\
&=& \delta_{\gamma\zeta}\delta_{\epsilon\delta}\delta_{\alpha\theta}\delta_{\eta\beta} \int_{\mathcal{C}} d\theta \int_{\mathcal{C}} d\theta' \rho_{A_{i}}(\theta,\theta') \frac{e^{-4\beta m}e^{i(\theta+\theta')}e^{2t \frac{(\theta'-\theta)}{\beta}}e^{i\frac{1}{2}(3\theta'+\theta)}}{\left(1+e^{-\beta m}e^{i\theta}\right)^2\left(1+e^{-\beta m}e^{i\theta'}\right)^2} \, .
\eea
with $\rho_{A_i}(\theta, \theta')$ given by~\ref{thetacorrelation}. To compute \eqref{OTOC} in the bulk scaling regime we change variables $\theta\rightarrow \theta^* + \frac{x_{1} 2\pi}{N}$ and  $\theta^{\prime}\rightarrow \theta^* + \frac{x_{2} 2\pi}{N}$, with $x_{1,2}\in\left[ -N/2,N/2 \right]$. Then the integral becomes ($\tau = i t$)
\bea\label{OTOCFS1}
F(\tau ; N) &=&  \int_{-N/2}^{N/2} d x_1 \int_{-N/2}^{N/2} d x_2  \, \rho_{A_{i}}^2 \left(\frac{ 2\pi x_{1}}{N}, \frac{2\pi x_{2}}{N} \right) \frac{e^{-4\beta m}e^{\frac{2\pi i (x_{1}+x_{2})}{N} } e^{- i \frac{2 \tau}{\beta}\frac{2\pi (x_{2}-x_{1})}{N}} e^{\frac{i\pi}{N}(x_{1}+3x_{2})}}{\left(1+e^{-\beta m} e^{i\frac{2 \pi}{N}x_{1}}\right)^2\left(1+e^{-\beta m} e^{i\frac{2 \pi}{N}x_{2}}\right)^2} \, . \nn \\
\eea 
At large-$N$ the kernel has the limit $\rho_{A_{i}}(\theta,\theta')  \rightarrow \rho_{sine}^2(x_{1},x_{2})$. The integral then simplifies and the universal leading contribution is given by
\be\label{OTOCS2}
F(\tau) \sim   \int_{-\infty}^{\infty} d x  \,  \left( \frac{\sin \left( \pi x  \right)}{\pi x} \right)^2 e^{-4\beta m}  e^{\frac{- 2 i \tau}{\beta}\frac{2\pi x}{N}} 
\ee
Let us note that this expression holds in the limit $N \rightarrow \infty$ together with $ \beta << \tau \sim  N c$ as can be seen from the first line of~\ref{OTOCFS1} and~\ref{OTOCS2}. By a change of variables this is also equivalent to
\be
F(\tau) \sim \lim_{N\rightarrow\infty} \int_{-1}^{1} d y  \,  \frac{\sin^2(N \pi y/2)}{\pi^2 y^2} e^{-4\beta m}  e^{- i \frac{\tau}{\beta} {2\pi y}}  
\ee
A similar universal expression appears in analyses of the spectral form factor~\cite{Cotler:2016fpe}, but in the present case it corresponds to a four-point function.
%We can reach the bulk scaling regime following~\ref{BulkScaling} to obtain
%\be\label{OTOCr}
%F(t)= \delta_{\gamma\zeta}\delta_{\epsilon\delta}\delta_{\alpha\theta}\delta_{\eta\beta} \int_{-\infty}^\infty d x_1 \int_{-\infty}^\infty d x_2  \, \rho_{sine}^2(x_1, x_2)  \frac{e^{-4\beta m}e^{i(\theta+\theta')}e^{2t \frac{(\theta'-\theta)}{\beta}}e^{i\frac{1}{2}(3\theta'+\theta)}}{\left(1+e^{-\beta m}e^{i\theta}\right)^2\left(1+e^{-\beta m}e^{i\theta'}\right)^2} \, .
%\ee
%where we defined the scaled time $\tilde{t} = t/N $. 
We notice the relative minus sign in~\ref{thetacorrelation}. This minus sign leads to an increase in the correlator with respect to the factorised value that is decreasing being the product of two two-point functions. This integral can be performed analytically and then rotated to real time $t = i \tau$. The resulting behaviour is presented in fig.~\ref{G4ramp}. It exhibits the well known ramp-plateau behaviour. The connected contribution to the correlator starts at a negative value and increases linearly in time up to $t =  N \beta/2$. After that time the connected correlator is zero and the total correlator reduces to its factorised value. Usually such a behaviour is found in an analysis of the \emph{spectral form factor} two point function, but here it corresponds to the four point function of gauge invariant operators made out of fundamental fields. We thus do not observe the exponential increase that is indicative of a fast scrambling chaotic behaviour in the $A_0$ phase, at least for such types of correlation functions. Moreover, the scaling limit that we take has an additional issue, since in this limit we cannot really probe the very small timescales $t \sim O(\beta)$ of the correlator.

%Another option is not to take the scaling limit that leads to the sine-kernel, but instead to keep the finite-N summation. In this approach we will need integrals of the form
%\be
%I_{m n}(\tau) = \int_{\mathcal{C}} d \theta M_m(\theta) M_n(\theta) e^{I A \tau \theta} 
%\ee
%The final result is expressed as
%\be
%F(\tau) = \sum_{m,n = 0}^{N-1} I_{m n}(\tau) I_{m n}(- \tau)
%\ee

On the other hand in the $A_1$ phase we have the emergence of an extra scale due to the finite support of $\rho(\theta)$. In particular we now have the new regime $\beta << t \sim N^{2/3} c'$ and the regime $N^{2/3} c' << t \sim N c $. In the second regime the correlator exhibits the behaviour of the $A_0$ phase discussed previously. In the first regime, we need to use the Airy kernel. 

For the Airy scaling regime we first find using $\theta = \theta_0 +  N^{- 2/3} x$
\bea
F(\tau ; N) =  \int_{- N^{2/3} \theta_0}^{N^{2/3} \theta_0} d x_1 \int_{- N^{2/3} \theta_0}^{N^{2/3} \theta_0} d x_2  \, \frac{\rho_{A_1}^2(\theta_0 +  \frac{x_1}{N^{2/3}}, \theta_0 +  \frac{x_2}{N^{2/3}}) e^{-4\beta m}e^{\frac{2 i (x_{1}+ 2 x_{2})}{N^{2/3}} } e^{- \frac{2 i \tau}{\beta}\frac{ (x_{2}-x_{1})}{N^{2/3}}} }{\left(1+e^{-\beta m} e^{i \frac{x_{1}}{N^{2/3}}}\right)^2\left(1+e^{-\beta m} e^{i \frac{x_{2}}{N^{2/3}}}\right)^2} \,  \nn \\
\eea
At large-N the kernel is now replaced by the Airy kernel $\rho_{A_1} \rightarrow \rho_{Airy}$ and the universal leading contribution is now
\be\label{Airyfourpoint}
F(\tau) \sim   \int_{-\infty}^{\infty} d x_1 \int_{- \infty}^{\infty} d x_2   \, \rho_{Airy}^2(x_1, x_2) \,   e^{-4\beta m}  e^{\frac{- 2 i \tau}{\beta}\frac{2\pi (x_2 - x_1)}{N}}
\ee
Unfortunately in this case we cannot obtain an analytic result for the integral, and hence we cannot perform an analytic continuation to real time consistently\footnote{Some integrals of Airy functions that might be useful to use are provided in the appendix~\ref{Airyformulae}.}.

%We find that in this window of time the Liapunov exponent maximises with the value. It then transitions to the behaviour bla. For $N \beta << t$ the late time OTOC correlator relaxes to its factorised value, but the really smooth plateau behaviour is not there once we include finite-$N$ effects, and we expect the existence of persistent oscillations. Let us now describe in more detail the new Airy regime.

%perform the following change of variables $\theta^{\prime}\rightarrow\theta+s2\pi/N $ then 
%\be\label{OTOCFS2}
%F(t)\sim\lim_{N\rightarrow\infty}\int_{-\pi}^{\pi}d\theta\int_{-N}^N ds  \rho_{sine}^2(s)  \frac{e^{-4\beta m}e^{i4\theta}e^{\frac{2t}{\beta}\frac{2\pi s}{N}}e^{\frac{i5\pi s}{N}}}{\left(1+e^{-\beta m}e^{i\theta}\right)^2\left(1+e^{-\beta m}e^{i\theta}e^{i\frac{\pi}{N}s}\right)^2} \, .
%\ee
%This works in this case because the sine kernel depends only in the difference of $\theta,\theta^{\prime}$. It is translation invariant.

%To compute the integral for the variable $s$ we change variables to $s\rightarrow s/N$ such that we can use a stationary phase approximation. The integral is from $s\in [-1,1]$. Then we perform two consecutive integrations by parts, mathematica cannot do the final integral...

\section{Connections with Liouville theory}\label{Liouvilleconnections}

\subsection{Liouville Theory and Long Strings}\label{longliouville}

In this section we briefly review the connection between FZZT branes, long strings and the adjoint representation of MQM. As we mentioned in the introduction, the FZZT brane is extending along the Liouville direction and can be thought of as a $D1$ brane with Neumann boundary conditions for the open strings in the Liouville direction. The Liouville action on a worldsheet with boundaries is~\cite{Fateev:2000ik,Teschner:2000md}  
\be 
S = \int_R d^2 z \sqrt{g} \left(\frac{1}{4 \pi} g^{a b} \partial_a \phi \partial_b \phi + \frac{1}{4 \pi} Q R \phi + \mu e^{2 b \phi} \right) + \int_{\partial R} d s g^{1/4} \left(\frac{Q K \phi}{2 \pi} + \mu_B e^{b \phi} \right)\, ,
\ee
with K the extrinsic curvature and the parameters $\mu, \mu_B$ the bulk-boundary cosmological constants. The length of the boundary (a loop) is $l = \oint ds  e^{b \phi}$. The following relations between the parameters hold ($\mu_{KPZ}$ the KPZ scaling parameter of correlation functions)
\bea\label{liouvilleparameters}
c_L=1+6 Q^2\, , \quad \quad Q= b + b^{-1}\, , \nn \\
\quad \mu_B = \frac{\Gamma(1-b^2)}{\pi}\sqrt{\mu_{KPZ}} \cosh(\pi b \sigma)\, , \quad \quad \mu_{KPZ} = \pi \mu \frac{\Gamma(b^2)}{\Gamma(1-b^2)}\, .
\eea
For $c_{matter}=1 \Rightarrow b=1$ and one finds a renormalization of $\mu_{KPZ}, \,  \mu_B$ such that
$\mu_B = \sqrt{\mu} \cosh(\pi \sigma)$ becomes the correct relation between physical parameters.

\begin{figure}[t!]
\begin{center}
\includegraphics[scale=0.5]{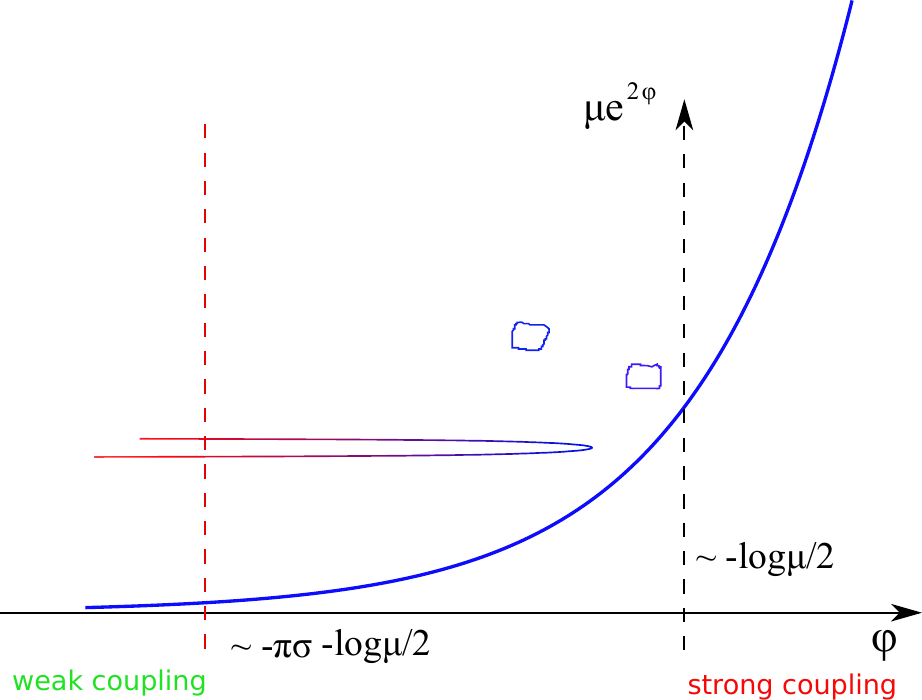}
 \end{center}
 \caption[]{The scattering of closed and open strings. Closed strings only see the bulk Liouville wall, while open strings have their endpoints pinned by the boundary potential.}
\label{scattering}
\end{figure}

We can now study qualitatively the dynamics of open and closed strings, see fig.\ref{scattering}. For more details the reader can consult~\cite{Maldacena:2005he,Gaiotto:2005gd,Fidkowski:2005ck,Kostov:2006dy,Bourgine:2007ua}. The dynamics of closed strings (tachyons) is governed by the bulk Liouville potential $\mu e^{2 b \phi}$. The tachyons start from the asymptotic region $\phi= - \infty$, move towards bigger values of $\phi$ until they reach the Liouville wall at $\phi \sim - \frac{1}{2 b}\log \mu$, where they get reflected and return back to the asymptotic region. The dynamics of open strings is more complicated since they also feel the boundary potential $\mu_B e^{b \phi}$. For large $\sigma$, one has two regimes. Let us consider a very energetic open string that starts at the asymptotic region. It will first reach the boundary potential wall at  $\phi \sim - \pi b \sigma - \frac{1}{2 b}\log \mu $, much before the bulk wall. It will then loose some of its energy, and get its end-points trapped due to the open string potential. If the string is energetic enough, the bulk of the string will nevertheless continue moving until it looses all the kinetic energy at $\phi \sim - \frac{1}{2b} \log \mu$ and returns back. It is then clear that for large $\sigma$, we have the formation of a long folded string, with endpoints stuck at the FZZT brane far away from the closed string scattering region. The reflection amplitude of this motion was computed in~\cite{Maldacena:2005he} both in Liouville theory and in the matrix model where it was found that the adjoint representation of MQM contains precicely one long folded string. In terms of the fermi-sea this long string should be thought of as an impurity interacting with the rest of the fermions via a Calogero interaction term. States containing $n$ folded strings should then be described by considering irreducible respresentations with a Young-Tableaux of n-boxes and n-anti-boxes. This brings forward the possibility of describing a large number of FZZT branes/impurities using the model we described in section~\ref{matrixmodel}. The statistics of the long-strings are then to be understood from the structure of the Young-Tableaux. Some further details on partitions and Young diagrams can be found in appendix~\ref{partitionssymmetric}.

\subsection{Cylinder partition function}\label{cylinderpartition}

One obstacle one needs to surpass when considering possible matrix quantum mechanics models of FZZT branes, is that there is no well established dictionary between the parameters on the two sides. We will now describe a computation at the Liouville theory side that will indicate such a connection with our matrix model\footnote{Some further relations between the boundary states of ZZ and FZZT branes are provided in appendix~\ref{ZZFZZTrelation}.}, that we analyse in the next subsection~\ref{2dBH}.
In the works of~\cite{Moore:1991ag,Ginsparg:1993is,Martinec:2003ka}, one finds the correlator of two macroscopic loops (cylinder partition function) with Neumann boundary conditions in Euclidean time to be
\be\label{loopcorrelator} 
\langle W(\sigma, p) W(\sigma' , -p) \rangle = \int_0^\infty d E \frac{\cos ( \sigma \pi E ) \cos ( \sigma' \pi E)  }{\sinh^2 (\pi E) (E^2 + p^2)}
\ee
with $p$ the momentum conjugate to Euclidean time and $\sigma$ is related to the boundary cosmological constant with the formulas~\ref{liouvilleparameters}. 
For winding modes around a compact Euclidean circle with Dirichlet boundary conditions, the analogous computation was performed in~\cite{Maldacena:2005he} (up to normalization)
\be\label{loopcorrelatorcompact}
Z_{cyl} = 2 \pi R \sum_n \int_0^\infty d E \frac{\cos^2 ( \sigma \pi E )   }{\sinh^2 (\pi E) \left(E^2 + (n R )^2\right)}
\ee
We can compute this integral by extending the integration range to $E \in (-\infty, \infty)$, writing the $\cos^2$ in terms of $e^{\pm 2 i \pi  \sigma}$ and picking up the appropriate poles. For the purely winding modes around the compact Euclidean direction  we need to pick the poles at $E= i n R $. One then finds
\be \label{winding1flavor}
Z_{wind} (\beta) = - 2 \pi^2 \sum_{n=1}^\infty \frac{e^{- 2 \pi n R \sigma} }{n} \frac{1}{\sin^2 (n \pi R)} =  - 2 \pi^2 \sum_{n=1}^\infty \frac{e^{-  \beta n  \sigma} }{n} \frac{1}{\sin^2 (\beta n/2)}  \, .
\ee
The first term has the interpretation of a factorised  vortex anti-vortex correlator of winding number $\pm 1$, the higher terms correspond to higher windings. It also seems that the extra poles at $E= i m$ correspond to other string excitations that exist both in the compact and non-compact case. In the compact case these might be interpreted as giving the contribution of interactions between these vortices.  The partition function of the first term in the expansion above, has been shown to be proportional to the matrix model partition function in the adjoint representation~\cite{Maldacena:2005he}. More precicely $Z_{\text{adjoint}}^{MQM} = Z_{\text{singlet}} Z_{\text{imp}}$ with $ Z_{\text{imp}}(\beta)$ the first term in the expansion~\ref{winding1flavor} describing the partition function of a single impurity (one extra box in the Young diagram). The total sum can also be interpreted as a grand-canonical free energy of impurities corresponding to vortices anti-vortices 
\bea\label{winding1flavorgrand}
Z_{wind} (\beta) & \sim &  \sum_{n=1}^\infty \frac{x^{2n} Z^{(1)}_{v-a}(n \beta)}{n} = \sum_{m, n =0}^\infty \log(1 - x^2 q^{n+m+1}) \nn \\
& = &  \sum_{m=0}^\infty (m+1) \log(1- x^2 q^{(m+1)})  \, ,
\eea
with $q=e^{i \beta}$, $x_v = x_a = x =  e^{-\beta \sigma/2}$ the fugacity and $Z^{(1)}_{v-a} =Z_v Z_a $ the one vortex, one anti-vortex partition function with $Z_v(\beta) = Z_a (\beta) = 1/i \sin(\beta/2)$, essentially the inverted harmonic oscillator partition function. This expression describes vortices - antivortices in terms of bosonic particles with no interactions between them. The extra poles would give rise to interactions among these bosonic excitations. For large $\sigma$ which is the regime of having long-strings, one can express the fugacity in more natural variables as $x=e^{- \beta \sigma/2} = \mu^{R/2} \mu_B^{-R}$. This expression also matches the genus zero, $\mu \rightarrow \infty$ limit of the expression given in~\cite{Mukherjee:2006hz} up to the expected leg-pole factor normalisation of the winding operators between the matrix model and Liouville side of the duality:
\be
Z_{wind} (\beta)  \sim  \sum_{n=1}^\infty \frac{\mu_B^{-2 n R}}{n^2} \frac{\Gamma(1+ n R)^2}{R^{n R}} \langle \mathcal{T}_{-n R} \mathcal{T}_{n R} \rangle \, , \quad   \langle \mathcal{T}_{-n R} \mathcal{T}_{n R} \rangle  = n (\mu R)^{n R} \left( \Gamma(- n R) \right)^2
\ee
with $\mathcal{T}_{n R}$ the Liouville vortex operator describing a string wound n-times around the thermal circle.

By introducing $N_f$ flavors of FZZT branes, one just multiplies~\ref{winding1flavor} by $N_f^2$ due to the Chan-Paton factors at the open string endpoints and it is then possible to take a double scaling limit such that the flavored FZZT branes backreact on the geometry produced by the ZZ branes
\be\label{scaling1}
N_f \rightarrow \infty \, , \, \, \sigma \rightarrow \infty \, , \, \,  N_f e^{-\beta \sigma/2} = \tilde{t} \, , \quad Z_{wind}^{(N_f)}  \rightarrow - \frac{{\tilde{t}}^2}{\sin^2 (\beta/2)} \, , 
\ee
keeping only the first winding mode. %In this limit and for $R<1$, the contribution from the poles at $E= i m$ vanishes as well, so it really decouples all the higher winding modes. 
One could also take a similar scaling limit, which has an advantage of separating the $\mu, \mu_B$ contributions~\cite{Maldacena:2005he}
\bea\label{scaling23}
N_f \rightarrow \infty \, , \, \, \quad \mu_B \rightarrow \infty \, , \, \, \quad  N_f \mu_B^{-R} = \tilde{t} \, , \quad Z_{wind}^{(N_f)}  \rightarrow - \frac{\mu^{R} {\tilde{t}}^2}{\sin^2 (\beta/2)}
\eea
These scaling limits have the additional property that relate the number of FZZT branes to the size of the thermal circle. One might then sensibly try to recover a semiclassical backreacted geometry for large $R$. This limit will be further discussed in the next chapters, where we will also perform a matching of the mass of the fundamental/antifundamental matrix model fields with the parameter $\sigma$ of the FZZT branes. We will also discuss the expectations that taking such a limit one can access the physics of two dimensional black holes, and corroborate this with an explicit computation of the partition function on the dual matrix model.

\begin{comment}
The scattering phase was found
\be
\delta(\epsilon) = - \int_{-\infty}^{\hat{\e}} d \e' \left(\frac{\pi \e'}{\tanh \pi \e'} + \pi \e'  \right) \, , \quad \quad \hat{\e} = \e +\frac{1}{2 \pi} \log \mu 
\ee

$Z_{adj}\sim Z_{sing} Z_{imp}$ with
\be
Z_{\text{imp}} = \int d \e \frac{\partial_\e \delta(\e)}{2 \pi} e^{- \beta \e} = - \mu^R \int_{- \infty}^\infty d \hat{\e} \half \left(\frac{\hat \e}{\tanh \pi \hat \e} + \hat \e \right) e^{- \beta \hat \e} = - \mu^R \frac{1}{4 \sin^2 \pi R }
\ee
\end{comment}

\subsection{Matching parameters and limit of the two dimensional Black Hole}\label{2dBH}

In order to compare computations between Liouville theory and the matrix model, one should perform a matching of the parameters on the two sides of the correspondence. A preliminary understanding can be provided through the study of winding modes present in our matrix model partition function eqns. \eqref{fermionwinding} and \eqref{bosonwinding}, that also allows to establish a connection with the matrix model of~\cite{Kazakov:2000pm}, conjectured to describe the physics of the two dimensional black hole.

We can now take again a double scaling limit that picks the first winding modes (assuming $m_\a = m,\, \forall \a$),
\be
N_f \rightarrow \infty, \quad m \rightarrow \infty, \quad  \text{with} \quad N_f e^{- \beta m} = {\tilde{t}}, \quad \text{finite}\, .
\ee
We thus see that the only winding modes surviving in this case, i.e. $\exp \left( {\tilde{t}} \tr U + {\tilde{t}} \tr U^\dagger \right)$, are identical to those studied in the matrix model of~\cite{Kazakov:2000pm}, conjectured to describe the physics of the $SL(2,R)/U(1)$ black hole coset via the black hole - sine Liouville correspondence of~\cite {Fateev}. %This limit can also be extended to a triple scaling limit of the Veneziano type by sending both $N, N_f \rightarrow \infty$ as in the previous section. 
We note that in this limit bosons and fermions behave in the same way, so it is universal for both realizations of open-closed string theory. In subsection~\ref{cylinderpartition}, we discussed the same limit in the Liouville theory side that again decouples higher windings. The Liouville and MQM limits turn out to produce exactly the same winding modes if we identify the parameters $\sigma=2 m$, so that the mass of the fermions is naturally related to the open string chemical potential. This leads to the identification of ${\tilde{t}}$ in terms of string theory parameters as
\be\label{identification}
{\tilde{t}}= N_f \mu^{R/2} \mu_B^{-R}\, .
\ee
This analogy can be extended further for all the higher winding modes. A further check of the matching $\sigma=2 m$ (for finite $\sigma$) goes through the relation between the ZZ and FZZT branes described in~\ref{ZZFZZTrelation} when $\sigma =i(n_1 + n_2)$. The matrix model parameter $e^{-\beta m}$ describing the winding mode fugacity becomes then $e^{-i \beta(n_1+n_2)/2}$ which for the $(1,1)$ brane corresponds precicely to the parameter $q = e^{- i \beta}$ of the inverted oscillator and hence the single FZZT brane partition becomes that of a single ZZ brane. It is also easy to check that this also holds for $N_f$ FZZT fundamentals/anti-fundamentals in \eqref{fullpartitionfunctionabstract} up to the fact that instead of having an adjoint representation, we describe all the reps contained in the tensor product $F \otimes \bar F$. 

In the next subsection we find that upon realising the partition function as a $\tau$ function of the Toda hierarchy, the natural Toda time variables are not ${\tilde{t}}_l = N_f e^{- l \beta m}/l$, but the rescaled $t_l = {\tilde{t}}_l / 2 i \sin (\pi n R) $, see eqn.~\ref{todacouplings}. The $\sin (\pi n R)$ in the denominator also appears in the Liouville description of the factorised vortex anti-vortex correlator eqn. \eqref{winding1flavor}. This then allows for a physical interpretation of the Toda time variables as the partition function of non interacting winding modes/vortices. We thus interestingly find that couplings/parameters of the partition function~\cite{Kazakov:2000pm} can be interpreted as the partition function of some more microscopic variables (wound strings). 

The only parameter thus left to be matched on the two sides is the Chern-Simons level $k$. Although we have discussed its relation to spacetime flux sourced by the FZZT branes, the formalism of integrable hierarchies described in Appendix~\ref{hierarchies} indicates that the partition function depends on the complex combination $\mu + i k$ that should be roughly thought of as an inverse complex string coupling $g_{st}^{-1} = \mu + i k$ along the lines of~\cite{Maldacena:2005he}. 
\\
\\
Let us finally note that it is expected that ``higher spin" generalizations of the 2-d black hole exist~\cite{Mukherji:1991kz}. In such backgrounds, remnants of the higher spin excitations present higher dimensions are excited together with the usual 2-d black hole operator. Such states are also related to the discrete states discussed in~\cite{Polyakov:1991qx,Klebanov:1991qa}. In~\cite{Mukherjee:2006zv} it was shown that a more general version of the FZZ correspondence relates these discrete states with the simultaneous perturbation of higher winding modes. In our case the fundamental fermions resulted into winding perturbations of arbitrary order and thus keeping all these modes, could mean that the model might be able to capture a higher spin generalisation of the 2-d black hole in a limit where one keeps all of them.

\subsection{The $\tau$-function and integrable hierarchies}\label{grandcanonicalhierarchies}

We now extend our analysis of the the model by passing to the grand-canonical ensemble. The reason for studying this ensemble, is because the connection to $c=1$ Liouville theory through the double scaling limit becomes more clear, whereby one tunes the level of the fermi-sea through the chemical potential $\mu$. One could in principle introduce chemical potentials both for $N_f$ and $N$. Since we do not understand the purpose it would serve for the FZZT branes very well, we will refrain from doing so in this work and therefore introduce only an extra chemical potential conjugate to $N$ to be later identified with the closed string cosmological constant $\mu$\footnote{A more detailed treatment can be found in the review~\cite{Klebanov:1991qa}.}. Later on we will also consider the double scaling limit of the previous section as $N_f \rightarrow \infty$. In our case one is thus lead to compute
\be
\mathcal{Z}^{(N_f)}_G = \sum_{N=0}^\infty  e^{\beta \mu N}  Z_{N, N_f} \, .
\ee
It is easy to pass over to the grand canonical ensemble using the Cauchy identity as in~\cite{Boulatov:1991xz,Kazakov:2000pm,Betzios:2016lne} to write the integrand as a determinant. The final result for the grand canonical ensemble takes the form of a Fredholm determinant
\be
\mathcal{Z}^{(N_f)}_G = \det (\hat I+ x \hat K^{(N_f)}) \, , \quad \, x=e^{\beta \mu}\, ,
\ee
where the integral kernel is defined via its action on test functions $f(z)$ as ($z=e^{i \theta}$)
\be
\left[ \hat K^{ (N_f)} f \right](z) = - \oint \frac{d z'}{2 \pi i} \frac{e^{u(z)+u(z')}}{q^\half z - q^{-\half} z'} f(z') \, ,
\ee
with $u(z)=-\frac{k}{2} \log z + \half   \sum_{n= -\infty}^\infty {\tilde{t}}_n z^n$ and ${\tilde{t}}_n = \frac{(\mp 1)^{n+1}}{|n|} z_w(|n| \beta )\,$, $z_w( \beta ) = \sum_\a^{N_f} e^{- \beta m_\a}$ and $\tilde{t}_0 = \beta \mu$. It is hard to compute the spectrum of such a kernel~\footnote{Although q-deformed polynomials might be of help here.}, nevertheless in appendix~\ref{energybasiskernel} we express the kernel in the energy basis. These expressions might prove useful in deriving the density of states for the partition function.
\\
\\
To proceed further it can be useful to exploit the integrable structure inherent in such models.
In particular, it was found~\cite{Kazakov:2000pm}, that in the case of $k=0$ and a finite number of $\tilde{t}_n$ this grand canonical partition function can be interpreted as a $\tau$ function of Toda integrable hierarchy with the rescaled couplings
\be\label{todacouplings}
t_n  =  \frac{{\tilde{t}}_n}{q^{-n/2}- q^{n/2}}\,  ,
\ee
playing the role of Toda ``times". In our example we find the presence of an extra conjugate-zero Toda ``time" term $k \log z$ arising essentially from the presence of the Chern-Simons term\footnote{The reason why $k$ is conjugate to $t_0$ can be very easily deduced from the form of Virasoro constraints, see appendix~\ref{virasoroappendix} and eqn.~\ref{L0constraint}.}.  Some more details about integrable hierarchies and free fermions and how we can still realise our partition function as a $\tau$ function for arbitrary $k$ are given in appendix~\ref{hierarchies}. The result is that 
\be
\mathcal{Z}^{(N_f)}_G(t; k, \mu) = e^{\sum_{n=1}^\infty n t_n t_{-n}} \tau(t; \mu + i k)\, ,
\ee
which is a $\tau$-function with a total charge dictated by the Chern-Simons level. In appendix~\ref{hierarchies} it is shown that shifts in the chemical potential are related to imaginary shifts in the charge $l$ of the $\tau$ function $\mu \rightarrow \mu +i l$, and the Chern-Simons level $k$ is directly related to the charge of the $\tau$ function.  This points to the possibility of defining a complex string coupling as $g_{str}^{-1} = \mu + i k$. 

We also observe  that the free energy normalization prefactor has some very interesting combinatorial interpretation, since it is also based on the completeness relation)~\ref{charactercompleteness} for Schur's symmetric polynomials (or characters)
\bea
\exp\left( \sum_{n \geq 1} n t_n t_{-n} \right) =\prod_{\a=1}^{N_f} \prod_{i, j \geq 0}^\infty \frac{1}{1- y^{(\a)}_i y^{(\a)}_j} =\prod_{\a=1}^{N_f} \sum_\l s_\l(y^{(\a)}) s_\l(y^{(\a)})  \, , \quad \text{with}\, , \nn \\
t_n =- t_{-n}=\sum_{\a=1}^{N_f} \frac{\sum_{i \geq 0} (y^{(\a)}_i)^n}{n}\, , \quad y^{(\a)}_i = e^{-\beta (m_\a +\omega \e_i)} \, ,
\eea
where the sum is over all partitions $\l$. In this formula we have $\e_i$ the oscillator energy levels (with $\omega = i$ for the inverted oscillator).

If we then restrict to the case of equal masses, we precisely  match the Liouville computation of the winding modes free energy~\ref{winding1flavor} and~\ref{winding1flavorgrand}, since
\be
\sum_{n=1}^\infty n t_n t_{-n} = -N_f^2 \sum_{n=1}^\infty \frac{e^{-2 \beta m n}}{n} \frac{1}{\sin^2(\beta n/2)} =N^2_f \sum_{n=0}^\infty (n+1) \log (1 - e^{-2 \beta m} q^{n+1}) \, ,
\ee
is precicely the grand free energy of non-interacting vortices~\ref{winding1flavor},~\ref{winding1flavorgrand} upon identifying $2 m= \sigma$. This piece of the free energy scales as $\sim N_f^2$, which points to the presence of the $N_f \times N_f$ fundamental/antifundamental degrees of freedom. Another interesting point is that it is the same irrespective of using fundamental bosons or fermions. We would like to understand in more detail the physical significance of this vortex free energy term in the large $N_f$ limit, since it has the correct scaling for a deconfined phase when $N_f \sim N \rightarrow \infty$ which means that it could account for a 2-d black hole entropy in a combinatorial microscopic fashion arising from the wound string degrees of freedom.
\\
\\
Since we have an infinite number of time variables turned on, the $\tau$ function or grand canonical partition function obeys the difference equation~\ref{differenceequation} discussed in appendix~\ref{hierarchies} which is actually a particular discrete analogue of the Toda equation (also known as Hirota-Miwa equation from the work of~\cite{Hirota:1981,Miwa:1982}). The dispersionless continuum limit of this equation would then suffice to obtain the genus-$0$ contribution to the grand-canonical free energy along with the associated spectral curve.

Taking the double scaling limit defined in the previous subsections \ref{cylinderpartition} and \ref{2dBH}, we find that only $t_{+ 1} = - t_{- 1} \,$, $t_0$ and $k$ remain. In this case the Hirota-Miwa equation reduces to the Toda differential equation (see~\cite{Kazakov:2000pm} and appendix~\ref{hierarchies})
\be\label{Todaequation2}
\half D_1 D_{-1} \tau_l \cdot \tau_{l} + \tau_{l+1} \tau_{l-1} = 0 \, \Rightarrow \quad \tau_l \frac{\partial^2 \tau_l}{\partial t_1 \partial t_{-1}} - \frac{\partial \tau_l}{\partial t_1 }  \frac{\partial \tau_l}{\partial t_{-1} } + \tau_{l+1} \tau_{l-1} = 0\, ,
\ee
which is obeyed $\forall k$. Both equations~\ref{Todaequation2} and~\ref{differenceequation} can be supplemented with extra conditions that relate the partition function for different values of $k$. From the integrability and $GL(\infty)$ point of view, it turns out that $k$ is a parameter of the point of Sato's Grassmannian, while the ``times" do not affect it. The appropriate $k$-dependent equations are the so-called Virasoro constraints, from which the most important is the so-called string equation which is the lowest of them~\cite{Morozov:1995pb}. From a practical viewpoint, since the Toda equation has derivatives of second order a single initial condition is not enough to fully determine the solution, one needs to supplement it with the Virasoro constraints for a unique solution to be found. To derive these constraints, one can study our partition function in the abstract form (here $\beta \mu = \tilde{t}_0 = t_0$)
\be 
e^{\tilde{t}_0 N}Z_N \sim \int \mathcal{D} U\, \det U^{-k} \exp\left( \sum_{n=-\infty}^\infty \tilde{t}_n \tr ( U^n) + \sum_{m=1}^\infty  \frac{q^m}{m} \tr (U^m) \tr( (U^{-1})^m)  \right)\, ,
\ee
under the variations $\delta_{+} U = \e^+ (U^{n+1} - U^{1-n}), \, n \geq 1$ and $\delta_{-} U =i \e^- (U^{n+1} + U^{1-n})\, n \geq 0$~\cite{Bowick:1990qc}. This is an extension of the known constraints in the case of fundamental plus adjoint representation. We derived these constraints that can be found in appendix~\ref{virasoroappendix}. The result is in terms of differential operators $\hat{L}_n$, obeying the centerless Virasoro algebra that annihilate the partition function.

Since we want to find their action in the grand-canonical ensemble, we note that we get
\be
\hat{L}_n \mathcal{Z}_G= \sum_{N=0}^\infty \hat{L}_n e^{\beta \mu N} Z_N = 0 \, , \quad \text{if} \quad \hat{L}_n e^{\beta \mu N} Z_N = 0\, ,
\ee
where we also included the ``zero-time'' term $e^{\tilde{t}_0 N} = e^{\beta \mu N}$, since the Virasoro constraints act on this term as well. When using these equations together with the Toda differential equation, one should be careful to express $\tilde{t}$ in terms of $t$ properly. The simplest constraint is $\hat{L}_0$ which reads
\bea\label{L0constraint}
\hat{L}_0 &=& \sum_{m=-\infty}^\infty  m t_m \frac{\partial}{\partial t_m} + k \frac{\partial}{\partial t_0}\, . 
\eea
\begin{comment}
Let us now list the lowest non-zero constraints we will need
\bea
L_0 &=& t_1 \frac{\partial}{\partial t_1} - t_{-1} \frac{\partial}{\partial t_{-1}}  + k\frac{\partial}{\partial t_0}\, , \nn \\
L_1 &=& \frac{1-q}{q^\half} t_{-1} \frac{\partial}{\partial t_0} + q^\half \frac{\partial^2}{\partial t_1 \partial t_0}  + \frac{q^\half}{1-q} k \frac{\partial}{\partial t_1}\, , \nn \\
L_{-1} &=&  \frac{1-q}{q^\half} t_{1} \frac{\partial}{\partial t_0} + q^\half \frac{\partial^2}{\partial t_{-1} \partial t_0}  -\frac{q^\half}{1-q}  k \frac{\partial}{\partial t_{-1}}\, , \nn \\
L_{2} &=&  t_{-1} \frac{\partial}{\partial t_{1}} + \frac{q}{1-q} \frac{\partial^2}{\partial t_{1} \partial t_{1}}  \, , \nn \\
L_{-2} &=&  -t_{1} \frac{\partial}{\partial t_{-1}} - \frac{q}{1-q}  \frac{\partial^2}{\partial t_{-1} \partial t_{-1}}  \, . \nn \\
\eea
\end{comment}
In case that $k=0$ the $\hat{L}_0$ constraint imposes a total level matching or winding conservation condition. When $k\neq0$ there is an imbalance between winding modes. It is also easy to see from this formula that $k$ is conjugate to the zero time $t_0$. 
\begin{comment}
In this case one also finds that there are the following two more independent constraints
\bea
\hat{A}\, \mathcal{Z}_G = \left[ \frac{1-q}{q^\half} t_{-1} + q^\half \frac{\partial}{\partial t_1 } \right]\, \mathcal{Z}_G = 0 \, \nn \\
\hat{B}\, \mathcal{Z}_G = \left[\frac{1-q}{q^\half} t_{1} + q^\half \frac{\partial}{\partial t_{-1}}\right]\,  \mathcal{Z}_G = 0\, ,
\eea
which in conjuction with $L_0$ reduce to a single constraint.
\bea
L_0 &=& t_1 \frac{\partial}{\partial t_1} - t_{-1} \frac{\partial}{\partial t_{-1}}  + k\frac{\partial}{\partial t_0}\, , \nn \\
L_1 &=& - t_{-1} \frac{\partial}{\partial t_0} +(1-q) \frac{\partial^2}{\partial t_1 \partial t_0}  + k \frac{\partial}{\partial t_1}\, , \nn \\
L_{-1} &=&  t_{1} \frac{\partial}{\partial t_0} +(q-1) \frac{\partial^2}{\partial t_{-1} \partial t_0}  + k \frac{\partial}{\partial t_{-1}}\, , \nn \\
L_{2} &=&  -t_{-1} \frac{\partial}{\partial t_{1}} +(1-q) \frac{\partial^2}{\partial t_{1} \partial t_{1}}  \, , \nn \\
L_{-2} &=&  t_{1} \frac{\partial}{\partial t_{-1}} +(q-1) \frac{\partial^2}{\partial t_{-1} \partial t_{-1}}  \, , \nn \\
\eea
\end{comment}
We hope to analyse some physically relevant solutions of the combined system of equations in the future. Some further help to reduce and solve such equations can be provided using KPZ-DDK scaling arguments along the lines of~\cite{Kazakov:2000pm}.

\section{Discussion}

In this work, we explored a model that captures the physics of non singlet sectors of matrix quantum mechanics dual to $c=1$ Liouville theory. The field content of the model is a $N\times N$ matrix that transforms in the adjoint of $SU(N)$ which describes the dynamics of $N$ unstable ZZ branes and $N_{f}\times N$ fundamental and anti-fundamental fields (either bosonic or fermionic) that describe the dynamics of open strings streched between the ZZ and FZZT branes. We also considered the presence of a $1d$ Chern-Simons term that enhances the symmetry of the model from $SU(N)$ to $U(N)$.   
We performed an analysis of the partition function as well as of retarded two-point functions and OTOC four point functions. The most interesting physics are obtained in a Veneziano-like limit of large $N, N_f$, where the bifundamentals backreact on the dynamics of MQM and non-trivial representations allow for the presence of a GWW phase transition.

This study also revealed a preliminary connection between matrix model and Liouville theory/ FZZT brane parameters. In a short summary we found that one can define a complex string coupling parameter through the combination $\mu + i k$ that takes into account fluxes sourced by the FZZT branes and that the mass $m$ of the extra fundamental fields are related to the boundary Liouville theory parameter $\sigma$ through $\sigma = 2 m$. 

One novelty of our work is that the parameters of the partition function/Toda times seem to correspond to partition functions themselves (of wound strings) when expressed in terms of more microscopic physical quantities of the theory. This opens the possibility to understand better the thermodynamical properties of this model in a microscopic fashion and elucidate possible connections with two dimensional black hole physics. To achieve this it might be enough to study the dispersionless limit of the difference/differential equations, that is known to reproduce the genus-$0$ contribution to the string partition function. 
\\
\\
Let us now, present in more detail what future directions one can follow:

\paragraph*{Thermodynamics:} We described an explicit connection between the proposed Matrix Quantum Mechanics description of the two dimensional black hole, our model and of a condensate of FZZT branes. As we briefly discussed in the previous sections, there is still some confusion about the thermodynamic interpretation of the matrix model of~\cite{Kazakov:2000pm}. The reason is two-fold. First in that work, it was not possible to fully solve the Toda equations but only the Toda equations with a single initial condition coming from the unperturbed circle partition function. The string equation was solved in the dispersionless limit in~\cite{Kostov:2001wv}, and provided an exact relation for the genus zero part of the free energy. Even more important is that the microscopic origin of the Toda time parameter $t$  that controls the strength of the vortex perturbations is obscure and one did not know whether it should be regarded as an independent physical parameter or if it should depend on $\mu,\, R$ or any other parameters when discussing the thermodynamics of the model. 

In this work we obtained some understanding on the possible microscopic origin of $t$ (through the presence of $N_f$ FZZT branes/ non-singlet representations of MQM) and elucidated its relation to Liouville theory quantities such as $N_f, \, \mu_B, \, \mu_{KPZ}$. We should also mention at this point that the FZZ correspondence in which the model~\cite{Kazakov:2000pm} is based on, is known to hold for the radius $R=3/2$ close to the so called black hole-string correspondence point~\cite{Giveon:2005mi}, which %means that the system might be better described in terms of a gas of strings and% 
poses some trouble into extrapolating this result to arbitrary radii. This is where we think that since our coupling is manifestly dependent on the size $R$ of the compact dimension and the number of flavors $N_f$, the thermodynamics might prove more reasonable to understand and it would be interesting to repeat a thermodynamic analysis and compare with the results of the free-energy computed via semiclassical effective action approaches~\cite{Kazakov:2001pj}. 
%To do this we should now solve the combined system of Virasoro constraints and Toda equation with the initial condition coming from the (universal) part of the grand canonical free energy of $c=1$ string theory on the circle of radius $R$.
%\bea
%\mathcal{F}_c(\mu, R) &=& - \frac{R}{2} \mu^2 \log \mu - \frac{1}{24} (R+\frac{1}{R}) \log \mu + R \sum_{n=2}^\infty \mu^{-2(n-1)} f_n(R)\, + \, O(e^{-2 \pi mu}) \, , \nn \\
%f_n(R) &=& \frac{(2n-3)!}{2^{2n}} \sum_{k=0}^n R^{-2k} \frac{(2^{2(n-k)}-2)(2)}{[2(n-k)!][2k]!} \, .
%\eea
%We believe that this is an important point to revisit in the future. 
%Since we found that $t= N_f \mu^{R/2} \mu_B^{-R}$ with the ratio $\sqrt{\mu}/\mu_B$ to be dimensionless. 

\paragraph*{Black holes - Integrability vs Chaos:}

Connected to the previous discussion is the possibility that such models could describe non-trivial target space states such as the two dimensional black hole. An immediate conceptual clash arises between the integrability of Spin-Calogero models and the apparent chaotic and thermal behaviour of black holes exemplified in the recent studies of four-point (OTOC) correlators. Even though we cannot provide a complete answer to this dichotomy we would like to make some preliminary comments.

A first possibility is simply that two dimensional black holes are special and correspond to fully integrable models. This could be true especially due to the fact that gravity (and even string theory) in two dimensions is very simple and does not display the intricacies and dynamics of the higher dimensional counterparts. From the point of view of the Spin-Calogero models, it is known that they exhibit quite interesting and highly degenerate spectra. As an example if we focus only on the spin degrees of freedom of our Hamiltonian~\ref{Spinchain}, this
Polychronakos-Frahm spin chain~\cite{Polychronakos:1993wc} has a highly degenerate (but equispaced spectrum), and for large number of sites $N$ the level density is approximated by a Gaussian distribution. Once one performs an unfolding of the energy spectrum, it is found that the unfolded spectrum statistics is neither of the Poissonian nor of the Wigner type indicative of the usual discrepancy between integrable and chaotic systems. For more details see~\cite{Barba:2008pc} and references within. To this end one can imagine the possibility of highly degenerate spectra acquiring a random ``chaotic" behaviour with the addition of small perturbations into the system. The highly degenerate levels start to split and form an approximate continuum around a given reference state. A correlator of probe operators computed on a high energy state that had a quite simple evolution law for the degenerate equidistant spectrum might then start to show approximate thermal characteristics~\footnote{We wish to thank Kyriakos Papadodimas for discussions on this point.}. In our model, such a behaviour might be provided by the external magnetic field coupling to the spin operators $\sum_{A i} B^A S_i^A$ with $i$ the N lattice sites and $A$ the $SU(2 N_f)$ generators. This magnetic field is proportional to the masses of the anti/fundamental degrees of freedom and points in relative opposite directions leading to a splitting of the degenerate levels.

One could further argue for some analogy with the well studied case of $AdS_2$, which admits no finite energy excitations but a ground state degeneracy~\cite{Maldacena:1998uz}. Nevertheless once one studies the case of nearly-$AdS_2$ through the SYK model~\cite{Maldacena:2016hyu}, where the conformal symmetry is broken spontaneously as well as explicitly by UV effects, one indeed finds an exponentially large number of states contributing to the low energy physics and a chaotic behaviour to emerge. So perhaps something similar could also happen for the model studied in this paper.

A preliminary picture of the formation of a black hole from a target space point of view, involves a large number of long-strings that start from infinity where the string coupling is small and they can be considered to be approximately free. As they move in the bulk they start interacting and slow down, and if they have enough energy it is then possible for them to form a condensate and backreact on the spacetime by curving it near the strongly coupled region in the bulk. If the backreaction is strong, we expect a geometric singularity with a horizon to form. A part of the long strings then remains outside the horizon while the tip condensate has formed the singularity. Extra closed strings can scatter off such a configuration.  This is reminiscent of the Susskind - Uglum picture of open strings with both ends stuck on the horizon~\cite{Susskind:1994sm}. The analogous process in the Spin-Calogero model, would involve a separation of two properties of the dynamics: the scattering off the inverted oscillator potential and the spin-chain part of the Hamiltonian. The first part is known not to provide any black hole characteristics by itself and is related to the Liouville wall on which closed strings scatter back to infinity. The magnetic field part due to the masses of the extra (anti)-fundamental fields is related to the open string Liouville wall, through the identification $\sigma = 2 m$. One should then consider the evolution of a state with a large number of (anti)-fundamental excitations on top of the vacuum equivalent to a large number of long strings. The limit of large $N_f$ is similar to the ``freezing'' limit of Polychronakos since the spin chain part of the Hamiltonian becomes parametrically large. From the point of view of representations, one is interested in the limit of large representations with a big number of (anti)-boxes. Since a magnetic field breaks the big degeneracy of the high energy state approximate chaotic evolution might then arise from the large number of energy states available near the original high energy state. It would be nice to check such a picture using the spin-chain dynamics at least numerically.

\paragraph*{State dependence:}

As a final comment, it would be interesting to consider the issue of state dependence~\cite{Papadodimas:2015jra} in the context of the bulk reconstruction of two dimensional string theory through the dual matrix model formulation. This is because on the one hand there still exist non trivial bulk geometries such as the two dimensional black hole and on the other hand one can hope for a much better handling of the duality with a matrix model at a perturbative and non-perturbative level. What the $c=1$ matrix model indicates is that the full Hilbert space decomposes into different sectors from which the singlet one is only able to capture the physics of the linear dilaton background and local excitations thereof. If one wants to describe other backgrounds one inevitably needs to enlarge the Hilbert space with the non-singlet sectors. There does not seem to be any fundamental need to employ any form of state dependence at the microscopic level of the matrix model. Nevertheless state dependence is expected to arise, for practical purposes from the non trivial mapping to target space quantities (or bulk reconstruction), once for example one wishes to describe local operators on the two dimensional black hole background in the interior and exterior regions. In fact the situation seems to be harder to analyse compared to higher dimensions, since it is very difficult to establish approximate bulk locality near the strongly coupled region of the Liouville direction. In addition the existence of the collective field theory description as a ``hydrodynamic" description of the tachyonic scattering indicates that a form of state dependence does arise once one uses more coarse grained variables, and currently it does not seem possible to uniquely define and describe bulk tachyons around an arbitrary background with this formalism. Related to this, the tachyonic field operators are related via a complicated non-local transform to the string theory loop operators, a relation which is known in detail only for the linear dilaton background and the singlet sector of the matrix model. One expects this construction to be modified for an arbitrary background and many further subtleties to arise. In order to clarify these issues it is very important to address the mapping between matrix model non-singlets and target space physics in more detail.

\acknowledgments

We wish to thank Johan van de Leur, Joseph Minahan, Kyriakos Papadodimas, Konstantinos Sfetsos and Miguel Tierz for stimulating discussions. We are especially grateful to Alexei Morozov and Alexios Polychronakos for many comments and clarifications related to Virasoro constraints and Spin-Calogero models and to Dionysios Anninos for his sharp comments while reading the draft. We also want to thank the organisers and the participants of the 9th Crete Regional Meeting in String Theory and especially Spenta Wadia for insightful comments. P.B. is supported by the Advanced ERC grant SM-grav, No 669288. O.P. is supported by the STFC Ernest Rutherford grants ST/K005391/1 and ST/M004147/1.

\appendix

\section{Gauge invariant excitations and Collective fields}\label{excitations}

\subsection{Chiral variables}\label{chiralvariables}
It turns out that one can simplify the description of MQM and elucidate the role of the constraint for the algebra of the simplest gauge invariant excitations by passing to the so-called chiral variables introduced in~\cite{Alexandrov:2002fh} and further studied in~\cite{Kostov:2002tk,Yin:2003iv,Maldacena:2005he,Kostov:2006dy}. We define
\be
\hat{X}_\pm = \frac{\hat{M} \pm \hat{P}}{\sqrt{2}}
\ee
with the commutation relations $[ (\hat{X}_+)_{i j}, \,  (\hat{X}_-)_{k l} ]  = - i \delta_{i l} \delta_{j k}  $.
The hamiltonian becomes
\be
\hat{H}_0= \half \tr \le(\hat{P}^2- \hat{M}^2 \ri) = - \half \tr \le(\hat{X}_+ \hat{X}_- + \hat{X}_- \hat{X}_+ \ri)
\ee
and the action for the usual MQM part in chiral variables can be written as
\be
S_{MQM}^{ch}= \int dt \,  \tr \le(i X_+ D_t X_- \ri) - \half \tr X_+ X_-
\ee
In this variables the action is first order in time derivatives and the normal ordered constraint acquires the form
\be\label{lightconeconstr}
 :[X_-, \, X_+]: - \sum_\a^{N_f} \left( \psi_\a \psi^\dagger_\a - \chi_\a^\dagger \chi_\a \right)  + k \mathbb{I} = 0
\ee
Similar first order actions had been proposed in studies of non-commutative Chern-Simons theory in relation to the quantum-Hall system~\cite{Polychronakos:2001mi,Morariu:2001qa} with a revived interest in the recent papers~\cite{Dorey:2016mxm,Dorey:2016hoj} that studied the connection with the $SU(N_f)_k$ WZW model in the large N-limit. Our model is a slight extension of these recent works, since we consider both fermionic and bosonic fundamental and antifundamental $U(N)$ matrices~\footnote{While completing this work, another model appeared in~\cite{Barns-Graham:2017zpv} that has both (anti)-fundamentals, but the physical context is again different.}, $\psi_{\a i}, \chi_{\a i}$ or $V_{\a i}, W_{\b j}$, instead of only complex fundamental bosons and most importantly the operators $\hat{X}_\pm$ are hermitean and not hermitean conjugates which in that case made them act as creation/annihilation operators. This is owed to the fact that we use an inverted harmonic oscillator potential. Had we used the usual harmonic oscillator we would have ended up with the term $\int dt \tr i Z^\dagger D_t Z - \half \tr Z^\dagger Z$ with $Z$ a complex matrix acting as an annihilation operator to physical states. The similarities of the two models though, pinpoint to the fact that our model is also naturally related to the WZW model at large N. At the level of the partition function, this is discussed in \ref{LargeNlimit} and appendix \ref{canonicalsymmetricpolynomials}.  We now turn to an analysis of the simplest gauge invariant excitations-currents and of the associated collective field theory description.

\subsection{Currents and Collective field theory}\label{chiralvariables}

Given the gauge invariant vacuum $|0\rangle$, one can build other $U(N)$ invariant states acting with the following operators-$U(N)$ singlets. To keep the discussion simple, we will describe the fermionic case in terms of the $2N_f$ variables $\Psi_{\tilde{\a} i} = \Psi_{m \a i}$ with $m = (\uparrow, \downarrow)$.
\begin{itemize}
\item One first defines the $U(1)$ currents
\be
J^n = \tr X_-^n , \quad \tilde{J}^n = \tr X_+^n\, .
\ee
These are known to correspond to tachyon vertex operators (closed strings) in the dual string theory.

\item Using $X_-$, one can then define the following $SU(2N_f)$ spin currents (positive graded) with $U(N)$ indices contracted 
\be
(S^n)^a_{\tilde \a} =  \tr \Psi^\dagger_{\tilde \a} X_-^n \Psi_{\tilde \a}\, .
\ee
In our original description one can split these into flavor, spin and flavor-spin currents
\bea
(T^n)^{{A}} &=& \tr \Psi^\dagger_{m \a} T^{ A}_{{\a}  \b} X_-^n \Psi_{m \b} \, , \nn \\
(S^n)^a_{ \a} &=&  \tr \Psi^\dagger_{m \a} \frac{\sigma^a_{m n}}{2} X_-^n \Psi_{n \a}\, , \nn \\
(G^n)^{a  A} &=& \tr \Psi^\dagger_{m \a} \frac{\sigma^a_{m n}}{2} T^A_{ \a  \b} X_-^n \Psi_{n \b}\, .
\eea

\item One can also define similar currents using $X_+$  
\bea
(T^n)^{{A}} &=& \tr \Psi^\dagger_{m \a} T^{ A}_{{\a}  \b} X_+^n \Psi_{m \b} \, , \nn \\
(S^n)^a_{ \a} &=&  \tr \Psi^\dagger_{m \a} \frac{\sigma^a_{m n}}{2} X_+^n \Psi_{n \a}\, , \nn \\
(G^n)^{a  A} &=& \tr \Psi^\dagger_{m \a} \frac{\sigma^a_{m n}}{2} T^A_{ \a  \b} X_+^n \Psi_{n \b}\, .
\eea
If we keep the two sets of currents separate, it is easy to check that they obey a form of the $SU(2N_f)$ K\v{a}c-Moody algebra without the central extension, for example $\left[(T^n)^{\tilde A} \, , (G^m)^{b \tilde B}\right]  =  i f_{\tilde A \tilde B \tilde C} (G^{n+m})^{b \tilde C}$. Upon identifying the negative grading currents as $J^{-m} = \tilde{J}^m$, one then expects that a central extension term will arise, coming essentially from terms containing commutators of the form $[(X_-)^n \, , (X_+)^m]$ when one substitutes the constraint~\ref{lightconeconstr}. We have not managed to show this in full generality due to technical complexity, but in~\cite{Dorey:2016hoj} it was shown that at large $N$ things simplify and one indeed recovers a central extension term from these commutators\footnote{This holds for physical states constructed by acting with $O(n)\ll O(N)$ operator insertions on the vacuum $|0\rangle$.}. Moreover, using the techniques in~\cite{Dorey:2016hoj} with some minor adjustments, we have also checked that the partition function is indeed related to the $SU(2N_f)_{\tilde{k}}$ WZW model at large N. This analysis is presented in appendix~\ref{canonicalsymmetricpolynomials}.
\end{itemize}
Currently we cannot directly prove the exact relation between these current excitations with the ones of the dual string theory, but $SU(N_f)$ singlets should correspond to open strings streched between the same FZZT brane, while non-singlets to open strings streched between different FZZT branes. The role of the sandwiched powers of $X_\pm$ seems to correspond to gravitational (tachyonic) dressing. A way to properly identify the corresponding open/closed string theory excitations that these operators correspond to, is to develop further the collective field theory description of this model. This will also indicate the symmetry algebra of the excitations in the double scaling limit, since in the simplest MQM case one can perturb the fermi sea with the operators $W_{n, m} =  x_+^n x_-^m$ which are known to obey a $W_{1+\infty}$ algebra~\cite{Minic:1991rk}
\be
[W_{n, m}\, , W_{n',m'}] = (n m' - m n') W_{n+n'-1, m+m'-1} \, .
\ee
The case of matrix-vector models has been treated in~\cite{Avan:1995sp}, where a very interesting non-linear algebraic structure was found that involves K\v{a}c-Moody and Virasoro subalgebras. A similar non-linear algebraic structure for all the non-singlet sectors is discussed in~\cite{Hatsuda:2006xr} and termed $\hat{\mathcal{W}}_\infty$ algebra, whose generators describe the joining and splitting of loop operators. We will now give some more details on the collective field theory of matrix-vector models.
\\
\\
One first defines the following collective fields
\bea
\phi(x, t) \, &=& \, \tr \delta(x - M(t)) \, = \alpha_+ (x, t) - \alpha_-(x, t) \, , \nn \\
\mathcal{J}_{\tilde \a \tilde \b}(x,t) \, &=& \, \tr \Psi_{\tilde \a}^\dagger \delta(x - M(t)) \Psi_{\tilde \beta} \, = \mathcal{J}_{\tilde \a \tilde \b}^+(x,t) - \mathcal{J}_{\tilde \a \tilde \b}^-(x,t)\, , 
\eea
The first is the usual bosonic collective field capturing the fermi sea eigenvalue density fluctuations while the second is a $SU(2N_f)$ current providing extra flavor like degrees of freedom to this fermi sea. $\alpha_\pm (x,t)$ are chiral collective fields, analogous to the matrix variables $X_\pm$, that are given by $\alpha_\pm (x,t) = \partial_x \Pi (x, t) \pm \pi \phi(x,t)$. We have also similarly split the current into chiral parts. These can be defined unambiguously only if the fermi-sea does not form crests, otherwise they are multivalued functions\footnote{One can overcome such obstacles with the formalism of~\cite{Dhar:1992hr}. The collective field theory is a ``hydrodynamic" approximation of that exact formalism.}. These definitions need to be supplemented with the appropriate Poisson brackets (that are replaced by commutators in the quantum theory)
\bea\label{commutators}
\lbrace \alpha_\pm(x) \alpha_\pm(y) \rbrace &=& \pm \partial_x \delta(x-y) \, , \nn \\
\lbrace \mathcal{J}^A_\pm (x) \mathcal{J}^B_\pm (y)  \rbrace &=& f^{ABC} \mathcal{J}^C_\pm (x) \delta(x-y)  \pm \frac{k}{2 \pi} \partial_x \delta(x-y)
\eea
where we used $ \mathcal{J}^A = T^A_{\tilde \a \tilde \b} \mathcal{J}_{\tilde \a \tilde \b}$, with $T^A_{\tilde \a \tilde \b}$ the $SU(2N_f)$ generators and allowed for the possibility of a central extension due to the presence of the CS-term.

If one neglects the currents the collective field Hamiltonian is 
\be
H_{coll} = \int d x \frac{1}{6} (\a_+^3 - a_-^3) + (\mu - \frac{x^2}{2}) (\a_+ - \a_-) 
\ee
Notice that even though the fermions are free, the bosonization of the theory results in an interacting string-field theory, with a cubic interaction term. The extra terms arising from the flavors are
\be
H_{coll}^f = \int d x \, (\a_+ T^{\mathcal{J}}_+  - \a_- T^{\mathcal{J}}_-) \, + \, \int dx dy \frac{\mathcal{J}_{\tilde \a \tilde \b} \mathcal{J}_{\tilde \b \tilde \a}}{(x-y)^2}\, + \, \int d x \, \left( W_3(\mathcal{J}_+) - W_3(\mathcal{J}_-) \right)\, ,
\ee
where $ T^{\mathcal{J}} = \frac{d^{\tilde A \tilde B} \mathcal{J}^{\tilde A} \mathcal{J}^{\tilde B}}{2(1+2N_f)}$ is the stress tensor arising from the currents ($d^{\tilde A \tilde B}$ is the $SU(2N_f)$ killing form). One needs to supplement the Hamiltonian with extra cubic terms due to the Jacobian of the transformation to collective fields. These involve the totally symmetric symbol $d^{A B C}$
\be
W_3(\mathcal{J}) = \frac{d^{A B C}}{6}  \mathcal{J}^A \mathcal{J}^B \mathcal{J}^C 
\ee
In a further development~\cite{Avan:1996vi}, these extra cubic terms were found to be related to higher spin operators\footnote{These are current tri-linear operators $W_3$ of spin-$3$.} and the total theory to possess a Yangian symmetry. One notices that this string field theory description is a-priori background independent, since we can find various solutions to the string field equations of motion. Furthermore it seems to describe the usual closed string fields interacting with a Sugawara stress tensor, with a cubic vertex for the closed string fields and a vertex where the closed string field is coupled to the stress tensor (or equivalently an interaction between closed and open string fields with the $SU(N_f)$ indices playing the role of Chan-Paton factors), the currents being also non-locally coupled. Let us now briefly analyse the equations of motion coming from the collective Hamiltonian. They read
\bea\label{collectiveEOMS}
 0 &=& \partial_t \alpha_\pm + \alpha_\pm \partial_x \alpha_\pm + \left( \partial_x V(x) + \partial_x T_\pm^\mathcal{J} \right) \, \nn \\
\partial_t \mathcal{J}_\pm^A &=& \frac{f^{A B C}}{2(1+ 2 N_f)} \mathcal{J}_\pm^B \mathcal{J}_\pm^C \alpha_\pm \pm \frac{k}{2 \pi} \partial_x \left( \alpha_\pm \mathcal{J}_\pm^A \right) 
\eea
The classical static solutions that correspond to time independent backgrounds satisfy
\be
\pi^2 \phi_0(x)^2 = \mu - \half x^2 + T^\mathcal{J}(x)\, ,
\ee
along with an equation for the $\mathcal{J}$'s. It is easy to see that in the case of no stress energy deformation, the only time independent solution is $\pi \phi_0(x) =  \sqrt{\mu - \half x^2}$, which is known to correspond to the linear dilaton background. The stress tensor can provide deformations of this fermi sea. It will be very interesting to further understand the properties of our model from the collective or bilocal fermionic field theory point of view and try to connect more general solutions/excitations with target space backgrounds/ perturbative fields. One drawback of the collective field theory approach is that typically one can find the target space metric up to a conformal factor, thus it is not clear if one can derive an exact metric for a specific fluid profile. We are thus leaving the development of a more systematic treatment for the future. 

We close this subsection with one final comment. In \eqref{collectiveEOMS}, the singlets are directly coupled to the non-singlet spin degrees of freedom. If we imagine that their fluctuations involve vastly different timescales, we can take a stochastic limit in which the correlations of the spin currents are fixed independently of those of the singlets. An interesting case then is the limit in which the variables $T(\mathcal{J})$ can be thought of as Gaussian-random variables (white-noise). The singlets then evolve according to a stochastic Burgers type equation \eqref{collectiveEOMS} in the bath of the non-singlet degrees of freedom. It is also possible to make a redefinition $\alpha_\pm = - \partial_x \, h_\pm$ to a KPZ type of equation
\be
\partial_t  h_\pm = (\partial_x h_\pm)^2 + V(x) + T_\pm^\mathcal{J} \, , \qquad \langle T_\pm^\mathcal{J}(x,t) T_\pm^\mathcal{J}(x',t') \rangle_{n.s.} = D_\pm \, \delta(x-x')\delta(t-t') \, ,
\ee
where the average is a non-singlet average and the fields $h_\pm(x,t)$ are ``height variables".

%\begin{comment}
\subsection{Relation between ZZ and FZZT}\label{ZZFZZTrelation}

In this appendix, we discuss a relation between the ZZ and FZZT boundary state, that provides a further check of our proposal for the matching between Liouville theory and Matrix model parameters. In the work of~\cite{Fateev:2000ik,Teschner:2000md,Zamolodchikov:2001ah} one finds two sets of natural boundary states, the FZZT brane boundary state ($|\nu \rangle$ are Ishibashi states)
\bea
|B_{\sigma}\rangle &=& \int_{-\infty}^\infty d \nu e^{2 \pi i \nu \sigma} \Psi_\nu (\sigma) | \nu \rangle\, , \nn \\
\Psi_\nu (\sigma) &=&  (\mu_{KPZ})^{-i\nu/b} \frac{\Gamma(1+2 i\nu b)\Gamma(1+2 i\nu/b) \cos (2 \pi \sigma \nu)}{2^{1/4}(-2 i \pi \nu)} \, ,
\eea	
and the ZZ-brane boundary state\footnote{The $(1,1)$ should be thought of as the ground state of the $D0$ brane and the other states as discrete excited states~\cite{Nakayama:2004vk}.} ($m,n$ are integers)
\bea
|m, n \rangle &=& \int_{-\infty}^\infty d \nu \Psi_\nu (m, n) | \nu \rangle\, , \nn \\
\Psi_\nu (m, n) &=&  \sinh(2\pi m \nu/b)\sinh(2\pi n \nu b)(\mu_{KPZ})^{-i\nu/b} \frac{\Gamma(1+2 i\nu b)\Gamma(1+2 i\nu/b)}{2^{-3/4}(-2 i \pi \nu)}  \, .
\eea
It was noticed in~\cite{Martinec:2003ka} that one can derive the ZZ-brane boundary state, from the FZZT one for a specific choice of imaginary $\sigma$
\bea\label{branerelations}
|m, n \rangle &=& |B_{\sigma(m,n)}\rangle - |B_{\sigma(m,-n)}\rangle\, , \nn \\
\sigma(m,n) &=&  i \left(m/b + n b \right)\, .
\eea
For $b=1$, since $\mu_B^2 = \mu \cosh^2 (\pi \sigma)$, this also results in a simple relation of the boundary and bulk cosmological constants $\mu = \mu_B^2$.
%\end{comment}

\section{Functional determinants}\label{functdet}
Let us define $Q$  a differential operator on a circle of length $\beta$.
\be
Q=- D_{\tau}^2+{\o}^2=-{\partial}_{\tau}^2 + 2 i\alpha{\partial}_{\tau}+{\alpha}^2+{\omega}^2\, ,
\ee
where $\alpha$ is a constant gauge field in the adjoint representation related to $\theta$ as $\theta _{i}=\alpha_{i}\beta$.
$Q$ acts on the matrices $M$ as
\be 
[Q, M]={\partial}_{\tau}M+i[\alpha ,M]
\ee
and
\be 
[\alpha ,M]_{ij}={\alpha}_{ij,kl}^{adj}M_{kl}={\alpha}_{ik}M_{kj}-M_{ik}{\alpha}_{kj}
\ee
\be 
(UMU^{\dagger})_{ij}=\exp[i\beta\alpha]^{adj}_{ij,kl}M_{kl}\, ,
\ee
with
\begin{eqnarray}
{\alpha}_{ij,kl}^{adj}={\alpha}_{ik}{\delta}_{jl}-{\alpha}_{lj}{\delta}_{ik}\, ,\qquad \exp[i\beta\alpha]^{adj}_{ij,kl}=U_{ik}U^{\dagger}_{lj}\, .
\end{eqnarray}
To evaluate the determinant, one wants to solve for the spectrum of the matrix equation $\sum_{k l}\hat{Q}_{ij , k l} f_{k l}(\t) = \lambda_{i j} f_{i j}(\t)$. It is convenient to expand the periodic functions in the fourier modes of $S^1$, i.e. $f(\t)= f_0 + \sum_n f_n e^{2 \pi n i \t/\beta}$. One then has a discrete set of eigenfunctions and $\det Q =\det_{matrix} \prod_n \lambda_n$. Using these modes, one can write
\begin{eqnarray}
\det\left(-D_{\tau}^2+{\o}^{2}\right)&=& \det_{matrix}\prod_{n=-\infty}^{\infty}\left[\left(\frac{2\pi n}{\beta}+{\alpha}\right)^2+{\o}^2\right]= \nn \\
\det_{matrix} \mathcal{N} \sin\left[\frac{\beta(\a + i \o)}{2}\right]\sin\left[\frac{\beta(\a -i \o)}{2}\right] &=&  \det_{matrix} \mathcal{N} \left(\cosh(\beta\o)-\cos(\beta\a)\right)\,,
\end{eqnarray}
where $\alpha$ is a matrix  and the determinant is with respect of this matrix structure. The normalization  can be set $\mathcal{N}=2$ to conform with the usual harmonic oscillator. Otherwise one can adopt a regularisation procedure and keep the finite piece\footnote{We thank Umut G\"ursoy for sharing his notes on regulating these determinants.}.
If the gauge field is $\a_{N\times N}=diag(\a_{1},\a_{2},...,\a_{N})$, then ${\alpha}_{ij,kl}^{adj}=({\alpha}_{i}-{\alpha}_{j}){\delta}_{ik}{\delta}_{jl}$. Thus we finally get
\be
\det\left(-D_{\tau}^2+{\o}^{2}\right) = \prod_{i j}^N \left(\cosh(\beta\o)-\cos(\theta_i - \theta_j)\right)
\ee

Similarly let us define the differential operator $\td Q$ acting on functions transforming in the fundamental representation (with a general mass $m$)
\be
\td Q = i\partial_\tau - \a +i m, \quad (\a^f \psi)_i = \a^f_{i j} \psi_j
\ee
Upon diagonalising the matrix $\a$ we get $\a^f_{i j}= \a_i \delta_{i j}$.
Similarly to the previous case we find (for periodic functions)
\be
\det \td Q = \prod_i^N \prod_{n = - \infty}^\infty \left[- \frac{2 \pi n}{\beta} - \a_i +i m \right] = \prod_i^N \mathcal{N}' \sin  \le(\half(\theta_i -i \beta m) \ri)\, . 
\ee
Since we are also interested in the Grassmann case (anti-periodic fermions), we find
\be
\det \td Q = \prod_i^N \prod_{n = - \infty}^\infty \left[- \frac{(2  n+1)\pi}{\beta} - \a_i +i m \right] = \prod_i^N \mathcal{N}' \cos  \le(\half(\theta_i -i \beta m) \ri)\, . 
\ee
The case of anti-fundamental can be treated in the same way and the result is obtained simply by sending $\theta \rightarrow - \theta$. We will also set $\mathcal{N}'=2$, with similar arguments as above.

\section{Partitions and symmetric polynomials}\label{partitionssymmetric}

\subsection{Partitions}

We provide some terminology on partitions. The reader can consult~\cite{Macdonald} for more details.
\begin{itemize}
\item
A partition $\l$ is a sequence of non-increasing integers such that
\be
\l_1 \geq \l_2 ....\l_{\ell(\l)+1} = 0
\ee
The number of non-zero elements $\ell(\l)$ is called the length of the partition. The sum of all the elements $|\l|= \sum_{i \geq 1} \l_i$ is called the weight of the partition.
\item
The multiplicity $m_j(\l)$ of the positive integer j is how many times the number j appears in the partition $\l$ (such that $\l_i = j$).
\item
The partitions are labelled graphically using Young diagrams. They are an array of boxes where the $i$'th row contains $\l_i$ boxes. This means that the number of rows is the length of the partition and the number of columns  is just $\l_1$. The total number of boxes is then equal to the weight $|\l|$.
\item
Another way of representing a partition is in the form $(2^{m_2}, 3^{m_3},...9^{m_9},...)$, which just means that the number j appears $m_j$ times.
\item
The conjugate or transpose of a partition $\l'$ or $\l^T$ is obtained by either reflecting the Young diagram along the diagonal exchanging rows and columns. As an example one obtains $\l_1' = \ell(\l)$. 
\end{itemize}
As a simple example to have in mind the partition $(7,5,3^2,1^2)$ corresponds to the following Young tableaux
\be \yng(7,5,3,3,1,1)\nn\ee

\subsection{Characters and Schur polynomials}\label{characterssymmetric}

In this section we will follow the review~\cite{Morozov:2009jv}. Schur polynomials $s_\l(X)$ are symmetric polynomials in $N$ variables that form a linear basis for the space of all symmetric polynomials. Seen from a representation theory point of view they are characters for the irreducible representations of the general linear groups.

To be more concrete the representations R of $GL(\infty)$ are labeled by Young diagrams, or ordered integer partitions $R: \l_1 \geq \l_2 ....\geq 0$ with $|R|$ the number of boxes in the diagram. We will thus use the representation index $R$ or the partition index $\l$ interchangeably in this case.
We will represent the characters either through the use of time or through auxiliary Miwa variables of a matrix $X$ as $\chi_R(t)$ or $\chi_R(X)$. The definition of Miwa variables is $t_k = \tr X^k / k = \sum_i x_i^k/k$ with $x_i$ the eigenvalues of the matrix $X$.
\\
\\
We define the Schur polynomials/characters as
\be\label{schurpoly}
s_\l(X) = \frac{\det [ x_j^{N - k + \l_{k}} ]_{j, k = 1,..,N}}{\det [ x_j^{N- k}]_{j, k = 1,..,N}} \, .
\ee
We have the following useful sum rule over all representations/partitions	
\bea\label{charactercompleteness}
\sum_R \chi_R(t) \chi_R(t') = \sum_\l s_\l(x) s_\l(x') = \exp\left( \sum_{n \geq 1} n t_n t'_n \right) = \prod_{i, j \geq 1} \frac{1}{1- x_i x'_j} \, .
\eea
We also have the generating functions of characters for the (anti)-symmetrised products of arbitrary copies of the representation $R$~\cite{Aharony:2003sx}
\bea 
\det(1+ t U_R) &\equiv& \sum_{n=0}^\infty t^n \chi_{anti^n(R)}(U) = e^{\sum_{l=1}^\infty (-1)^{l+1} t^l \chi_R (U^l)/l} \, , \nn \\
\det(1- t U_R)^{-1} &\equiv& \sum_{n=0}^\infty t^n \chi_{sym^n(R)}(U) = e^{\sum_{l=1}^\infty t^l \chi_R (U^l)/l} \, .
\eea
These expressions describe the formulae~\ref{fermionwinding},~\ref{bosonwinding} and~\ref{adjointmatter} in terms of character expansions of the fundamental/antifundamental and adjoint representations of $U(N)$.

\section{Canonical ensemble and symmetric polynomials}\label{canonicalsymmetricpolynomials}

It is possible to compute the partition function using the technology of Hall-Littlewood symmetric polynomials. This method is powerful enough to treat the case of different masses/chemical potentials for each flavor and simplifies the method of~\cite{Boulatov:1991xz} in terms of characters, since one needs to perform less summations over partitions. The canonical reference is~\cite{Macdonald}. This method was also used to compute the partition function of the matrix model in~\cite{Dorey:2016hoj}. 
\\
\\
We start by first defining the q-Hall inner product for two symmetric functions $f(Z) , g(Z)$
\be\label{inner}
\langle f \, , g \rangle_q = \frac{1}{N!} \left(\prod_{i=1}^N \frac{1}{2 \pi i} \oint \frac{d z_i}{z_i} \right) \frac{\prod_{i \neq j} (z_i - z_j)}{\prod_{i \neq j} (z_i - q z_j) } f(Z) g(Z^{-1})\, .
\ee
The orthogonal polynomials with respect to this measure are the Hall-Littlewood polynomials
\be
P_\lambda (Z ; q) = \frac{1}{\mathcal{N}_\lambda} \sum_{\sigma \in S_N} \sigma \left[ z_1^{\lambda_1}...z_N^{\lambda_N} \prod_{i<j} \frac{z_i-q z_j}{z_i - z_j} \right] \, ,
\ee
where $\lambda \in \mathcal{P}$ denotes the partition and the normalization is
\be
\mathcal{N}_\lambda = \frac{\phi_{N - \ell(\lambda)} \prod_{j \geq 1} \phi_{m_j(\lambda)}}{(1- q)^N} \, , \quad \quad \phi_m = \prod_{j = 1}^m (1- q^j) \, ,
\ee
with $m_j(\lambda)$ the multiplicity of the positive integer j in the partition $\lambda$ and $m_0 = N - \ell(\l) \geq 0$. One also defines the Q-Hall polynomials as $Q_\l (X; q) =b_\l(q) P_\l (X; q)$ with $b_\l (q) = \langle P_\l , P_\l \rangle_q^{-1} = \prod_{j \geq 1} \phi_{m_j(\lambda)} $. The orthogonality relation can then be written as
%\be\label{orth1}
%\langle P_\mu \, , Q_\l \rangle_q =  \delta_{\mu, \l}\, 
%\ee
%Another orthogonality relation we found in the literature is {\bf which is correct?}
\be\label{orth2}
\langle P_\mu \, , P_\l \rangle_q = \frac{1}{\mathcal{N}_\mu} \delta_{\mu, \l}\, .
\ee 
We next define the Schur polynomials through eqn.~\ref{schurpoly},
which are also a limit of the Hall-Littlewood polynomials for $q=0$ and orthonormal under the inner product~\ref{inner} upon setting $q=0$. One also has the relation
\be
s_\l(Z) = \sum_\mu K_{\l , \mu}(q) P_\mu (Z ; q) 
\ee
with $K_{\l , \mu}(q)$ the Kostka-Foulkes polynomials. The inverse relation defines the Modified Hall-Littlewood polynomials. 
\be
Q'_\mu (Z ;q) = \sum_\l K_{\l , \mu} (q) s_\l(Z) \, , \quad \quad  \langle P_\l \, , Q'_\mu \rangle_{q=0} = \delta_{\l , \mu}
\ee
There is also a relation between the Modified Hall and the Q-Hall polynomials that reads
\be
Q'_\l (Z ; q) = Q_\l \left(\frac{Z}{1-q} ; q \right) 
\ee
Let us note some useful properties of the Kostka polynomials
\begin{itemize}
\item $K_{\l , \mu}(q)=0$ unless $|\l|=|\mu|$. All the non-zero coefficients are positive.
\item They reduce to Kostka numbers for $q=0$, $\forall \l, \mu$. 
\item $K_{\l, \mu}(0) = \delta_{\l, \mu}$.
\end{itemize}
We will also use the following Cauchy identities 
\bea
\sum_\lambda s_\l (X) s_\l (Z) &=& \sum_{\l , \rho} s_\l(X) K_{\l , \rho}(q) P_\rho(Z ; q) = \sum_\rho   Q'_\rho(X ; q) P_\rho(Z ; q)\, \nn \\
 &=& \prod_{\a=1}^{N_f} \prod_{j=1}^N (1- x_\a z_j)^{-1} \, , \nn \\
\sum_\lambda s_{\l'} (X) s_{\l} (Z) &=&  \sum_\lambda Q_{\l'} (X) Q_{\l} (Z) = \sum_\lambda P_{\l'} (X) P_{\l} (Z) = \prod_{\a=1}^{N_f} \prod_{j=1}^N (1+ x_\a z_j)  \, , \nn \\
\sum_\l P_\l(X ; q) Q_\l(Y ; q) &=& \prod_{\a, \b \geq 1}^{N_f} \frac{1 - q x_\a y_\b}{1 - x_\a y_\b}\, ,
\eea
with $\l'$ the conjugate partition to $\l$. These should be thought of as completeness relations with respect to the inner product~\ref{inner}, the first two for $q=0$ and the last for non-zero $q$ (q-Hall-inner product). The summands need to vanish unless $\ell(\l) \leq min \lbrace N, N_f \rbrace$.
\\
\\
To exploit these identities, it is convenient to use the representation in terms of $2N_f$-fundamental fields only. One needs to remember the shifts  $\tilde{k}=k \mp N_f$ of the Chern-Simons level. For fundamental bosons one can write down their contribution in terms of symmetric functions as follows 
\be
Z_b(z, x) = \prod_{\tilde{a}=1}^{2N_f} \prod_{i=1}^N (1- x_{\tilde{a}} z_i)^{-1} = \sum_\l Q'_\l(X ; q) P_\l(Z ; q)  \, ,
\ee
where $x_{\tilde{a}} = e^{-\beta m_{\tilde{a}} }$ behave like fugacities for each flavor.
The Chern-Simons term can be written as
\be
\prod_{i=1}^N z_i^{-\tilde{k}} = P_{(\tilde{k}^N)}(Z^{-1} ; q)
\ee
with $(\tilde{k}^N)$ the partition with N non-zero parts equal to $\tilde{k}$.
\\
\\
The full partition function in the case of bosons is then
\bea
\mathcal{Z}_N^{b} &=& \frac{q^{N^2/2}}{N!} \left(\prod_{i=1}^N \frac{1}{2 \pi i} \oint \frac{d z_i}{z_i^{\tilde{k}}} \right) \frac{\prod_{i \neq j} (z_i - z_j)}{\prod_{i,  j} (z_i - q z_j) } \prod_{{\tilde{a}}=1}^{2N_f} \prod_{i=1}^N \frac{1}{(1- x_{\tilde{a}} z_i)} \, \nn \\
 &=& \frac{q^{N^2/2}}{N!} \left(\prod_{i=1}^N \frac{1}{2 \pi i} \oint \frac{d z_i}{z_i} \right) \frac{\prod_{i \neq j} (z_i - z_j)}{\prod_{i \neq j} (z_i - q z_j) } \frac{P_{\left(\tilde{k}^N \right)}(Z^{-1} ; q)}{(1-q)^N} \times \, \nn \\ &\times& \sum_{\l} Q'_\l(X ; q) P_\l(Z ; q)    \,  \nn \\
&=& \sum_{\l } Q'_\l(X ; q)   \frac{q^{N^2/2}}{(1-q)^N} \langle  P_\l  \, , P_{\left(\tilde{k}^N \right)} \rangle_q  \, = \frac{q^{N^2/2}}{\phi_N (q)} \sum_\l K_{\l, (\tilde{k}^N)}(q) s_\l (X) \, .
\eea
For fermions we will need to use the dual cauchy identity to simplify
\bea
\mathcal{Z}_N^{f} &=& \frac{q^{N^2/2}}{N!} \left(\prod_{i=1}^N \frac{1}{2 \pi i} \oint \frac{d z_i}{z_i^{\tilde{k}}} \right) \frac{\prod_{i \neq j} (z_i - z_j)}{\prod_{i , j} (z_i - q z_j) } \prod_{{\tilde{a}}=1}^{2N_f} \prod_{i=1}^N(1+ x_{\tilde{a}} z_i) \,  \nn \\
&=& \sum_{\l } P_{\l'}(X ; q)   \frac{q^{N^2/2}}{(1-q)^N} \langle  P_\l  \, ,  P_{\left(\tilde{k}^N \right)} \rangle_q \, =\frac{q^{N^2/2}}{\phi_N (q)} P_{(N^{\tilde k})}(X) \, .
\eea
The bosonic partition function can also be written in terms of characters of $A_{2N_f - 1}$ representations $R_\Lambda$ with $\Lambda$ the dominant integral weight $\Lambda \in \mathcal{L}_W^+$. If we set $N = 2 L  N_f + C$ we find
\bea\label{strangepfform}
\mathcal{Z}^b_{N} &=& w^{\tilde{k} L} q^{E_0} \prod_{j=1}^N \frac{1}{1-q^j} \sum_{\Lambda \in \mathcal{L}_W^+(\tilde{k}C)} q^{-E_0} K_{\l(\Lambda),(\tilde{k}^N)}(q) \chi_{\Lambda}(W)\, , \nn \\
w_1 &=& x_1 w^{-1/2N_f}\, , \quad w_2 = x_1 x_2 w^{-2/2N_f}\, , \quad ...  \quad \text{with} \quad w=x_1 x_2 ... \, x_{2N_f}
\eea
with a similar expression for the fermionic case.
\\
The ground state energy for the bosonic expression was first computed in~\cite{Dorey:2016hoj} and upon using the parametetrization $N = 2 L  N_f + C$ and keeping the leading term in $|q|<<1$ we find
\be
E_0^{b}(\tilde{k}, L, N_f, C) \,=\, \tilde{k} L(L-1)N_f + \tilde{k} L C + \half N^2
\ee

A limit one could take in expression~\ref{strangepfform} is $L \rightarrow \infty$ with $N_f , C$ fixed. This is the standard large $N$ limit. In this limit the Kostka polynomial becomes a branching function that relates the $\hat{A}_{2N_f-1}$ affine characters with the $A_{2N_f-1}$ characters, see~\cite{Nakayashiki:1995bi}. The partition function in the large $L$ limit corresponds to
\be
\mathcal{Z}^b_{L \rightarrow \infty} = q^{E_0} \prod_{j=1}^N \frac{1}{1-q^j} \chi_{R_{\tilde{k}, C}} (q, X)
\ee
where $\chi_{R_{\tilde{k}, C}}$ is the character of the $\hat{A}_{2 N_f}$ affine Lie algebra at level $\tilde{k}$ associated to the rep $R_{\tilde{k}, C}$ of $SU(2N_f)$. This is the $\tilde{k}$-fold symmetrization of the $C^{th}$ antisymmetric rep. For $C=0$ the character can be also identified with the vacuum character corresponding to the partition function of the $WZW$ model~\cite{Dorey:2016hoj}. The partition function for the fermions is quite similar and contains the character of the $\tilde{k}$-fold antisymmetrization of the $C^{th}$ symmetric rep.

\section{Virasoro constraints}\label{virasoroappendix}

In this appendix we briefly discuss the derivation of Virasoro constraints for our model that contains fundamental and adjoint characters. It is based on the thorough derivation of Virasoro constraints for the case of fundamental characters~\cite{Bowick:1990qc}.

We start with the general action (one should remember that these $t$'s are actually $\tilde{t}$'s to match with the main text)
\be
S[t_m, p_m] = \sum_{m=-\infty}^\infty t_m \tr U^m \, + \sum_{m=1}^\infty q_m \tr U^m \tr (U^\dagger)^m \, , 
\ee
and then perform variations of the matrix $U$ consistent with the Unitary symmetry of the model 
\bea
\delta_{+} U &=& \e_n^+ (U^{n+1} - U^{1-n}) \, , \quad n \geq 1\, , \nn \\
\delta_{-} U &=& i \e_n^- (U^{n+1} + U^{1-n})\, , \quad n \geq 0\, .
\eea
One then needs to pinpoint the transformation of the various terms in the path integral and write them in terms of differential operators acting on the action $S$. In particular for the fundamental character term one finds
\bea
\delta S_{f}\,[t_m] \, &=& \, \left( \e_n^+ L_n^+ + i \e_n^- L_n^- \right) S\, , \quad \text{with} \, , \nn \\
L_n^{+ f} &=& \sum_{m=-\infty}^\infty m t_m \left( \frac{\partial}{\partial t_{m+n}} - \frac{\partial}{\partial t_{m-n}}\right)\, , \quad \text{for} \quad n \geq 1\, , \nn \\
L_n^{- f} &=& \sum_{m=-\infty}^\infty m t_m \left( \frac{\partial}{\partial t_{m+n}} + \frac{\partial}{\partial t_{m-n}}\right)\, , \quad \text{for} \quad n \geq 0\, .
\eea
One also finds from the variation of the measure that it is expressible in terms of second derivatives of $e^{S}$ 
\bea
L_n^{+ me} &=& \sum_{m=1}^n \left( \frac{\partial^2}{\partial t_{n-m} \partial t_{m}} + \frac{\partial^2}{\partial t_{m-n} t_{-m}}\right)\, , \quad \text{for} \quad n \geq 1\, , \nn \\
L_n^{- me} &=& \sum_{m=1}^n \left( \frac{\partial^2}{\partial t_{n-m} \partial t_{m}} - \frac{\partial^2}{\partial t_{m-n} t_{-m}}\right)\, , \quad \text{for} \quad n \geq 0\, .
\eea
We now come to the adjoint character terms which are interestingly again quadratic similarly to the terms in the measure
\bea
L_n^{+ a} &=& \sum_{m=1}^\infty m q_m \left( \frac{\partial^2}{\partial t_{n+m} \partial t_{-m}} - \frac{\partial}{\partial t_{m-n} t_{-m}} - \frac{\partial^2}{\partial t_{n-m} \partial t_{m}} + \frac{\partial^2}{\partial t_{-n-m} \partial t_{m}}\right)\, , \quad \text{for} \quad n \geq 1\, , \nn \\
&=&  \sum_{m=-\infty}^\infty m q_m \left( \frac{\partial^2}{\partial t_{n+m} \partial t_{-m}} - \frac{\partial}{\partial t_{m-n} t_{-m}} \right)\, , \quad \text{for} \quad n \geq 1\, , \quad q_{-m}=q_m \, , \nn \\
L_n^{- a} &=& \sum_{m=1}^\infty m q_m \left( \frac{\partial^2}{\partial t_{n+m} \partial t_{-m}} + \frac{\partial}{\partial t_{m-n} t_{-m}} - \frac{\partial^2}{\partial t_{n-m} \partial t_{m}} - \frac{\partial^2}{\partial t_{-n-m} \partial t_{m}}\right)\, , \quad \text{for} \quad n \geq 0\, , \nn \\
&=&  \sum_{m=-\infty}^\infty m q_m \left( \frac{\partial^2}{\partial t_{n+m} \partial t_{-m}} + \frac{\partial}{\partial t_{m-n} t_{-m}}\right)\, , \quad \text{for} \quad n \geq 0\, , \quad q_{-m}=q_m \,.
\eea
Finally there is the Chern-Simons term  with level $k$, $\det U^k$ for which the appropriate operators read
\bea
L_n^{+ k} &=& k \left( \frac{\partial}{\partial t_{n}} - \frac{\partial}{\partial t_{-n}}\right)\, , \quad \text{for} \quad n \geq 1\, , \nn \\
L_n^{- k} &=& k \left( \frac{\partial}{\partial t_{n}} + \frac{\partial}{\partial t_{-n}}\right)\, , \quad \text{for} \quad n \geq 0\, .
\eea
The total Virasoro constraints are simply the sum of the various terms $L_n = L_n^a+L_n^f + L_n^{me} + L_n^k$. Taking linear combinations of $L^\pm_n$ one can show that they indeed obey the Virasoro algebra. This means that imposing the lowest constraints one automatically satisfies the higher ones. In particular one needs to satisfy $L_0$ and $L_{\pm 1}, \, L_{\pm 2}$. For $L_0$ we get
\bea
L_0 &=& \sum_{m=-\infty}^\infty  m t_m \frac{\partial}{\partial t_m} + k \frac{\partial}{\partial t_0}\, . 
\eea
\\
\\
The model we study is a particular reduction of this general case since the $q$'s and $t$'s are related
due to the fact that they are expressed in a Miwa-like parametrization $q_m = q^m/m\,$, $t_m= (\pm)^{m+1}N_f x^m/m $. This means that we need to impose correctly the constraints in the one dimensional subspace of possible $q$'s and $t$'s respectively.

\section{Integrable Hierarchies}\label{hierarchies}

In this appendix we provide a brief discussion on the Toda integrable hierarchy and how one identifies the grand canonical partition function as a Toda $\tau$ function. We will follow the conventions in~\cite{Kazakov:2000pm,Alexandrov:2003ut}. We extended the discussion of those references in two ways. First we included the possibility of turning on the ``conjugate zero-time'' $k$, and second we cover also the case of having singularities in the measure for $z \neq 0, \infty$, coming from terms of the form $(1 + a z)^{N_f} (1 + a/z)^{N_f}$, which lead to discrete equations for the $\tau$ function as we will see.  Our focus is on enabling the reader to perform the calculations step by step and not into giving a rigorous definition of the construction. Some more complete introductions to $\tau$ functions and free fermions are the classic review~\cite{Jimbo:1983if} and the modern~\cite{Alexandrov:2012tr}.

\subsection{Fermionic Algebra}

To set up our conventions we define $\psi_n\, , n \in \mathbb{Z}+\half$ to be free fermionic operators satisfying
\be 
[\psi_n\, , \psi^*_m]_+ = \delta_{n m} \,, \quad [\psi^*_n\, , \psi^*_m] = [\psi_n\, , \psi_m] = 0\, ,
\ee
and the fermion fields $\psi(z) = \sum_{n \in \mathbb{Z}+\half} \psi_n z^{-n-\half}$ and $\psi^*(z) = \sum_{n \in \mathbb{Z}+\half} \psi^*_{-n} z^{-n-\half}$. We then define the vacuum with charge $l$ as follows
\bea
\psi_m | l \rangle = \langle l | \psi^*_m = 0\, \quad m>l \nn \\
\psi^*_m | l \rangle = \langle l | \psi_m = 0\, \quad m<l\, ,
\eea
which in particular means that $\psi^*$ creates particles and $\psi$ creates holes and the fermi sea has particle-states filled up to charge $l$. With these definitions one finds the fermionic correlator
\bea
\langle l | \psi(\z_1)...\psi(\z_m) \psi^*(z_m)...\psi^*(z_1) | l \rangle &=& \prod_{k=1}^m z_k^l \z_k^{-l}  \frac{\prod_{i<i'} (\z_i - \z_i') \prod_{j<j'} (z_j - z_{j'})}{\prod_{i j}(\z_i - z_j)} \nn \\
&=& \prod_{k=1}^m z_k^l \z_k^{-l} \det_{i j} \frac{1}{\z_i - z_j}\, .
\eea
Bilinear expressions of the fermions generate an infinite dimensional Lie algebra $GL(\infty)$ and a generic element of the group will take the form
\be
{\bf G} = \exp \sum_{m, n \in \mathbb{Z}+\half} b_{m n} \psi_m \psi^*_n\, .
\ee
We will be interested in the specific case of $b_{m n} = q^{i \mu + n } \delta_{m, n}$ (case with zero Chern-Simons level) and $b_{m n}^{(k)} = q^{i \mu + n } \delta_{m, n-k}$ for the case of  C.S. level $k$.
One also defines the current operators (or Hamiltonians) that generate the Toda time flows as
\be
J_n = \sum_{r \in \mathbb{Z}+\half} \psi^*_{r-n} \psi_r \, ,   
\ee
which obey the following
\be
[J_n\, , J_m] = n \delta_{n+m, 0}\, ,  \quad J_n | l \rangle = \langle l | J_{-n} = 0 \, , \quad n>0 \, .
\ee
A special case of these operators is the total charge operator
\be
Q = \sum_{r \in \mathbb{Z}+\half} \psi^*_{r} \psi_r
\ee
One can pass to a bosonic description via the bosonization formulas 
\be
\psi(z) = :e^{-\phi(z)}:\, \quad \psi^*(z)= :e^{\phi(z)}:\, \quad \partial \phi(z) = :\psi^*(z) \psi(z): 
\ee
The bosonic field has the modes ($[\hat Q \, , \hat P] = 1$)
\be
\phi(z) = \hat P + \hat Q \log z + \sum_{n\neq 0} \frac{1}{n} J_n z^{-n}
\ee
and the normal ordering is defined by putting all $J_n\, , n>0$ to the right and $J_n\, , n<0$ to the left as well as $:\hat P \hat Q: = :\hat Q \hat P: = P Q$. The bosonic vacuum is defined as
\be
\hat Q | s \rangle = s | s \rangle\, , \quad J_n| s \rangle = 0 \, , \quad (n>0)\, ,
\ee
so consistently with our definitions the operator $\hat Q$ measures the total charge of the state\footnote{Notice that in string theory notation one labels this operator with $\hat p$, since the charge then would be the momentum of the string.}. The operator $\hat P$ is a shift operator that transitions between vacua of different charge $e^{\pm \hat P} | l \rangle = |l \pm 1 \rangle$.
\\
One can then define the vertex operator 
\be
V_q(z) =\psi(q^{-\half} z) \psi^*(q^{\half} z)\, .
\ee
\\
Using this, we will now define a specific $GL(\infty)$ operator with charge-$\hat Q$, ${\bf G}^{(\hat Q)}$ that takes the form 
\be
\label{gloperator}
{\bf G}^{(\hat Q)} = \exp \left( e^{\beta \mu} \oint \frac{dz}{2 \pi i} z^{-\hat Q} V_q(z) \right)
\ee
\\
Given a $GL(\infty)$ operator ${\bf G}$ one can write down the $\tau$ function as
\be
\tau_l[t] = \langle l | e^{J_+ [t]} {\bf G} e^{- J_-[t]} | l \rangle\, 
\ee
with
\be
J_+[t] = \sum_{r>0} t_r J_r\, , \quad J_-[t] = \sum_{r<0} t_r J_r
\ee
where $t$'s are Toda ``time" parameters. In our case, the analogous formula for a $\tau$ function where the $GL(\infty)$-element carries the charge $\hat Q$ operator is
\be
\tau_l[t] = \langle l | e^{J_+ [t]} {\bf G}^{(\hat Q)} e^{- J_-[t]} | l \rangle\, 
\ee
\\
Expanding the $GL(\infty)$ operator~\ref{gloperator} in a series and commuting the fermions past the currents $J_\pm$ and the charge operator $\hat Q$ using eqn.~\ref{fermioncomm}, the $\tau$ function becomes
\bea
\tau_l[t]&=& e^{- \sum_{1}^\infty n t_n t_{-n}} \sum_{N=0}^\infty \frac{ e^{\beta \mu N}}{N!} \oint  \prod_{n=1}^N \frac{d z_n \, }{2 \pi i} e^{\sum_{m \neq 0} (q^{-m/2} - q^{m/2}) t_m z_n^m}\, \times \,  \nn \\
&\times& \langle l |\prod_{n=1}^N \psi(q^{\half} z_n) \psi^*(q^{-\half} z_n) z_n^{-\hat Q} | l \rangle \nn \\
&=& e^{- \sum_{1}^\infty n t_n t_{-n}} \sum_{N=0}^\infty \frac{q^{l N} e^{\beta \mu N}}{N!} \oint  \prod_{n=1}^N \frac{d z_n \, z_n^{-l}}{2 \pi i} e^{\sum_{m \neq 0} \tilde{t}_m z_n^m} \det_{i j} \frac{1}{q^{\half} z_i - q^{-\half} z_j} \, \nn \\
&=& e^{- \sum_{1}^\infty n t_n t_{-n}} \mathcal{Z}_G (t, \mu+i l ; l \equiv k ) \, , \nn \\
\eea
where we related the correct Toda times  $t_m = \frac{\tilde{t}_m}{q^{-m/2}- q^{m/2}}\,$ in terms of the apparent $\tilde{t}$ as in the main text. The chemical potential is related to a shift to the charge $l$, that is real for the usual oscillator but imaginary for the case of the inverted oscillator ($q= e^{i \beta}$). Notice also that this formula proves that the Chern-Simons level $k$ can be identified with the charge $l$ of the $\tau$ function and therefore allows to define the complex string coupling $g_{str}^{-1} = \mu + i k$.

\subsection{Hirota equations}

The $\tau$ function satisfies an infinite set of difference-differential equations known as Hirota equations. We now give the main formulas for their derivation.
\\
\\
We define the Casimir operator for the diagonal subgroup of $GL(\infty) \times GL(\infty)$
\be
\label{casimir}
\mathcal{C} = \sum_{r \in \mathbb{Z}+\half} \psi^*_r \otimes \psi_r = \oint \frac{dz}{2 \pi i} \psi^*(z) \otimes \psi(z)\, , \quad \mathcal{C} ({\bf G} \otimes {\bf G}) = ({\bf G} \otimes {\bf G}) \mathcal{C}\, .
\ee
Note also the fact that $\psi_r | l \rangle \otimes \psi^*_r | l' \rangle\ = 0$ for $ l>l'$. Now one multiplies the second equation of~\ref{casimir} by
\bea
\langle l+1 | e^{J_+[t]} \otimes \langle l'-1| e^{J_+[t']} \left[ \mathcal{C} ({\bf G} \otimes {\bf G}) \right] e^{- J_-[t]} | l \rangle \otimes e^{- J_-[t']} | l' \rangle\, = \nn \\
\langle l+1 | e^{J_+[t]} \otimes \langle l'-1| e^{J_+[t']} \left[({\bf G} \otimes {\bf G}) \mathcal{C}  \right] e^{- J_-[t]} | l \rangle \otimes e^{- J_-[t']} | l' \rangle\,
\eea
and commute all the fermion operators until they hit the vacuum.
To perform the commutations, we define the sequences 
\be
\tilde \zeta_+ = (...,0,z^{-1},z^{-2}/2, z^{-3}/3 , ...)\, , \quad \tilde \zeta_- = (...,z^{3}/3,z^{2}/2, z, 0, ...)
\ee
With the bosonization formulas one can prove
\bea
\label{fermionaction}
\psi^*(z) | l \rangle &=& z^l e^{J_-[\tilde \zeta_-]} | l+1 \rangle\, , \quad \langle l+1 | \psi^*(z) = \langle l | z^{l} e^{-J_+[\tilde \zeta_+]}\, , \nn \\
\psi(z) | l \rangle &=& z^{-l} e^{-J_-[\tilde \zeta_-]} | l-1 \rangle\, , \quad \langle l-1 | \psi(z) = \langle l | z^{-l} e^{J_+[\tilde \zeta_+]}
\eea
One also finds the following commutation relations
\bea
\label{fermioncomm}
e^{- J_\pm [t]} \psi^*(z) e^{J_\pm [t]} &=& \exp \left( - \sum_{n>0} t_{\pm n} z^{\pm n} \right) \psi^*(z), \nn \\
e^{ J_\pm [t]} \psi(z) e^{-J_\pm [t]} &=& \exp \left( - \sum_{n>0} t_{\pm n} z^{\pm n} \right) \psi(z)\, , \nn \\
e^{- s \hat Q} \psi^*(z) e^{s \hat Q} &=& e^{-s} \psi^*(z), \nn \\
e^{s \hat Q} \psi(z) e^{- s \hat Q} &=& e^{- s} \psi(z)\, , 
\eea
which can be thought of as the ``time" evolutions of the functions $\psi(z)\, , \psi^*(z)$.
\\
Using these, one finds that
\bea
\langle l+1 |e^{J_+[t]}  {\bf G} \psi^*(z) e^{-J_-[t]} | l \rangle &=& z^l \exp \left(\sum_{n<0} t_n z^n \right) \langle l+1 |e^{J_+[t - \tilde \z_+]} {\bf G} e^{-J_-[t]} | l+1 \rangle \, , \nn \\
\langle l+1 |e^{J_+[t]} \psi^*(z) {\bf G} e^{-J_-[t]} | l \rangle &=& z^l \exp \left(\sum_{n>0} t_n z^n \right)\langle l |e^{J_+[t - \tilde \z_+]} {\bf G} e^{-J_-[t]} | l \rangle \, , 
\eea
Combining them with the analogous equations for $\psi(z)$ one finds the following Hirota equations
\bea\label{Hirotaeqns}
\oint_{C_\infty} dz z^{l-l'} \exp \left(\sum_{k>0} (t_k - t'_k) z^k \right) \tau_l[t-\tilde \z_+] \tau_{l'} [t' + \tilde \z_+] = \nn \\
\oint_{C_0} dz z^{l-l'} \exp \left(\sum_{k<0} (t_k - t'_k) z^k \right) \tau_{l+1}[t-\tilde \z_-] \tau_{l'-1} [t' + \tilde \z_-] \, ,
\eea 
where the contour $C_0$ encircles all the singularities of $z^{l-l'} \exp \left(\sum_{k<0} (t_k - t'_k) z^k \right) $ and similarly the contour $C_\infty$ those coming from $z^{l-l'} \exp \left(\sum_{k>0} (t_k - t'_k) z^k \right)$. In particular when a finite number of $t$'s are turned on, one should enclose the essential singularities either at $\infty$ or $0$. Such cases lead to differential equations. In the case that the essential singularity splits into poles, we need to take the residues around these poles and this will result in discrete equations. As a general rule, the differential/difference equations that the $\tau$ function obeys are uniquely determined from the explicit form of the measure, or in other words from the evolution of the fermionic operators $\psi(z)$.
\\
\\
One can also write Hirota's equations in a more explicit form (essential singularity case) by introducing the Schur polynomials $s_j$
\be
\sum_{k=0}^\infty s_k[t] z^k = \exp \left(\sum_{n=1}^\infty t_n z^n \right)
\ee
and Hirota's bilinear operators
\be
D_n f[t] \cdot g[t] = \partial_x f(t_n +x) g(t_n - x) |_{x=0}
\ee
Using the notation
\bea
\tilde D_\pm &=& (D_\pm , D_{\pm 2}/2 , D_{\pm 3}/3, ....)\, , \nn \\
y_\pm &=& (y_\pm , y_{\pm 2}, y_{\pm 3}, ....)\, , \nn \\
y_n &=& \half (t'_n - t_n)\, ,
\eea
one can rewrite Hirota's equations as a hierarchy of partial differential equations
\bea
\sum_{j=0}^\infty s_{j+i}(-2 y_+) s_j(\tilde D_+) \exp \left(\sum_{k \neq 0} y_k D_k \right) \tau_{l+i+1}[t] \cdot \tau_l[t] = \nn \\
\sum_{j=0}^\infty s_{j-i}(-2 y_-) s_j(\tilde D_-) \exp \left(\sum_{k \neq 0} y_k D_k \right) \tau_{l+i}[t] \cdot \tau_{l+1}[t]
\eea
The Hirota equations can also be thought of as equations for the correlators of operators generating the Toda flows. In this form they make practical sense if a finite number of Toda times are turned on as we discussed.
\\

As an example the first Toda equation of the hierarchy is obtained by setting $i=-1$ and extracting the coefficient in front of $y_{-1}$ to get
\be\label{Todaequation}
\half D_1 D_{-1} \tau_l \cdot \tau_{l} + \tau_{l+1} \tau_{l-1} = 0 \, \Rightarrow \quad \tau_l \frac{\partial^2 \tau_l}{\partial t_1 \partial t_{-1}} - \frac{\partial \tau_l}{\partial t_1 }  \frac{\partial \tau_l}{\partial t_{-1} } + \tau_{l+1} \tau_{l-1} = 0\, .
\ee
This corresponds to having only the two first Toda-times turned on. 
\\
\\
As we have seen in the main text we are also interested for the general case which corresponds to the case of (anti)-fundamentals prior taking the double scaling limit. In this case we have an evolution of the form $\psi(z) \rightarrow z^{-k} (1-a z)^{-m}(1-b/z)^{-n} \psi(z)$ and in eqn.~\ref{Hirotaeqns} we need to pick the residues at $b, 1/a, 0$. This specific case is also discussed also in~\cite{Date:1982gj}. The resulting discrete equation is
\bea\label{differenceequation}
\tau_{(l,m,n+1)}\tau_{(l,m+1,n)} - \tau_{(l,m+1,n+1)}\tau_{(l,m,n)} \, \, \nn \\
+ a b\left(\tau_{(l,m+1,n+1)}\tau_{(l,m,n)} - \tau_{(l-1,m+1,n)}\tau_{(l+1,m,n+1)} \right) = 0 \, , \nn \\
\frac{u(l,m,n)}{u(l,m+1,n)}-\frac{u(l,m,n+1)}{u(l,m+1,n+1)} + a b \left(\frac{u(l-1,m+1,n)}{u(l,m+1,n+1)} - \frac{u(l,m,n)}{u(l+1,m,n+1)} \right)=0\, . \nn \\
\eea
with $u(l,m,n)= \tau_{(l,m,n)}/\tau_{(l+1,m,n)}\,$. This is the simplest difference equation in this category of deformations and relates the partition function for different values of $N_f$ and complex string coupling $g_{str}^{-1}$ and a specific version of the general Hirota's DAGTE~\cite{Hirota:1981}. One can again take a continuous limit that sends eqn.~\ref{differenceequation} to eqn.~\ref{Todaequation}, see~\cite{Date:1982gj}. 

%\subsection{Dispersionless limit}

%Given the discrete equation

%\end{itemize}

\section{Kernel in the energy basis}\label{energybasiskernel}
It will be useful to study also the integral kernels in the energy basis. This is interesting, since  the information about the spectrum of the theory should be in principle extractable from such a representation. A paper where kernels with a similar structure are studied is~\cite{Borodin} and a detailed review can be found in the lectures by van Moerbecke in~\cite{Harnad}. The context of those studies is \emph{``Probability on Partitions and the Plancherel measure''}.

In our context, one can understand the energy basis from the basis of polynomials on the circle. It is a discrete basis where the energy levels are conjugate variables to the $\theta$'s. In order to change basis one needs to use $\langle n| \theta \rangle= e^{-i n \theta}$ or $\langle n| z \rangle = z^{-n}, \, \, z=e^{i \theta}$. Using the resolution of the identity $\hat I = \oint \frac{d z}{2 \pi i z} |z\rangle \langle z|$, a generic vortex perturbed Kernel can be written in the energy basis as:
\bea
\langle m| \hat K |n \rangle  & = &  \oint \frac{ d z }{2 \pi i} \oint \frac{ d z' }{2 \pi i} \langle m| z \rangle \langle z| \hat K |z' \rangle \langle z' | n \rangle  \\ \nn
& = & - \oint \frac{ d z}{2 \pi i} \oint \frac{ d z'}{2 \pi i} z^{-m-1} \frac{e^{u(z)+ u(z')}}{q^{\half} z - q^{-\half} z'} {z'}^n \, ,
\eea
with the most general $u(z)=-\frac{k}{2} \log z + \half  \sum_{l} {\tilde{t}}_l z^{l}$ ($k$ is the Chern-Simons coupling). One can expand the denominator in powers of $q$ and perform the integrals term by term
\be
\langle m| \hat K |n \rangle   =\sum_{l=0}^\infty q^{l + \half} \oint \frac{ d z }{2 \pi i} z^{l-m-1} e^{u(z)}  \oint \frac{ d z' }{2 \pi i}  {z'}^{n-l-1} e^{u(z')}\, = \sum_{l=0}^\infty q^{l + \half} I_{m-l} I_{l - n} \, ,
\ee
with the basic integral $I_l(t)=\oint \frac{ d z }{2 \pi i} z^{-l-1} e^{u(z)} $.

The simplest case is when $u(z)=0$ i.e. no vortex perturbations. Then one finds the result $\langle m| \hat K |n \rangle  =\delta_{m n} q^{m+\half}$, the harmonic oscillator propagator in the energy basis.
Turning on just the Chern-Simons coupling $k$, one finds 
\be
\langle m| \hat K |n \rangle  = \sum_{l=0}^\infty q^{l+\half} \delta_{l-m-1-\frac{k}{2}, 0} \delta_{n-l-1-\frac{k}{2}, 0} = q^{n- \frac{k+1}{2}} \delta_{n-m-k,0}\, ,
\ee
the kernel is no longer diagonal in the energy basis and acts more like a lowering/raising operator depending on the sign of $k$ (one should remember that it is incosistent to just add the C-S term).

To cover the deformations caused by the presence of the fermions, in a case where $e^{u(z)}= (1+ \xi z)^b (1+ \xi z^{-1})^{b'}, \, \, |\xi|<1$ and $l \geq 0$ we can find~\cite{Borodin}
\bea
I_{l} (\xi , b, b') &=& \frac{(-b)_l}{l!} (-\xi)^l (1-\xi^2)^{-b'} {_2F_1} (1+b,b'; l+1; \frac{\xi^2}{\xi^2-1}) \nn  \\
&=& \frac{(-b)_l}{l!} (-\xi)^l  {_2F_1} (l-b,b'; l+1; \xi^2) \nn  \\
I_{-l} (\xi , b, b') &=& \frac{(-b')_l}{l!} (-\xi)^l (1-\xi^2)^{-b} {_2F_1} (b,1+b'; l+1; \frac{\xi^2}{\xi^2-1}) \nn \\
&=& \frac{(-b')_l}{l!} (-\xi)^l  {_2F_1} (b,l-b'; l+1; \xi^2)\, ,
\eea
with $(a)_x= \Gamma(x+a)/\Gamma(x)$ the rising factorial and we are interested in particular for the symmetric case with ($b=b'=N_f/2 \, , \xi=e^{- \beta m}$). Fundamental bosons can be treated in a similar fashion, just inverting the signs of $\xi , b , b'$.
In the case of the 2d black hole $u(z)= \half t (z+z^{-1})$, one can use the generating function for Bessel functions to find
\be
I_{l}(t) = i^{-l} J_{-l}(-i t)
\ee
with $J$ the Bessel-J function. Upon setting $\xi=t/N_f$ and taking the limit $N_f \rightarrow \infty$, we can obtain the Bessel-kernel from the Hypergeometric one. The double scaling limit here is just the limit that connects Hypergeometric with Bessel functions. 
\\
\\
To connect with the work of~\cite{Borodin}, one instead defines the rescaled variables $\z= q^{1/2} e^{i \theta}, w= q^{-1/2} e^{i \theta'}$, %Moreover one passes to the natural variables as explained in the Toda hierarchy appendix~\ref{hierarchies}. This requires the change $\tilde{t}_n \rightarrow t_n (q^{-n/2}-q^{n/2}) $ 
by which
\bea
u(z)\, \rightarrow \, V(\z):&=& \sum_{k>0}(T_k \z^k + S_k \z^{-k})\, , \quad T_k =\half q^{-\half k} \tilde{t}_k \, , \quad S_k =  \half q^{\half k} \tilde{t}_k \, \nn \\
u(z')\, \rightarrow \, \tilde{V}(w):&=& \sum_{k>0}(S_k w^k + T_k w^{-k})  \, . 
\eea
It is easy then to see that the kernel  $\langle m| \hat K |n \rangle$ is written as
\bea
\langle m| \hat K |n \rangle &=& -q^{\frac{m+n+1}{2}}\left(\frac{1}{2\pi i}\right)^2
 \oint_{|\z|=q^{\half}<1}
 \oint_{|w|=q^{-\half}>1}
  \frac{d\z~
dw}{\z^{m+1}w^{-n}} \frac{e^{V(\z) + \tilde V(w)}}{\z-w}\, ,
 \no\\  \no\\
&=& -\frac{q^{\frac{m+n+1}{2}}}{m-n}\left(\frac{1}{2\pi
i}\right)^2\oint\!\!\oint _{{|\z|<1}\atop
{|w|>1}}\frac{d\z~dw}{\z^{m+1}
w^{-n}}\frac{\z\frac{d}{d\z}V(\z)+ w\frac{d}{dw}\tilde V(w)}{\z-w}
 ~e^{V(\z)+V(w)}\, ,
\no\\
&& \hspace*{9cm} \mbox{with $m\neq n$}\, .
 \no\\
 \label{kernel-non-diagonal}%
\eea
The second line of \ref{kernel-non-diagonal}, is derived by rescaling $z\rightarrow t \z , \, w \rightarrow t w$ and taking derivatives with respect to t. The above integrals are around the annular region with the width dictated by $\beta$, or upon considering the inverse oscillator where $q=e^{i \beta}$, one should consider a small regulator $ \epsilon$ such that $|q|<1$. Unfortunately in contrast with the cases considered in~\cite{Borodin} where $T_k = - S_k$, our kernel does not take the form of a Christoffel-Darboux kernel. Nevertheless, it might be useful to study further the kernel in this basis to extract information about the spectrum of the theory.

\section{Orthonormal Mathieu}\label{OrthoMathieu}

We fix here our conventions for the orthonormal Mathieu functions. Starting with the GWW action 
\be
S_{GWW} = \frac{1}{g^2} \left( \tr U + \tr U^\dagger \right) \, .
\ee
The single particle Hamiltonian is given by
\be
- 2 g^2 \frac{\partial^2 \psi(\theta)}{\partial \theta^2} + \frac{2}{g^2} \left( 1 -\cos \theta \right) \psi(\theta) = E \psi(\theta)
\ee
Our orthonormal Mathieu functions with (anti)-periodic boundary conditions are eigenfunctions of this equation.

\section{Airy integrals}\label{Airyformulae}

Here we collect some useful integrals of Airy functions found in the main text. First an integral expression for the Airy kernel
\be
\rho_{Airy}(x_1, x_2) = \int_0^\infty Ai(x_1 + s) Ai(x_2 + s) d s \, .
\ee
A fourier integral:
\be
\int_{-\infty}^\infty d s e^{i 2 P s} Ai(x-s)Ai(x+s) \, = \, 2^{-1/3}Ai(2^{2/3} (x+P^2))
\ee
More generally one finds the following integral:
\be
\frac{1}{|A B|} \int_{-\infty}^\infty Ai(\frac{x + a}{A}) Ai(\frac{x+ b}{B}) e^{i \lambda x} d x  \,= \, 
\ee
$$
\, = \, \frac{e^{	- i \frac{\lambda}{B^3 - A^3} (\lambda^2 A^3 B^3/3 + a B^3 - b A^3)}}{(B^3 -A^3)^{1/3}}  Ai \left(\frac{b-a - (\lambda^2 A^3 B^3)/(B^3-A^3)}{(B^3 -A^3)^{1/3}} \right) \, ,
$$
which for $B=A$ becomes
\be
I_\lambda = \frac{1}{2 \sqrt{\pi}|\lambda|^{1/2}|A|^{2/3}} e^{- i \left(\frac{A^3 \lambda^3}{12} - \frac{(a-b)^2}{4 A^3 \lambda} + \frac{\lambda(a+b)}{2} + \frac{\pi}{4} sgn(A \lambda) \right)}
\ee

\end{document}